\documentclass[review,numbers,sort&compress]{elsarticle}

\usepackage[utf8]{inputenc}
\usepackage{amsmath}
\usepackage{amsfonts}
\usepackage{enumitem}
\usepackage{etoolbox}

\usepackage{setspace}

\usepackage{lineno,hyperref}
\usepackage[letterpaper,%
            left=1in,right=1.75in,top=1in,bottom=1in,%
            footskip=.25in]{geometry}

\journal{Elsevier}

\usepackage{subfig}
\usepackage{caption}
\usepackage{graphicx}
\usepackage{todonotes}
\usepackage{nomencl}
\usepackage{bm}
\usepackage[final]{changes}
\usepackage{xcolor}
\usepackage{etoolbox}
\usepackage{comment}
\graphicspath{{./}}
\definechangesauthor[name={Reviewer 1}, color = blue]{R1}
\definechangesauthor[name={Reviewer 2}, color = red]{R2}
\definechangesauthor[name={All reviewers}, color = purple]{All}
\definechangesauthor[name={Editor}, color = brown]{Editor}
\definechangesauthor[name={Xinlei Zhang}, color = olive]{Author}
\definechangesauthor[name={Heng Xiao}, color = red]{hx}

\renewcommand\nomgroup[1]{%
  \item[\itshape%\bfseries
  \ifstrequal{#1}{A}{Symbols}{%
  \ifstrequal{#1}{B}{Roman Letters}{%
  \ifstrequal{#1}{C}{Greek Letters}{%
  \ifstrequal{#1}{D}{Abbreviations}{}}}}%
]}

\newcommand{\sx}{\mathsf{x}} %
\newcommand{\oy}{\mathsf{y}} %

\begin{document}

\begin{frontmatter}
\title{Assimilation of disparate data for enhanced reconstruction of turbulent mean flows}

  \author[cas,ucas]{Xin-Lei Zhang}
  \author[vt]{Heng Xiao} 
  \author[cas,ucas]{Guo-Wei He\corref{mycor}}
  \cortext[mycor]{Corresponding author}
  \ead{hgw@lnm.imech.ac.cn}
  \author[cas,ucas]{Shi-Zhao Wang}

  \address[cas]{The State Key Laboratory of Nonlinear Mechanics, Institute of Mechanics, Chinese Academy of Sciences,
Beijing 100190, China}
  \address[ucas]{School of Engineering Sciences, University of Chinese Academy of Sciences, Beijing 100049, China}
  \address[vt]{Kevin T. Crofton Department of Aerospace and Ocean Engineering, Virginia Tech, Blacksburg, VA 24060, USA}

\begin{abstract}
    Reconstruction of turbulent flow based on data assimilation methods is of significant importance for improving the estimation of flow characteristics by incorporating limited observations.
    Existing works mainly focus on using only one observation data source, e.g., velocity, wall pressure, lift or drag force, to reconstruct the flow.
    In practical applications observations are disparate data sources that often vary in dimension and quality.
    Simultaneously incorporating these disparate data is worth investigation to improve the flow reconstruction.
    In this work, we investigate the disparate data assimilation with ensemble methods to enhance the reconstruction of turbulent mean flows.
    Specifically, a regularized ensemble Kalman method is employed to incorporate the observation of velocity and different sources of wall quantities (e.g., wall shear stress, wall pressure distribution, lift and drag force).
    Three numerical examples are used to demonstrate the capability of the proposed framework for assimilating disparate observation data.
    The first two cases, i.e., a one-dimensional planar channel flow and a two-dimensional transitional flow over plate, are used to incorporate both the sparse velocity and wall friction.
    In the third case of the flow over periodic hills,
    the wall pressure distribution and the lift and drag force are regarded as observation in addition to velocity, to recover the flow fields.
    The results demonstrate the merits of incorporating various disparate data sources to improve the accuracy of the flow-field estimation.
    The ensemble-based method can assimilate disparate data non-intrusively and robustly without requiring significant changes to the model simulation codes.
    The method demonstrated here opens up possibilities for assimilating realistic experimental data, which are often disparate.
\end{abstract}

\end{frontmatter}

% \linenumbers

\section{Introduction}

\subsection{Turbulent flow reconstruction}
Estimation of turbulent flow field is an important yet challenging subject for both academic and industrial investigations.
Computational fluid dynamics (CFD) simulations and experimental measurements are primary approaches to estimate the states of turbulent flows.
However, CFD simulations have to balance the trade-off between computational cost and predictive accuracy.
For example, the low fidelity methods such as Reynolds-averaged Navier-Stokes (RANS) simulation can provide fast but inaccurate prediction, while high fidelity simulations such as large eddy simulations and direct numerical simulations can make satisfactory predictions but at prohibitive computational costs~\cite{witherden2017future,xiao2019quantification}.
On the other hand, the experimental measurements face the challenges of limited view domain, measurement noises, and insufficient resolution~\cite{sciacchitano2013piv,wang2018error}.
Due to these limitations, it is appealing to combining computational models and experimental measurements of various techniques for reconstructing turbulent flows.

Reconstruction of turbulent flows essentially involves the minimization of the discrepancy between model prediction and observation data.
The objective function~$J$ can be written as
\begin{equation}
     J = \parallel \oy - \mathcal{H}[\sx] \parallel \text{,}
    \label{eq:cost}
\end{equation}
where $\sx$ is the state (e.g., model parameters), $\oy$ is the measurement data,~$\mathcal{H}[\sx]$ can be surrogate models (e.g., linear regression model or neural network) or physics-based models (e.g., RANS model),~and $\parallel \cdot \parallel$ indicates a norm in a Hilbert space.
The experimental data~$\oy$ is often sparse, only providing limited information of the flow field. 
On the other hand, the model can predict the flow field in the full computational domain but likely with large discrepancies.
The reconstruction of turbulent flow aims to complement the low fidelity model with experimental data.
Specifically, based on the experimental data, we can infer optimal state~$\sx$ or model operator~$\mathcal{H}$ by minimizing the cost function~\eqref{eq:cost}.
Further, we can reconstruct the flow in the entire computational domain with the inferred state~$\sx$ or operator~$\mathcal{H}$.
Different strategies have been explored for reconstructing turbulent flows from data, including sparse representation~\cite{callaham2019robust}, machine learning~\cite{brunton2020machine}, and data assimilation~\cite{xiao2019quantification}.

The sparse representation method \replaced[id=R2]{assumes that a dynamic flow evolution can be represented based on a reduced-order model such as proper orthogonal decomposition~(POD) by promoting the sparsity.}{assumes that the observed field can be represented by a reduced-order model.} Specifically, the computational model~$\mathcal{H}[\sx]$ is reformulated as a linear combination of basis functions~$\bm{\Phi}$, e.g.,~$\mathcal{H}[\sx] = \bm{\Phi}^\top \bm{\omega}$, where~$\bm{\Phi}$ is \replaced[id=Author]{POD}{proper orthogonal decomposition~(POD)} modes and $\bm{\omega}$ is the mode coefficients.
This method needs to build a library of modes and infers optimal mode coefficients~$\bm{\omega}$ by reducing the data misfit and enforcing sparsity constraint simultaneously.
The merits of the sparse representation method have been demonstrated for the reconstruction of some canonical flows, e.g., the vortex shedding in cylinder wakes and the mixing layer at low Reynolds numbers~\cite{callaham2019robust}.
However, the method requests a large library that must contain a sufficiently extensive collection of example flow fields, and it is still not clear how to build the suitable library for various types of flows.

Machine learning techniques have developed rapidly in the past few years and are also increasingly used to address challenges in fluid mechanics, including flow reconstruction~\cite{milano2002neural, fukami2018super, guastoni2020prediction}.
Several machine learning techniques have been used in reconstructing turbulent flows, e.g., random forest~\cite{2017Physics}, neural networks~\cite{milano2002neural}, and sparse regression~\cite{2019Discovery}.
Here we take neural network-based machine learning as an example.
The neural network can be considered as a surrogate model~$\mathcal{H}[\sx]$, where $\sx$ is the input features.
To be used for the reconstruction of turbulent flows, this method trains a model operator/mapping (i.e., $\mathcal{H}$ in Eq.~\eqref{eq:cost}) based on reference data by inferring optimal weights of each layer in neutral network.
The trained model can be further used to map other available low-resolution data to high-resolution data~\cite{fukami2018super, wu2020enforcing}.
In contrast to the sparse representation, the neural network can represent turbulent flows more flexibly.
However, it usually needs a large set of data to train the functional mapping.
The high fidelity data is not straightforward to acquire, particularly in CFD applications at high Reynolds numbers.
Moreover, the trained model is often very sensitive to data with weak generalization ability of extrapolating to new cases.

Data assimilation is another data-driven method that combines physical model and limited observation data to reconstruct turbulent flows.
\added[id=R2]{
It can be also applied for the reduced-order dynamic systems to improve the computational efficiency~\cite{artana2012strong}, but here we focus on the conventional data assimilation method based on the physical model.
}
In the objective function defined in Eq.~\eqref{eq:cost}, the data assimilation framework regards $\mathcal{H}[\sx]$ as the physical model, e.g., the RANS model, \replaced[id=R2]{where $\mathcal{H}$ represents a composition of the model operator and the observation operator, and $\sx$ represents the flow state that depends on the initial condition and}{and $\sx$ as} the uncertain terms in the model, e.g., the Reynolds stress or eddy viscosity.
The flow-field reconstruction is to reduce the discrepancy between model prediction~$\mathcal{H}[\sx]$ and observation data~$\oy$ by inferring optimal~$\sx$.
\added[id=R2]{The cost function in Eq.~\eqref{eq:cost} is often ill-posed, and hence a background term, e.g., $\parallel \sx - \sx^0 \parallel_{\mathsf{P}^{-1}}$, is added in the cost function to penalize the departure from the initial condition~$\sx^0$ in the data assimilation method. The term is weighed by a given covariance matrix~$\mathsf{P}$ to impose the spatial correlation of the solution.}
In contrast to the sparse representation, data assimilation is more flexible in representing  turbulent flow fields, since it incorporates the physical model instead of the reduced-order model, i.e., the linear combination of basis functions.
Compared to machine learning, data assimilation can use very limited observation data to reconstruct the flows in the entire computational domain and ensure the reconstructed flow states conform to physical models.
In other words, the data assimilation integrates the physical model and the data so that the physical model can reduce the data requirement and the data can reduce the model uncertainties.
For this reason, the data assimilation method has emerged as a practical tool for flow-field reconstruction.
As such, here we review in more detail recently developed data assimilation methods for the flow reconstruction.

\subsection{Data assimilation for flow reconstruction}
Data assimilation is widely used for state estimation of chaotic systems such as ocean and atmosphere.
It can be categorized into adjoint-based methods (e.g., variational data assimilation method~\cite{cummings2013variational}) and ensemble-based methods (e.g., ensemble Kalman method~\cite{evensen2009data} \added[id=R2]{and ensemble-based variational method~\cite{yang2015enhanced}}), depending on how the cost function is minimized.
The adjoint-based methods search for the optimal solution based on the derivative of cost function that is estimated through solving an adjoint equation.
This method \added[id=R2]{enables the recovering of smaller scales than that is possible with the ensemble-based methods using limited sample sizes~\cite{yang2015enhanced} and} has been used for flow reconstruction~\cite{foures2014data, kato2015data, mons2016reconstruction, symon2017data, he2018data, chandramouli20204d}.
However, the adjoint method needs intrusive modification of the CFD solver, leading to a very time-consuming and laborious process for development.
This makes the non-intrusive methods, e.g., ensemble-based data assimilation method, very appealing for the reconstruction of turbulent flows.

Ensemble Kalman filtering (EnKF)~\cite{evensen2003ensemble} is a widely used ensemble-based data assimilation method, which is developed from Kalman filter~\cite{welch1995introduction}.
The Kalman filter is a derivative-free data assimilation method, and the gradient information is obtained by evolving the linearized system dynamics.
Nevertheless,
the standard Kalman filter is computationally prohibitive when used for high-dimensional problems, such as flow-field reconstruction, because it needs to propagate and store in time a high-rank error covariance matrix.
Hence, it usually requires reduced-order techniques to make it practical for turbulence problems~\cite{farrell2001state, meldi2017reduced, meldi2018augmented}.
Ensemble technique can be regarded as one of the reduced-order methods and has been introduced in the Kalman filter~\cite{evensen2003ensemble}.
It leverages the ensemble realizations to estimate the error covariance at each iteration, thus avoiding the storage and propagation of the full rank error covariance as in the Kalman filter.
From this point of view, EnKF inherits the advantage of non-intrusiveness from the Kalman filter and also is feasible to be applied for the high-dimensional problems, e.g., flow reconstruction.
A number of researchers have utilized the EnKF and its variants for the reconstruction of flow fields.
For instance,
Colburn et al.~\cite{colburn2011state} used the EnKF method to estimate the turbulent near-wall flows with only skin friction and pressure at the wall.
Xiao et al.~\cite{xiao2016quantifying} adopted an iterative ensemble Kalman method to estimate the entire flow velocity field in periodic hills and square duct from sparse velocity observations.
Mons et al.~\cite{mons2016reconstruction} investigated various ensemble-based methods, including ensemble Kalman filtering, to assimilate different observations, e.g., the velocity, pressure coefficient, drag and lift coefficient.
They evaluated the performance of different observations for the reconstruction of unsteady viscous flows around a cylinder.
Zhang et al.~\cite{zhang2019bayesian} applied an ensemble-based data assimilation method to construct the flow over a bump by assimilating the observation of the wall friction coefficient.
The recent developments on this topic can be found in
Refs.~\cite{2015A,Jin2016A,Deng2018Recovering, 2020Parameter, liu2020new}.

The works mentioned above use either wall measurements or sparse velocity to reconstruct the flow fields.
\replaced[id=R2]{The}{It was observed that the} wall information could be used to reconstruct the near-wall flow \added[id=R2]{and much more convenient to obtain than PIV measurements.} \replaced[id=R2]{However, they}{but} have limited effects on the flows away from the wall, likely due to the decaying correlation between the local flow and the wall observation with increasing spatial distance~\cite{mons2016reconstruction}.
On the other hand, the sparse velocity observation away from the wall can provide improved local estimation around observed locations but may not affect the near-wall flow.
Hence, the use of all these available information sources is a promising method for improving the performance of flow reconstruction.
In practice, various disparate data sources are available from the experimental measurements with different techniques including, among others, 
\begin{enumerate}[label=({\alph*})]
    \item volume data sources \added[id=R2]{(low data sparsity)}, e.g., the velocity, which can be measured with hot-wire anemometry, laser doppler velocimetry (LDV), and particle image velocimetry (PIV) inside the flow domain~\cite{tropea2007springer},
    \item surface data sources \added[id=R2]{(medium data sparsity)}, e.g., the pressure and wall shear stress distribution along the wall, which can be measured with pressure and micro-pillar shear stress sensors~\cite{grosse2007mean}, respectively, and
    \item integral data sources \added[id=R2]{(high data sparsity)}, e.g., lift and drag force measured with force-moment sensor cell~\cite{hu2008flexible}, and acoustic noise measured from microphones~\cite{glegg2017aeroacoustics}.
\end{enumerate}
All the data enumerated above are measurable from experiments and can be used to improve the accuracy of flow reconstruction.
These data are often heterogeneous in quality and dimension.
It would be of significant interest to assimilate these disparate data sources, thereby enhancing the reconstruction of turbulent flows.
Recently, He and Liu~\cite{he2020time} used the adjoint-based method to assimilate both the velocity and wall pressure.
They combined the POD technique and linear stochastic estimator~(LSE) to correct the velocity measurements based on the pressure signal as a pre-processing, and then they applied the adjoint-based data assimilation method to reconstruct the flow by assimilating the corrected velocity field.
They showed that the mean velocity, wall pressure coefficient, and the normal Reynolds stress could be significantly improved by considering the data of velocity and wall pressure.
However, 
the sensitivity to the choice of the modes and the absence of the small-scale fluctuations after the LSE process need to be addressed to further improve the accuracy of the wall pressures.
The ensemble-based data assimilation method is able to assimilate disparate data without the specific pre-processing and intrusive code modifications due to its non-derivative nature.
Hence, the ensemble method warrants further investigations in CFD applications.

\subsection{Proposed approach and contributions of present work}
A regularized ensemble Kalman filtering (REnKF) method was proposed by Zhang et al.~\cite{zhang2020regularized} to incorporate general regularization terms during the data assimilation process.
This method makes only minor modifications to the standard ensemble Kalman filter, leading to a derivative-free method that is able to enforce regularizations.
Hence, it incorporates the regularizations in the inference without requiring the derivation of adjoint equations.
This method can also be used for assimilating disparate data sources to improve the flow reconstruction.
It is achieved by regarding the discrepancy of model predictions with some data sources as regularizations.
For instance, the sparse velocity data are assimilated as observation, and the wall quantities are used as the regularization.

The present work aims to enhance the reconstruction of turbulent mean flow by assimilating disparate data sources with the REnKF method~\cite{zhang2020regularized}.
Different types of data sources are incorporated in this work, including velocity, wall friction coefficient, wall pressure distribution, and lift and drag force.
In contrast to the adjoint-based method, the REnKF method is able to incorporate disparate data non-intrusively because of its derivative-free nature.
Admittedly, the standard EnKF can also be used to assimilate multiple data sources by embedding all the available data in the observation, but it requires to compute the augmented Kalman gain matrix involving an inverse of a large matrix, which may result in high computational costs. 
The REnKF method can avoid the computation of the augmented Kalman gain matrix by considering some data sources the regularization instead of the observation, thereby improving the efficiency of the assimilation.
In fact, we will show in Section~$2.2$ that the EnKF and REnKF are equivalent in the context of disparate data assimilation.
The proposed disparate data assimilation framework is a useful and convenient tool to combine CFD simulations and multiple experimental data sources for the reconstruction of turbulent flow fields.

The rest of the paper is structured as follows.
Section~$2$ presents the data assimilation framework used for disparate data assimilation and algorithm for practical implementation.
Section~$3$ showcases the capability of the proposed data assimilation framework to enhance the reconstruction of turbulent flows in three different configurations.
Finally, Section~$4$ concludes the paper.

\section{Methodology}
In this section, we present the framework of the ensemble-based data assimilation method to assimilate disparate data sources.
The objective in the context of flow reconstruction is to reduce the data mismatch between CFD predictions and reference data by inferring optimal state~$\sx$ that can be model uncertain parameters or ambiguous boundary conditions.
We consider two disparate data sources~$\oy_1$~(e.g., velocity) and~$\oy_2$~(e.g., wall shear stress), which are different from each other in dimensionality, physical quantity, and measurement quality.
To designate the relationship between the state and the observed quantities, we define the observation models of~$\oy_1$ and~$\oy_2$ as
\begin{equation}
\begin{aligned}
    \oy_1 &= \mathsf{Hx} + \epsilon \text{,} \\
    \oy_2 &= \mathsf{Dx} + \eta \text{,}
\end{aligned}
\end{equation}
respectively.
\added[id=All]{The observation errors of data $\oy_1$ and $\oy_2$ are assumed to be uncorrelated with each other.}
In the equations above,~$\mathsf{H}$ and~$\mathsf{D}$ are the observation operator mapping the state~$\sx$ onto the observation spaces where $\oy_1$ and~$\oy_2$ are in, respectively; $\epsilon$ and~$\eta$ are random observation errors, conforming to a Gaussian distribution of~$\epsilon \sim \mathcal{N}(0, \mathsf{R})$ and~$\eta \sim \mathcal{N}(0, \mathsf{Q})$; $\mathsf{R}$ and $\mathsf{Q}$ are the corresponding error covariances.
It is noted that multiple data sources can be incorporated in the data assimilation framework, and here we only focus on the scenario of two disparate data sources for simplicity.
However, the method can be straightforwardly extended to multiple data sources.

\subsection{Conventional EnKF method}
The ensemble Kalman filtering method is a widely used ensemble-based data assimilation method, where the optimal state can be searched with an explicit analysis scheme. 
The scheme is formulated as
\begin{equation}
    \sx_j^\text{a}  = \sx_j^\text{f} + \mathsf{K} (\oy_j - \mathsf{H} \sx_j^\text{f}) \text{,}
\end{equation}
where~the superscript~$\text{a}$ and~$\text{f}$ represent the analysis and forecast,~$j$~indicates the sample index,~$\mathsf{K} = \mathsf{PH}^\top (\mathsf{H} \mathsf{P} \mathsf{H}^\top + \mathsf{R})^{-1}$ is known as Kalman gain matrix, and $\mathsf{P}$ is the model error covariance, which is estimated with ensemble samples~$\mathsf{X} = \{\sx_j\}_{j=1}^M$ through
\begin{equation}
    \mathsf{P} = \frac{1}{M-1} \left( \mathsf{X} -  \bar{\mathsf{X}} \right) \left( \mathsf{X} -  \bar{\mathsf{X}} \right)^\top \text{,}
\end{equation}
where the sample mean~$\bar{\mathsf{X}} = \frac{1}{M} \sum_{j=1}^M \sx_j$.
When used to assimilate the disparate observation data of different physical quantities, the observation is augmented to include both~$\oy_1$ and~$\oy_2$.
Specifically, we can reformulate the observation model as
\begin{equation}
    \oy_\text{aug} = 
    \begin{bmatrix}
    \oy_1\\
    \oy_2
    \end{bmatrix} = \mathsf{H}_\text{aug} \mathsf{x} + \epsilon_\text{aug},
\end{equation}
where $\epsilon_\text{aug}$ is subject to a Gaussian process~$\mathcal{GP}(0, \mathsf{R}_\text{aug})$ with
\begin{equation}
    \mathsf{R}_\text{aug} = 
    \begin{bmatrix}
    \mathsf{R} & 0\\
    0 & \mathsf{Q} 
    \end{bmatrix}\text{, and}
    \quad
    \mathsf{H}_\text{aug} = 
    \begin{bmatrix}
    \mathsf{H} \\
    \mathsf{D} 
    \end{bmatrix} \text{.}
\end{equation}
Accordingly, the update scheme need to be modified to be
\begin{equation}
\begin{aligned}
    \sx^\text{a}_j &= \sx^\text{f}_j + \mathsf{K}_\text{aug}\left(\oy_{(\text{aug},j)} - \mathsf{H}_\text{aug}[\sx^f_j] \right), \\
    \label{eq:aug_EnKF}
    \mathsf{K}_\text{aug} &= \mathsf{PH}_\text{aug}^\top \left(\mathsf{H}_\text{aug} \mathsf{P} \mathsf{H}_\text{aug}^\top + \mathsf{R}_\text{aug}\right)^{-1} \text{,}
\end{aligned}
\end{equation}
where the subscript~$j$ of $\mathsf{y}$ is omitted for brevity in the following context.
Note that the Kalman gain matrix~$\mathsf{K}$ involves the inverse of the matrix~$\mathsf{H} \mathsf{P} \mathsf{H}^\top + \mathsf{R}$ that has the same rank as the dimension of observation space.
Hence, directly embedding disparate data in the observation may cause a daunting computational burden in case of a large volume of data~$\oy_2$
\added[id=R1]{, e.g., for time-resolved PIV data. }
Another approach to assimilate different data sources is performing multiple standard EnKF steps to assimilate each data sequentially.
However, it is also computationally inefficient since it requires to rerun simulation after assimilating each disparate data.

\added[id=All]{
Efficient data assimilation techniques have been investigated in ensemble-based methods to reduce computational costs.
For instance,
the ensemble transform Kalman filter~\cite{bishop2001adaptive} uses the square of ensemble covariance matrix (i.e., the anomaly matrix) to avoid the high computational cost.
Additionally, the localization techniques such as domain localization~\cite{asch2016data} have been employed to handle the large data set through performing local Kalman analysis with local observations.
These approaches have been extensively used in weather forecasting and  in geoscience.
In addition, recently a multi-grid ensemble Kalman filter strategy~\cite{moldovan2020multigrid} has been proposed to facilitate the efficient data assimilation for unsteady flows.
It is achieved by using the ensemble of realizations from low resolution simulations to generate the Kalman correction and then projecting on high-resolution grids to correct the flow state.
In this work we use a regularized ensemble Kalman filtering method to assimilate the disparate data, which is a promising alternative assimilate large sets of data.
}

\subsection{Regularized EnKF framework}

The regularized ensemble Kalman filtering (REnKF) method is proposed to empower the conventional EnKF to incorporate additional regularizations or constraints~\cite{zhang2020regularized}.
EnKF can be derived from the minimization of an objective function, as discussed in the literature~\cite{evensen2018analysis,zhang2020evaluation}.
Compared to the EnKF, the REnKF is derived in a similar manner, i.e., by minimizing an objective function but with a regularization term.
The objective function involves an additional regularization term~$\mathcal{G}[\mathsf{x}]$ as
\begin{equation}
     \mathop{\arg \min}_{\sx} J = \parallel \mathsf{x}_j^\text{a} - \mathsf{x}_j^\text{f} \parallel_{\mathsf{P}^{-1}}^2 + \parallel \mathsf{y}_1 - \mathsf{H}\mathsf{x}_j \parallel_{\mathsf{R}^{-1}}^2 + \parallel \mathcal{G}[\mathsf{x}_j^\text{a}] \parallel_{\mathsf{Q}^{-1}}^2 \text{,}
     \label{eq:cost_renkf}
\end{equation}
where the third term on~$\mathcal{G}[\sx_j]$ is added to represent the regularization in contrast to cost function of EnKF.
\added[id=R2]{The added term can be considered from a Bayesian viewpoint~\cite{zhang2020regularized}.
That is, the posterior distribution is conditioned on the two measurements sequentially, i.e., $P(\mathsf{x}|\mathsf{y}_1, \mathsf{y}_2) \propto P(\mathsf{x})P(\mathsf{y}_1 | \mathsf{x})P(\mathsf{y}_2 | \mathsf{x})$, where $P$ indicates the probabilistic distribution.
On the other hand, the regularization term can be interpreted as a weak constraint strategy in the classical data assimilation framework.
Specifically, the model error is usually imposed in the cost function as the weak constraint to alleviate the ill-posedness of the problem~\cite{chandramouli20204d}.
The added regularization term in the proposed method can be regarded as the weak constraint to penalize the discrepancy from the disparate data~$\oy_2$ instead of the model error.}
By minimizing the cost function~\eqref{eq:cost_renkf}, an explicit update scheme can be formulated as two steps: a pre-correction step and a standard Kalman correction step as
\begin{subequations}
\begin{align}
    \tilde{\sx}_j^\text{f} & = \sx_j^\text{f}  - \mathsf{P} \mathcal{G'}[\sx_j^\text{f}]^\top \mathsf{Q}^{-1} \mathcal{G}[\sx_j^\text{f}] \text{,}
    \label{eq:pre-correction} \\
    \sx_j^\text{a} & = \tilde{\sx}_j^\text{f} + \mathsf{K} (\oy_j - \mathsf{H} \tilde{\sx}_j^\text{f}) \text{.}
    \label{eq:renkm_update_2}
\end{align}
\label{eq:renkf}
\end{subequations}
The readers are referred to Ref.~\cite{zhang2020regularized} for further details.

To use this method for assimilating disparate data, we formulate the misfit between model prediction and extra observation data $\oy_2$ as the regularization term. That is,
\begin{equation}
    \mathcal{G}[\mathsf{x}_j] = \mathsf{Dx}_j - \oy_2 \text{.}
\end{equation}
The REnKF framework requires the derivative of the regularization $\mathcal{G}[\sx]$ with respect to~$\sx$.
However, direct computation of the derivative $\mathcal{G}'[\sx]$ is not straightforward, which usually requires intrusive modifications with adjoint methods.
Here we estimate the sensitivity matrix with the tangent linear operator~$\mathsf{D}$ as the standard EnKF~\cite{evensen2003ensemble}.
The weight~$\mathsf{Q}$ is constructed as a diagonal matrix based on the disparate data noise.
Further, the original pre-correction step~\eqref{eq:pre-correction} can be formulated as
\begin{equation}
\begin{aligned}
    \tilde{\sx}_j^f &= \sx_j^\text{f} - \mathsf{P} \mathcal{G'}[\sx_j^\text{f}]^\top \mathsf{Q}^{-1} \mathcal{G}[\sx_j^\text{f}]  \\
           &= \sx_j^\text{f} - \mathsf{P} \mathsf{D}^\top \mathsf{Q}^{-1} {(\mathsf{Dx}_j^\text{f} - \oy_2)}
    \text{.}
\end{aligned}
\label{eq:reg_step}
\end{equation}
This method only makes a small modification on the conventional EnKF to account for additional observations or regularizations, and it is very straightforward to implement.

Although the motivations of EnKF and REnKF are inherently different, the connection between them for assimilating disparate data is worthy of further discussions.
The conventional EnKF method consider the disparate data~$\oy_2$ as same as the data~$\oy_1$, while the REnKF method regards $\oy_1$ as primary observation and the disparate data~$\oy_2$ as a regularization or a secondary observation usually with larger data noise.
However, it can be derived that the EnKF and REnKF for disparate data assimilation are equivalent with practical implementation. 
The details of derivation are presented in~\ref{sec:Append_A}.
\replaced[id=All]{
In contrast to the conventional EnKF, REnKF can avoid the computation of the large matrix in case of the additional data with high dimension, since it only needs to compute the inverse of a diagonal matrix~$\mathsf{Q}$.
This method can be considered as an alternative method for computationally efficient assimilation of large data set.
}{The merit of REnKF, in contrast to EnKF, is to avoid the computation of the large matrix in case of the additional data with high dimension.
Taking a large matrix of $10^5$ rank for example, the inverse of this matrix is very costly to deal with.
As for the REnKF, it only needs to compute the inverse of a diagonal matrix~$\mathsf{Q}$, where the computational cost is negligible.}
Hence in this work, we aim to demonstrate the assimilation of disparate data sources with the REnKF method to enhance the reconstruction of turbulent mean flows.
It is noted that the ensemble Kalman method is also able to be used for disparate data assimilation.
We run the test cases in this paper with conventional EnKF, and the results agree with what we obtained with REnKF and are thus omitted for brevity.
\added[id=R1]{We provide the comparison between the EnKF and the REnKF in terms of the error of the reconstructed fields in Appendix.~B.
}

\subsection{Implementation}
\label{sec:implementation}
Here we show the practical implementation to apply the REnKF method for assimilating the disparate data sources.
% add the state augmentation
\added[id=R1]{
The state augmentation is employed for the joint state and parameter estimation, i.e., $\sx^{\text{(aug)}} = [\sx, \mathcal{H}(\sx)]$.
Moreover, the iterative ensemble Kalman filter~\cite{iglesias2013ensemble} recasts the steady state inverse problem as artificial dynamic data assimilation problem.
That is, $\sx_{i+1}^\text{(aug)} = [\sx_i^\text{(aug)}, \mathcal{H}(\sx_i^\text{(aug)})]^\top$, where the linear observation operator is given as~$\mathsf{H}=[0, \mathsf{I}]$, and a single update is done at each observation time. Hereafter, the state is taken to the augmented version and thus the superscript ``(aug)'' is omitted for brevity.}

In this work, we focus on assimilating two disparate data sources, but it is noted that the framework can be extended to multiple disparate data assimilation by formulating the pre-correction step~\eqref{eq:pre-correction} as
\begin{equation}
    \tilde{\sx}_j^\text{f} = \sx_j^\text{f} - \sum_{p=1}^n \mathsf{P} \mathcal{G}'_p[\sx_j^\text{f}]^\top\mathsf{Q}^{-1}\mathcal{G}_p[\sx_j^\text{f}],
\end{equation}
where $p$ indicates the index of observation data source.
The sample collapse is a common issue for iterative ensemble Kalman method~\cite{zhang2020evaluation}, i.e., the samples often converge to the sample mean after a few iterations.
This would lead to a very small error covariance~$\mathsf{P}$,
% which can hold the equivalence between the REnKF and EnKF as discussed in subsection~$2.2$.
and the pre-correction step involving the regularization term is not effective.
To avoid the effects of sample collapse and keep the regularization term active, we weigh the pre-correction term with the Frobenius norm of~$\mathsf{P}$ to retain only the direction of $\mathsf{P}$.
That is, we reformulate the weight matrix as
\begin{equation}
    \mathsf{Q}^{-1} = \frac{\chi}{\parallel \mathsf{P} \parallel_F} \bar{\mathsf{Q}}^{-1} \text{,}
\end{equation}
where~$\chi$ is constant parameter and $\bar{\mathsf{Q}}$ is normalized such that its largest diagonal element is~$1$.
\added[id=R2]{
In steady-state scenarios as investigated in this work, the variance of the samples may be underestimated due to the repeated use of the data~\cite{zhang2020evaluation}, and consequently the norm of error covariance~$\parallel \mathsf{P} \parallel_F$ can become very small.
This can cause the filter to diverge.
To avoid the possibly overlarge regularization term, the convergence criteria are introduced based on the discrepancy principle~\cite{schillings2018convergence}.
Specifically, the convergence criteria are set as $\parallel \mathsf{Hx} - \mathsf{y}_1 \parallel \geq 2 \sqrt{\text{trace}(\mathsf{R})}$ and $\parallel \mathsf{Dx} - \mathsf{y}_2 \parallel \geq 2 \sqrt{\text{trace}(\mathsf{Q})}$.}
\added[id=R2]{
As with the ensemble Kalman filter, the REnKF method can diverge  when applied to high dimensional space and non-Gaussian distributions with small ensemble sizes~\cite{evensen2003ensemble}.
}
Additionally, we gradually enhance the strength of the regularization term to ensure the robustness.
Specifically, the parameter~$\chi$ is adjusted dynamically with a ramp-up function as
\begin{equation}
    \chi = 0.5 \left( \tanh \left( \frac{i - S}{d} \right) + 1 \right) \text{,}
\end{equation}
where $i$ denotes the iteration step, the parameters $S$ and $d$ control the slope of the ramp curve and are chosen as $5$ and $2$, respectively, in this work.

Given the disparate data sources~$\oy_1$ and $\oy_2$ with the data error covariance $\mathsf{R}$ for $\oy_1$ and $\mathsf{Q}$ for $\oy_2$ and the prior state~$\sx^0$ with the prior covariance function~$\mathcal{K}$, we can implement the disparate data assimilation framework as follows:
\begin{enumerate}
    \item \textbf{Sampling step}: \\ 
     Draw initial ensemble samples $\{ \mathsf{x}^0 \}_{j=1}^n$ from the Gaussian process~$\mathcal{GP}(\mathsf{x}^\text{0}, \mathcal{K})$ 
    \item \textbf{Prediction step}: \\
    i) For each sample, perform the model prediction in next iteration $\sx^{(i-1)} \rightarrow \sx^{(i)}$ \\
    ii) Compute the sample mean~$\bar{\sx}$ and the model error covariance $\mathsf{P}$ as
    \begin{subequations}
    \begin{align}
    \bar{\mathsf{X}}^{(i)} &= \frac{1}{M} \sum_{j=1}^M \sx_j^{(i)} \\
    \mathsf{P}^{(i)} &= \frac{1}{M-1} \left( \mathsf{X}^{(i)} -  \bar{\mathsf{X}}^{(i)} \right) \left( \mathsf{X}^{(i)} -  \bar{\mathsf{X}}^{(i)} \right)^\top \text{.}
    \end{align}
    \end{subequations}
    \item \textbf{Regularization step}: \\
    Perform the regularization step to assimilate the data~$\oy_2$ and obtain the regularized state~$\tilde{\sx}$ based on
    \begin{subequations}
    \begin{align}
    \delta_j^{(i)} &= - \mathsf{P}^{(i)} \mathsf{D}^\top \mathsf{Q}^{-1} (\mathsf{D} \sx^{(i)} - \oy_2) \\
    \tilde{\sx}_j^{(i)} & = \sx_j^{(i)} + \delta_j^{(i)} \text{.}
    \end{align}
    \end{subequations}
    \item \textbf{Kalman update step}: \\
    Compute the Kalman gain matrix and update the regularized state based on
    \begin{subequations}
    \begin{align}
    \mathsf{K}^{(i)} &= \mathsf{P}^{(i)}\mathsf{H}^\top (\mathsf{H} \mathsf{P}^{(i)} \mathsf{H}^\top + \mathsf{R})^{-1} \\
    \sx_j^{\text{a}, (i)} & = \tilde{\sx}_j^{(i)} + \mathsf{K}^{(i)} (\oy_1 - \mathsf{H} \tilde{\sx}_j^{(i)}) \text{.}
\end{align}
\end{subequations}
    \item Return to step~2 until the ensemble is statistically converged or the maximum iteration number is reached.
\end{enumerate}
The REnKF method is implemented in our software suite DAFI for data assimilation and field inversion~\cite{dafi_github,dafi_rtd,carlos2020dafi}.

\section{Test cases}

In this section, we showcase the superiority of the proposed disparate data assimilation framework for the reconstruction of turbulent flows.
Data assimilation involves the physical model describing the system state. 
For the simulation of turbulent flows particularly at high Reynolds numbers, the  most used physical approach is still the RANS method due to its computational efficiency and tractability.
Hence, we consider the RANS equation involving the turbulent mean flow as the physical model in the data assimilation framework.
It is well known that the RANS method is usually not confident to predict complex turbulent flows in the presence of mean pressure gradient, mainly due to the modeling of closure term, i.e., Reynolds stress.
Based on the linear eddy viscosity model of Reynolds stress, the RANS equation can be formulated as
\begin{equation}
     \begin{subequations}
      \begin{aligned}
        \quad \frac{\partial U_i}{\partial x_i} & = 0  \\
        U_j \frac{\partial U_i}{\partial x_j} & =  -\frac{\partial {p^*}}{\partial x_i} + \frac{\partial}{\partial x_i} \left[ \left(\nu + \nu_\text{t} \right) \left( \frac{\partial U_i}{\partial x_j} + \frac{\partial U_j}{\partial x_i} \right) \right],
      \label{eq:rans-momentum}
      \end{aligned}
    \end{subequations}
\end{equation}
where $U$ is velocity, $i, j$ indicate the spatial direction, $x$ is spatial coordinate, $p^*$ is pressure term, $\nu$ is the fluid viscosity, and $\nu_t$ is the eddy viscosity.
The eddy viscosity can be estimated with various turbulence models, e.g., Spalart-Allramas model~\cite{spalart1992one-equation}, $k$--$\varepsilon$ model~\cite{launder1974application}, and $k$--$\omega$ model~\cite{wilcox1998turbulence}.
However, there is still no universal model that is able to make accurate predictions in all flow conditions.
For this reason, we consider the uncertainty within the eddy viscosity.
The turbulent flow fields, e.g., velocity or pressure, can be recovered by inferring the optimal eddy viscosity with the proposed data assimilation method.

To represent the uncertainty within the eddy viscosity, we assume the prior~\replaced[id=R2]{$\log \nu_t$}{$\nu_t$} conforms to a Gaussian process as
\begin{equation}
    \log \nu_t \sim \mathcal{GP}(\log \nu_t^0, \mathcal{K}) \text{.}
\end{equation}
where $\nu_t^0$ is the prior mean and $\mathcal{K}$ is the kernel function.
The logarithm ensures the non-negativity of $\nu_t$.
A Gaussian kernel~$\mathcal{K}$ is used in this work as
\begin{equation}
    \mathcal{K}(x, x') = \sigma^2 \exp \left(-\frac{(x - x')^2}{l^2} \right) ,
    \label{eq:kernel}
\end{equation}
where $x$ and $x'$ indicate two different spatial locations, $\sigma$ represents variance, and $l$ is the correlation length scale.
With this kernel, we can generate ensemble realizations based on Karhunen–Lo\`eve~(KL) expansion~\cite{le2010spectral} to guarantee the smoothness of samples.
Specifically, we build KL modes with~$\phi_i = \sqrt{\lambda} \hat{\phi}_i$ where $\lambda$ and $\hat{\phi}$ are the eigenvalue and eigenvector of~$\mathcal{K}$.
Further, we truncate the modes~$\{\phi_i\}_{i=1}^n$ to cover more than~$99\%$ variance and draw the random coefficients~$\{\omega_i\}_{i=1}^n$ from normal distribution~$\mathcal{N}(0, 1)$.
With the KL modes~$\bm{\phi}$ and KL coefficient~$\bm{\omega}$, the eddy viscosity field can be constructed based on
\begin{equation}
    \log \nu_t = \log \nu_t^0 + \sum_{i=1}^n \phi_i \omega_i \text{.}
\end{equation}
\added[id=R1]{
Thus, in this work the coefficients~$\bm{\omega}$ are the uncertain model parameters used to reconstruct the eddy viscosity field $\nu_t(\bm{x})$. 
Further the proposed REnKF framework is used to infer the optimized coefficients~$\bm{\omega}$ through incorporating the observation data as illustrated in Section~\ref{sec:implementation}.}

The near-wall flow is of great importance to account for the shear and related local turbulence production as well as the energy dissipation~\cite{jimenez2013near}.
However,
the near-wall region of wall-bounded flows is usually challenging to be measured accurately due to the decreasingly small size of this region at higher Reynolds number~\cite{mathis2011predictive}.
On the other hand, the near-wall flow is significantly affected by the wall where the information is measurable with different techniques.
For instance, the pressure distribution along the wall can be measured with wall tapping or static tube~\cite{tropea2007springer}, the force can be measured with force balance instrument~\cite{tropea2007springer},
and wall shear stress can be measured with pillar sensors~\cite{grosse2007mean} or micro PIV~\cite{kahler2006wall}.
Moreover, the velocity away from the wall is relatively straightforward to be measured along straight lines with the planar PIV or at sparse locations with the LDV technique.
The local velocity data can improve the model prediction around the observed positions,
while the wall measurements are able to enhance the reconstruction of near-wall flows.
Therefore, it is promising to enhance the reconstruction of turbulent mean flows by combining these different disparate data sources with the RANS equation.
\added[id=R2]{It is not straightforward to use 3D PIV data for 3D flow reconstructions. Chandramouli et al.~\cite{chandramouli20204d} recently attempted reconstructing 3D turbulent flows based on the PIV data on two orthogonal planes with a variational data assimilation method.
Here we focus on the reconstruction of turbulent mean flows in 2D.}

We use three different cases to show the enhancement of the flow reconstruction by assimilating various disparate data sources with the REnKF method.
The first case is a fully developed turbulent flow in 1D channel, where the sparse velocity away from the wall and the wall friction velocity are given as disparate observation data.
In the second case, we test on a more challenging case, i.e., 2D transitional flow over flat plate where the sparse velocity away from the wall and friction coefficient along the wall are considered.
In the first two cases, we mainly consider the disparate data of wall friction that is related to the velocity gradient adjacent to the wall.
In the third case, we consider disparate data from the pressure at the wall.
We use the sparse velocity measurements and the pressure distribution along the wall to reconstruct both the velocity and pressure fields.
Moreover, we incorporate the integration of pressure along the bottom wall, i.e., lift and drag forces, to improve the flow reconstruction.
In all the cases, we provide the REnKF results assimilating both sparse velocity and wall information compared to the EnKF results assimilating either wall information or sparse velocity alone.
The summary of the case setup is shown in Table.~\ref{tab:summary_setup}.
\begin{table}[!htbp]
    \centering
    \begin{tabular}{c|c|c|c|c|c}
    \hline
         Case & Geometry & $\oy_1$ & dim($\oy_1$) & $\oy_2$ & dim($\oy_2$)  \\
         \hline
        1 & Channel &  & 2 & \replaced[id=R1]{$u_\tau$}{$U^*$} & 1\\
        2 & Flat plate & $U_1$ & 17 & $C_f$ & 10 \\
        3a & Periodic hills &  & 10 & $p_w$ & 10 \\
        3b & Periodic hills &  & 10 & $\mathsf{F}$  & 1 \\
    \hline
    \end{tabular}
    \caption{Summary of case setups. The dim($\oy$) indicates the number of the observed positions for $\oy$; \replaced[id=R1]{$u_\tau$}{$U^*$} is the friction velocity at wall; $C_f$ represents the friction coefficient along the wall; $p_w$ is the pressure along the wall; $\mathsf{F}$ is the lift and drag force.}
    \label{tab:summary_setup}
\end{table}

In this work, we use OpenFOAM
to simulate the incompressible, steady-state turbulent flows.
The SIMPLE (Semi-Implicit Method for Pressure Linked Equations) algorithm
is used to solve the RANS equations.
Second-order spatial discretization schemes are applied to discretize the equations on an unstructured mesh.
The prior mean and synthetic truth are both obtained from RANS simulations using the built-in
\textit{simpleFOAM} solver but with different turbulence models.
We created a modified solver~\textit{nutFOAM} that uses a given eddy viscosity field instead of using a turbulence model.
This modified solver is used as the forward model that propagates the specific eddy viscosity field to the velocity and pressure fields.

\subsection{Channel flow}

\begin{figure}[!htb]
    \centering
    \subfloat[channel domain]{\includegraphics[width=0.4\textwidth]{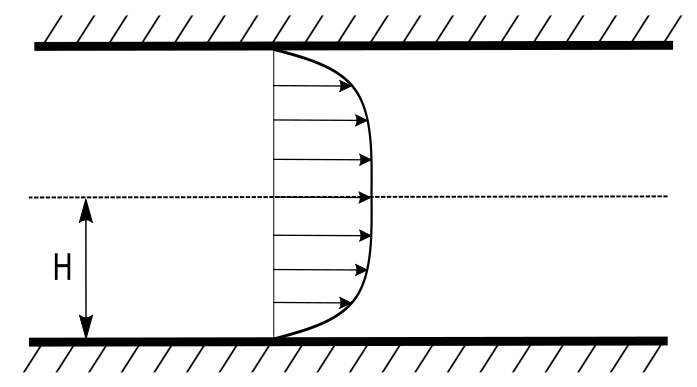}}
    \subfloat[channel mesh]{\includegraphics[width=0.35\textwidth]{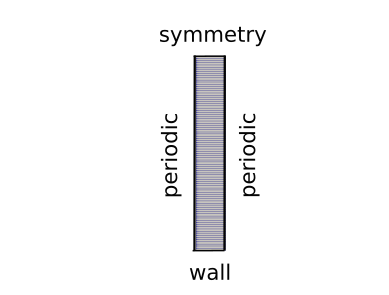}}
    \caption{ Computational setup of channel case. Figure (a) presents the channel domain where half of the channel height is $H$. Figure (b) shows the mesh and boundary setup. Only bottom half channel need to be computed because of the symmetry.}
    \label{fig:mesh_chan}
\end{figure}

\begin{figure}[!htb]
    \centering
    \includegraphics[width=0.7\textwidth]{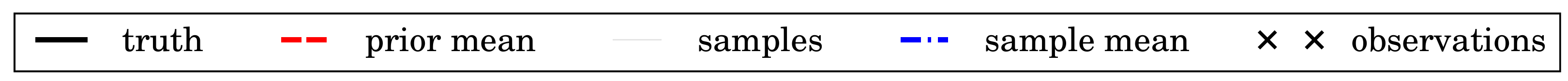}\\
    \subfloat[$\nu_t$]{\includegraphics[width=0.45\textwidth]{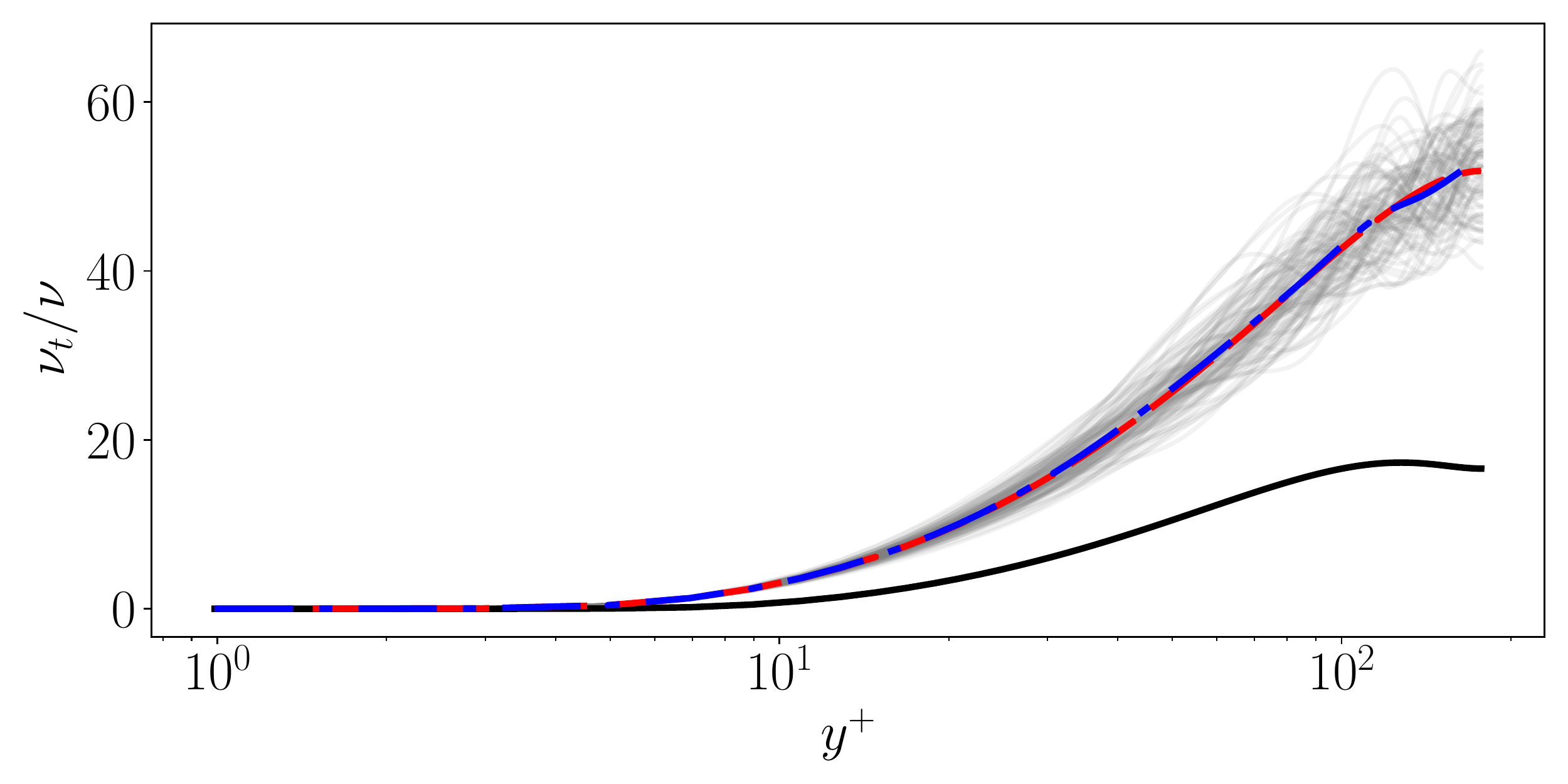}}
    \subfloat[$U_1$]{\includegraphics[width=0.45\textwidth]{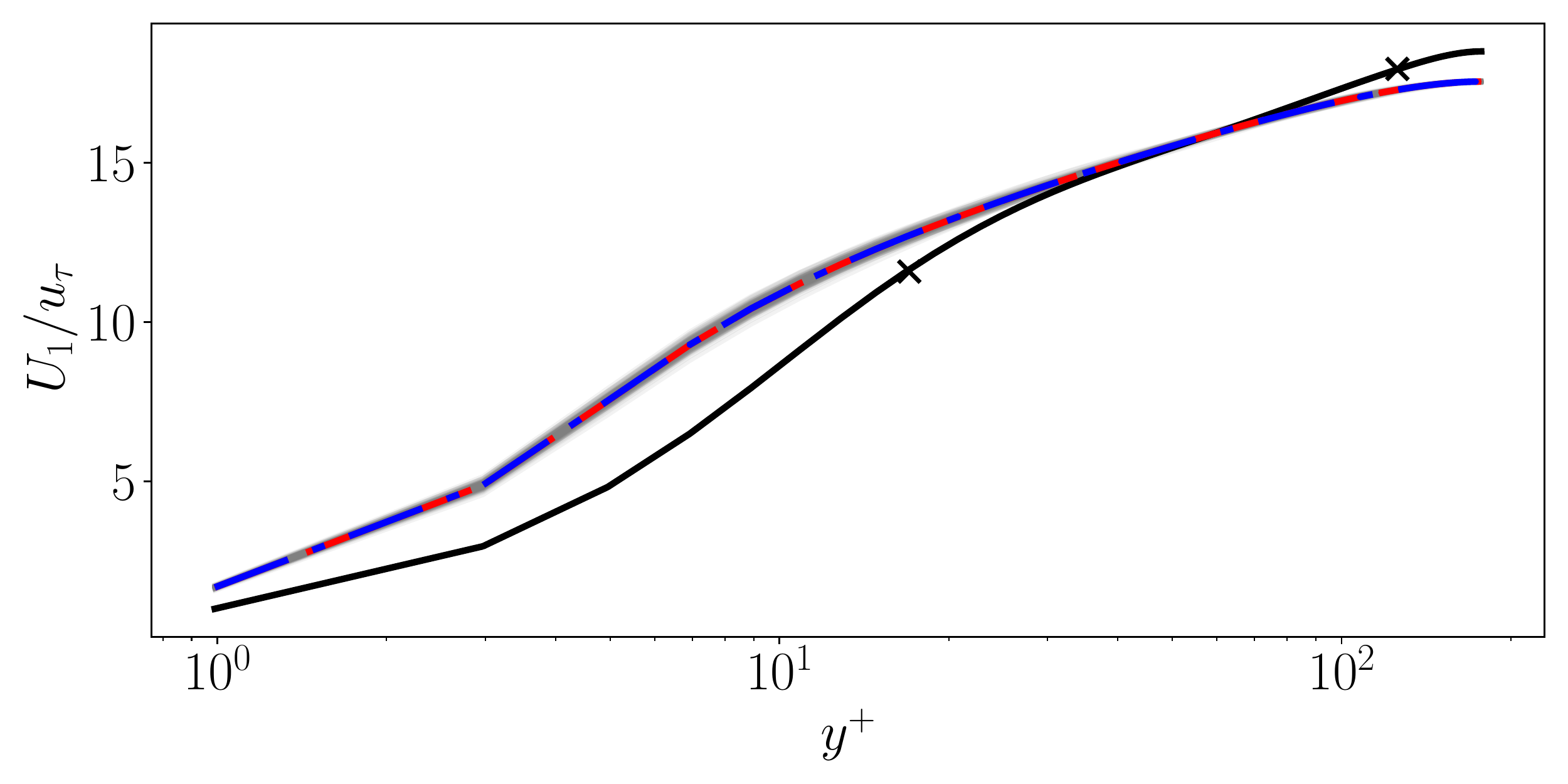}}
    \caption{Plots of the prior samples of \added[id=R1]{eddy viscosity~$\nu_t$ and} propagated velocity~$U_1$ for channel case. The observed position of sparse velocity~$\oy_1$ is indicated with crosses~$\times$.}
    \label{fig:chan_prior}
\end{figure}

The first case is the turbulent flow in a planar channel, which is extensively used for the validation of CFD solvers.
The computational domain is one-dimensional as presented in Fig.~\ref{fig:mesh_chan}. 
The mesh is evenly distributed with $90$ cells with the dimensionless distance~$y^+$ of the first cell adjacent to the wall around~$1$.
Periodic boundary conditions are imposed on the inlet and outlet.
\replaced[id=R1]{A no-slip boundary condition is imposed on the bottom wall,}{The bottom and top are solid walls with the no-slip boundary condition } \added[id=R1]{and a symmetry condition is imposed on the top boundary.}.
The Reynolds number based on the friction velocity and half channel height is $180$.
We regard the DNS results~\cite{moser1999direct} as the reference data.
The prior mean is applied from RANS simulation with $k$--$\omega$ model but using a reduced model coefficient $C_\mu=0.45$ to have an apparent difference from DNS in the velocity profile.
The number of samples is set as $100$.
Both the length scale $l$ and the variance $\sigma$ in Eq.~\eqref{eq:kernel} are set as $0.1$.
The number of modes used to generate ensemble is $20$ to cover more than $99\%$ of variance.
The prior samples of \added[id=R1]{the eddy viscosity and} the propagated velocity are shown in Figure.~\ref{fig:chan_prior}.
The disparate data we consider here are the sparse velocity~$U_1$ and friction velocity~\replaced[id=R1]{$u_\tau$}{$U^*$}.
We place the sparse velocity observation at two different locations, one in the buffer layer ($y=0.1H$) and one in the outer layer ($y=0.8H$) as shown in Fig.~\ref{fig:chan_prior}.
The relative observation error of the velocity is $10^{-3}$, and that of friction velocity is~$0.1$.

\begin{figure}[!htb]
    \centering
    \includegraphics[width=0.8\textwidth]{RANS-legend_Ux_profile.pdf}\\
    \subfloat[$U_1$: assimilate $u_\tau$]{
    \includegraphics[width=.33\textwidth]{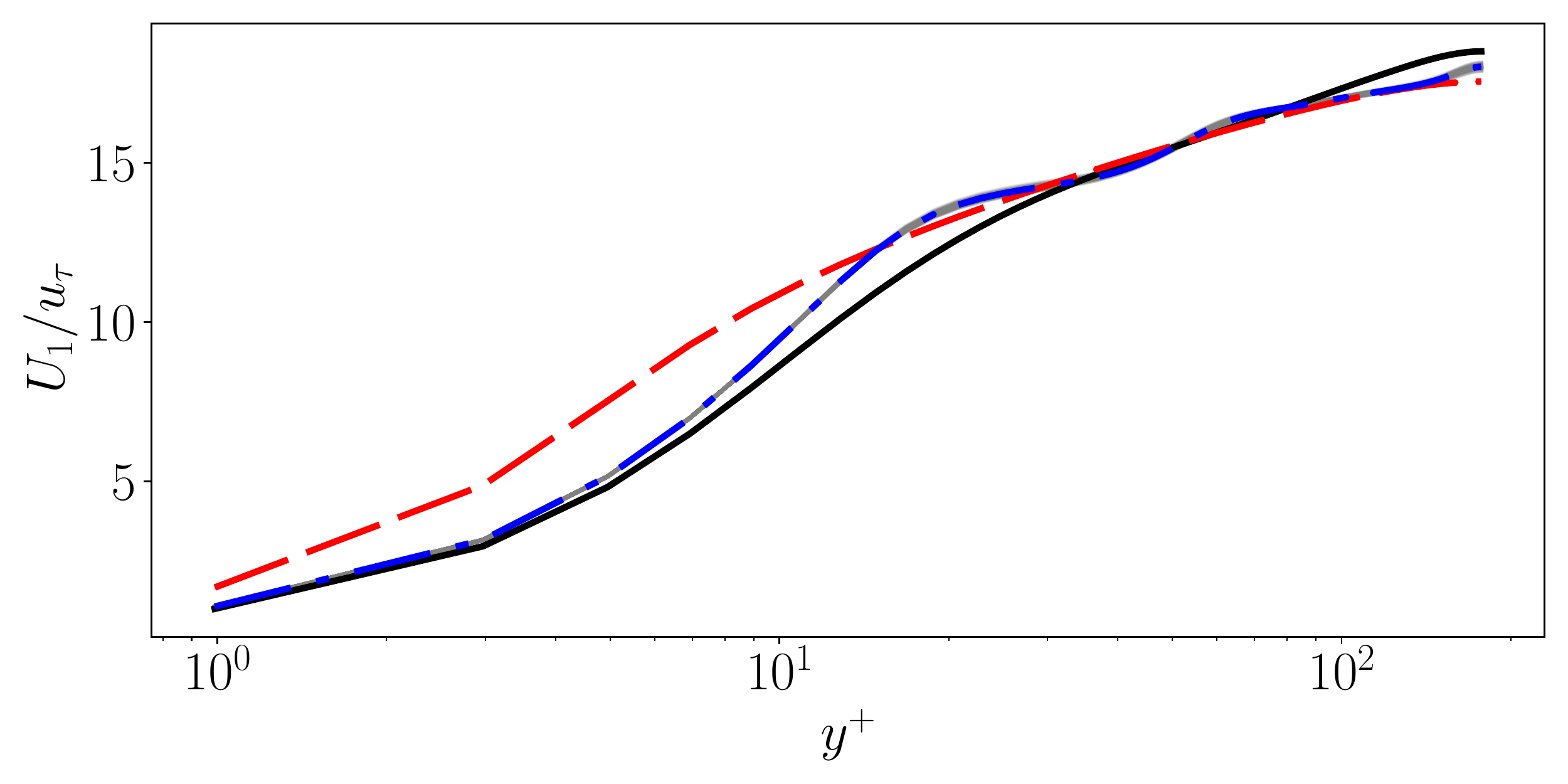}}
    \subfloat[$U_1$: assimilate $U_1$]{
    \includegraphics[width=.33\textwidth]{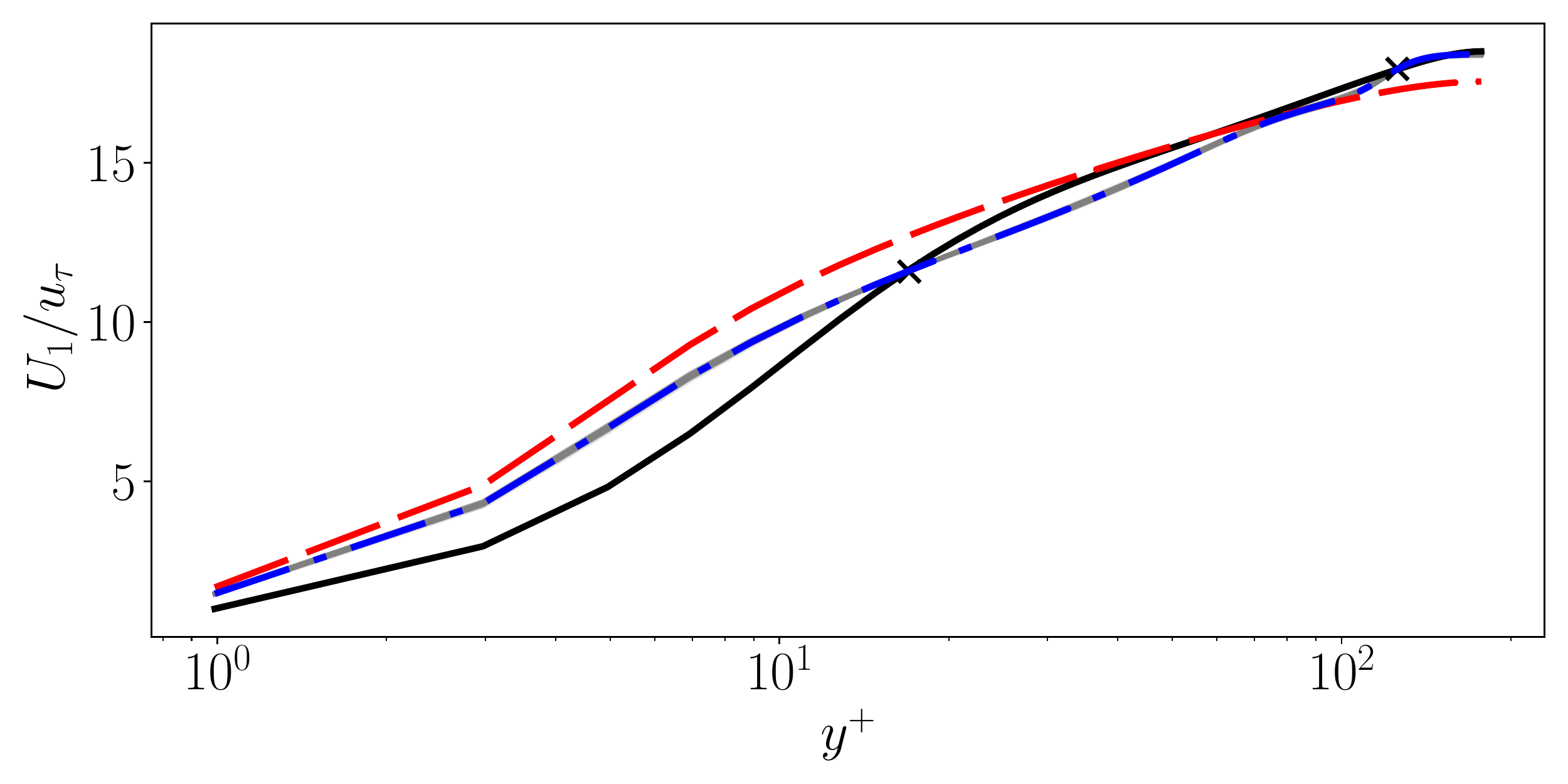}}
    \subfloat[$U_1$: assimilate both $U_1$ and $u_\tau$]{
    \includegraphics[width=.33\textwidth]{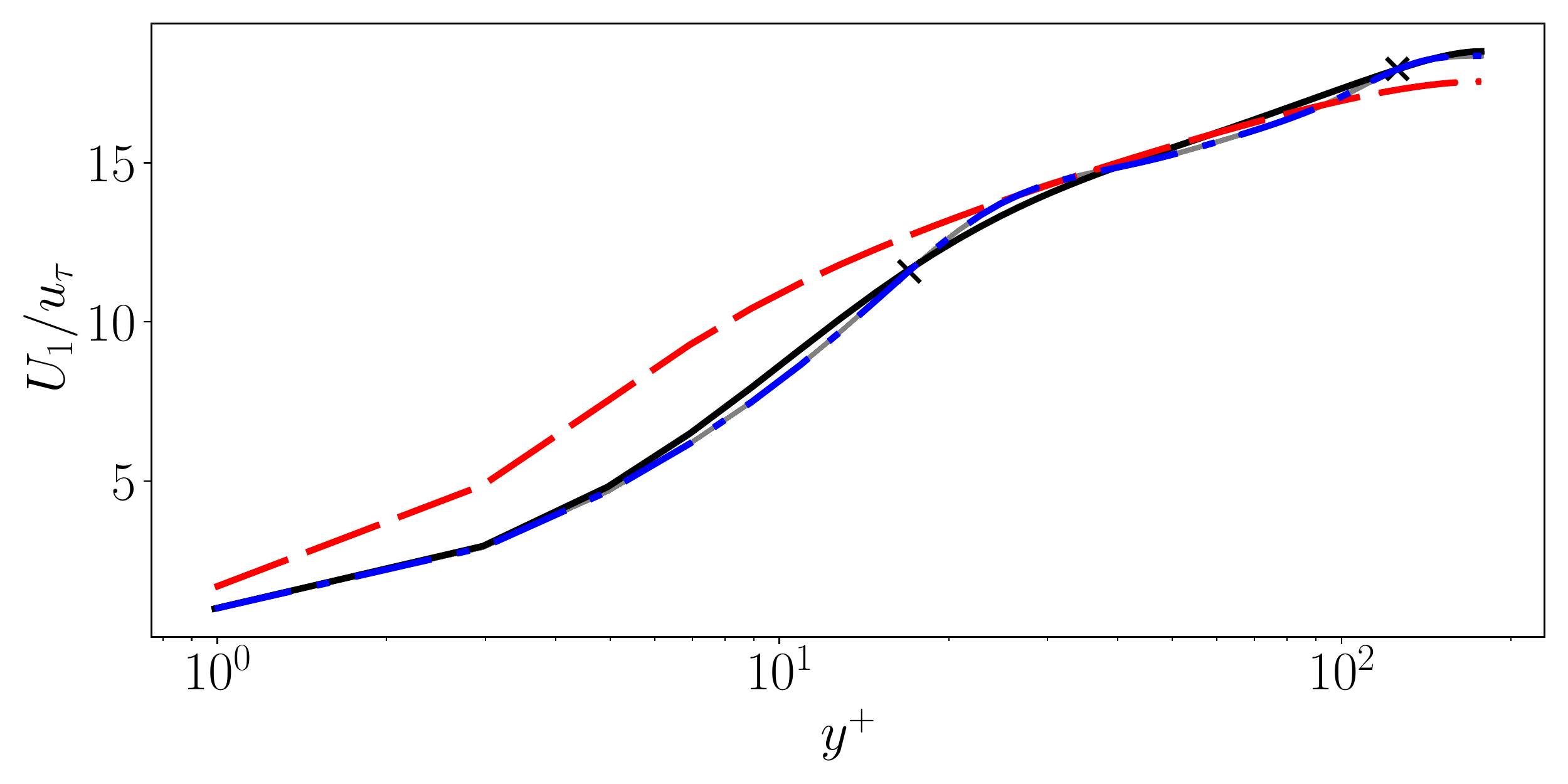}} \\
    \subfloat[$\nu_t$: assimilate $u_\tau$]{
    \includegraphics[width=.33\textwidth]{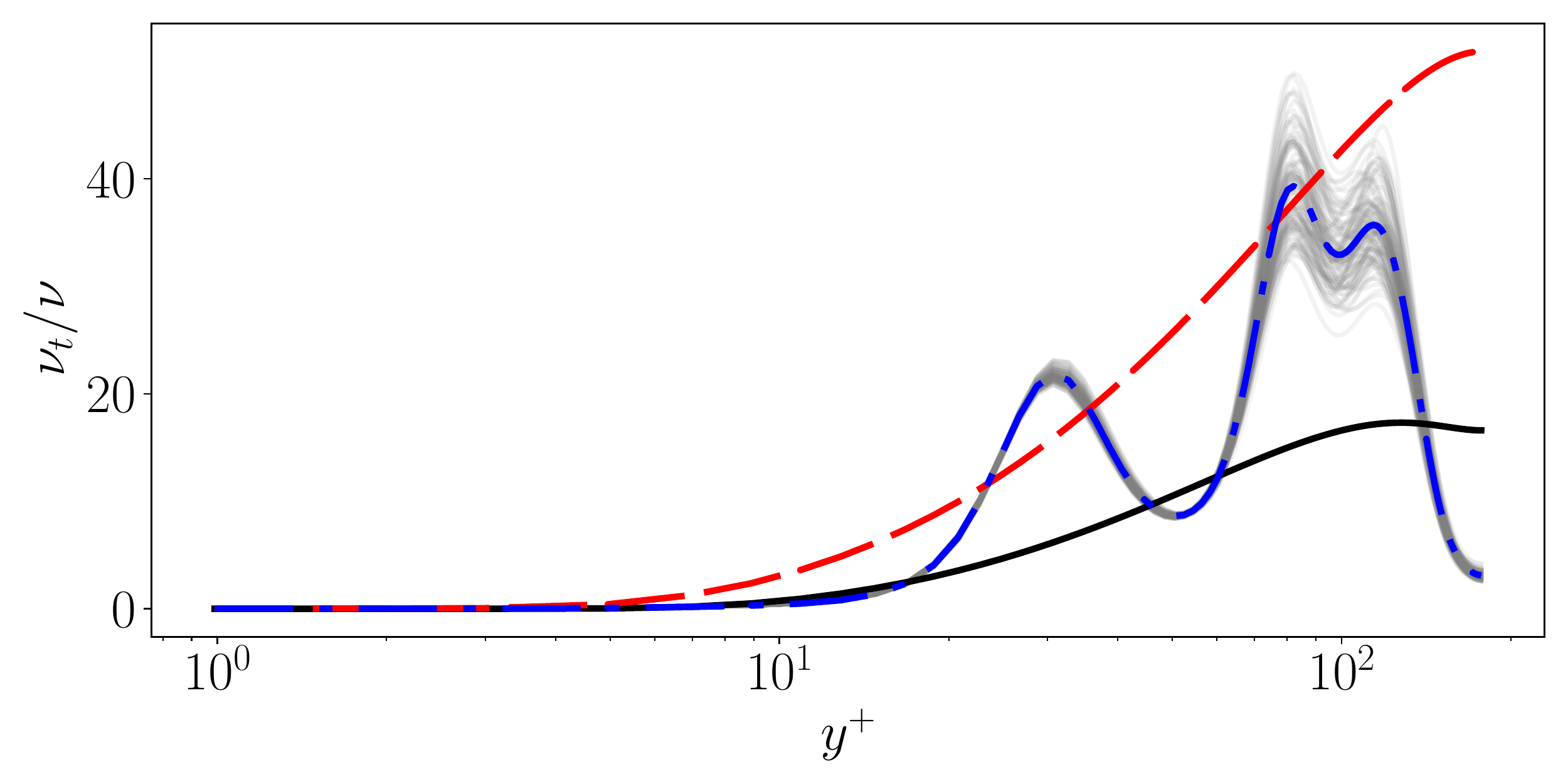}}
    \subfloat[$\nu_t$: assimilate $U_1$]{
    \includegraphics[width=.33\textwidth]{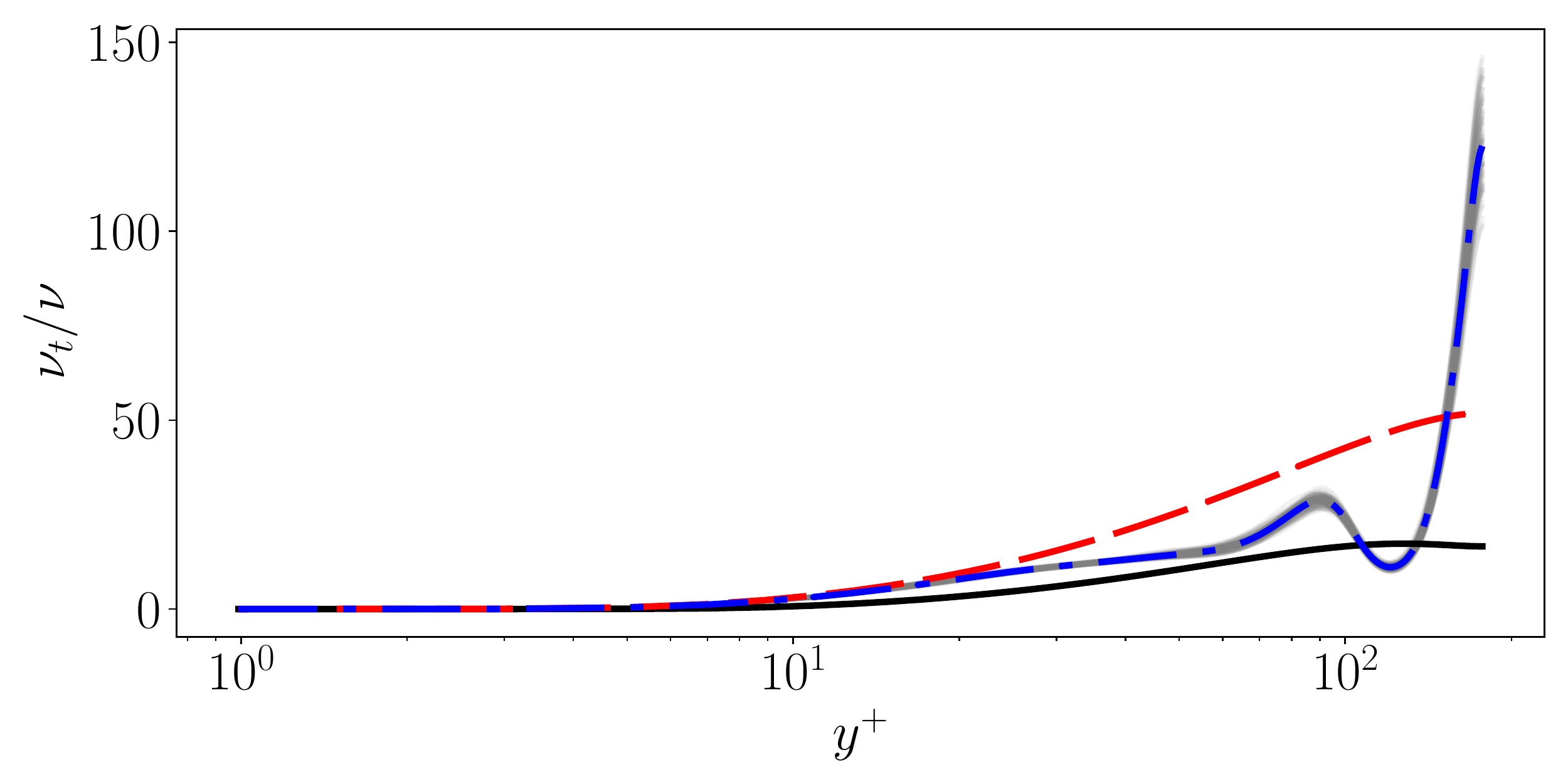}}
    \subfloat[$\nu_t$: assimilate both $U_1$ and $u_\tau$]{
    \includegraphics[width=.33\textwidth]{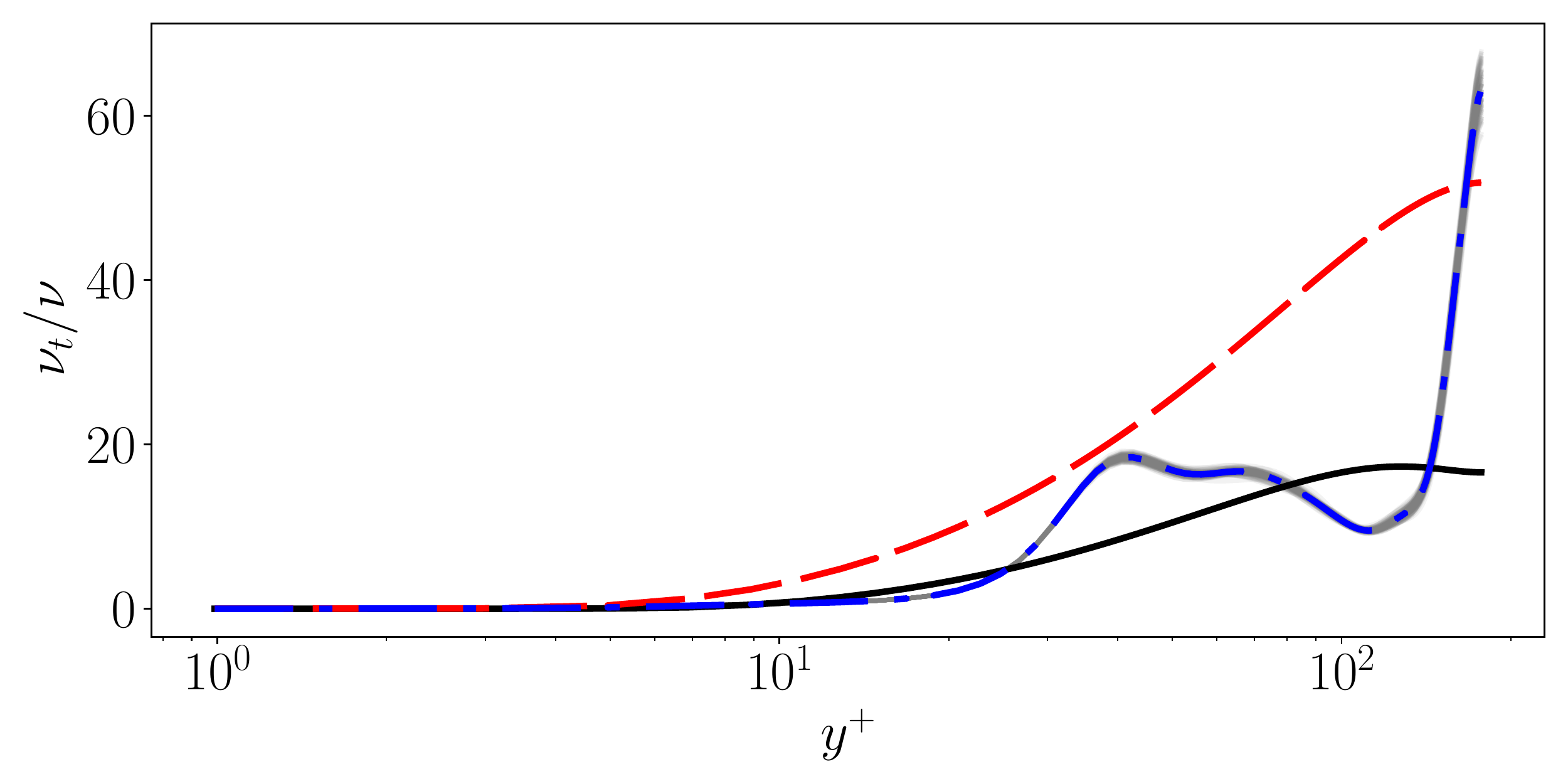}} \\
    \caption{Data assimilation results of the streamwise velocity \added[id=R1]{and eddy viscosity} by incorporating different observation data for channel case.}
    \label{fig:channel_yle5}
\end{figure}

The comparison results of assimilating different observations are plotted in Fig.~\ref{fig:channel_yle5}.
It is noticeable in Fig.~\ref{fig:channel_yle5}a that EnKF with only wall information can provide the velocity of the near-wall flow in a good agreement with the DNS data, since the friction velocity can inform the velocity gradient in the viscous layer ($y^+ < 5$).
While in the buffer layer and log layer ($y^+ > 5$), the flow velocity cannot be recovered accurately and exhibit a relatively large difference likely due to a lack of local information.
We also perform a test with only velocity observation, and the results are shown in Fig.~\ref{fig:channel_yle5}b. 
It can be seen that the velocity can only match in the outer region where we have local observation but cannot recover the near-wall region.
That is likely due to the large spatial distance leading to a low correlation between the observation and the near-wall region.
Finally, when considering both the friction velocity and the velocity measurements, we can reconstruct the velocity profile accurately in the entire domain, as shown in Fig~\ref{fig:channel_yle5}c.
The friction velocity can reconstruct the velocity in the viscous layer, and the two sparse velocity observations, one in the log layer and one in the outer layer, can recover the flow velocity regionally.
Clearly, with a combination of these two disparate data sources, the best data fit can be achieved even in the region where we do not provide observation data.
\added[id=R1]{
The inferred eddy viscosity with different observations are shown in Figs.~\ref{fig:channel_yle5}d, e and f.
It can be seen that all the inferred eddy viscosity in the three cases have a large discrepancy with the truth particularly near the center of channel.
That is not surprising since around the center region of the channel, the mean rate of strain tensor is very small, and thus the velocity is insensitive to the eddy viscosity, making the inference problem ill-posed.
Additional constraints in the eddy viscosity such as its smoothness can be  used to regularize the problem and to improve the inference of the eddy viscosity.
}

\begin{figure}[!htb]
    \centering
    \includegraphics[width=0.6\textwidth]{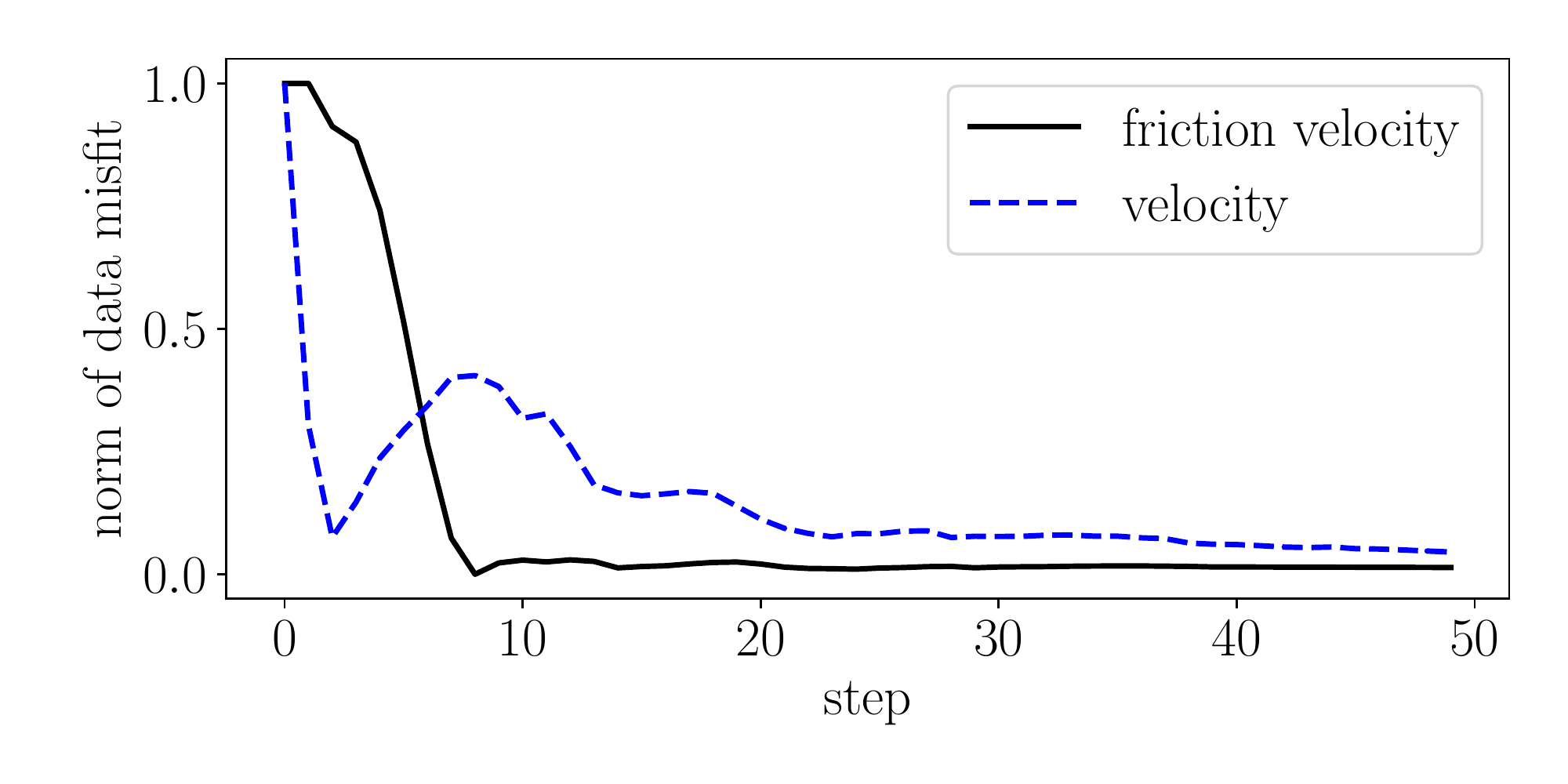}
    \caption{Convergence feature of REnKF in the normed data misfit for channel case. The data misfit is normed by the initial misfit to keep the convergence curve in the same magnitude.}
    \label{fig:convergence_chan}
\end{figure}

For better illustration, the error of quantity~$q$ between the model estimate and truth in the entire computational domain is defined based on
\begin{equation}
    \text{error} = \frac{\parallel q_\text{truth} - q_\text{estimate} \parallel}{\parallel q_\text{truth} \parallel} \text{.}
    \label{eq:error_def}
\end{equation}
The error in friction velocity and velocity is summarized in Table~\ref{tab:summary_results}.
It can be seen that assimilating both sparse velocity and friction velocity can achieve the best data fit in both~\replaced[id=R1]{$u_\tau$}{$U^*$} and $U_1$.
The evolution of relative error normalized by the initial error is provided in Fig.~\ref{fig:convergence_chan}, showing that the assimilation process is very robust.

\subsection{Flow over flat plate}

\begin{figure}[!htbp]
    \centering
    \subfloat[Computational setup]{\includegraphics[width=0.45\textwidth]{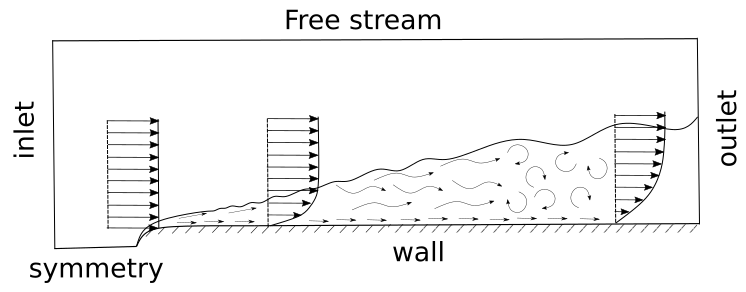}}
    \hfill
    \subfloat[Zoomed mesh around plate]{\includegraphics[width=0.45\textwidth]{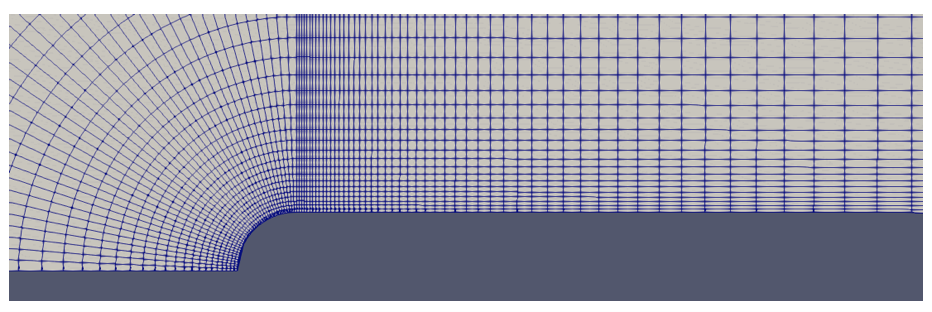}}
    \caption{Computational setup of flat plate case. Figure (a) presents the by-pass transition process and the computational domain with boundary setup. Figure (b) shows the mesh information around the plate.}
    \label{fig:mesh_T3A}
\end{figure}

\begin{figure}{!htb}
    \centering
    \includegraphics[width=0.8\textwidth]{RANS-legend_Ux_profile.pdf}\\
    \subfloat[eddy viscosity]{\includegraphics[width=0.33\textwidth]{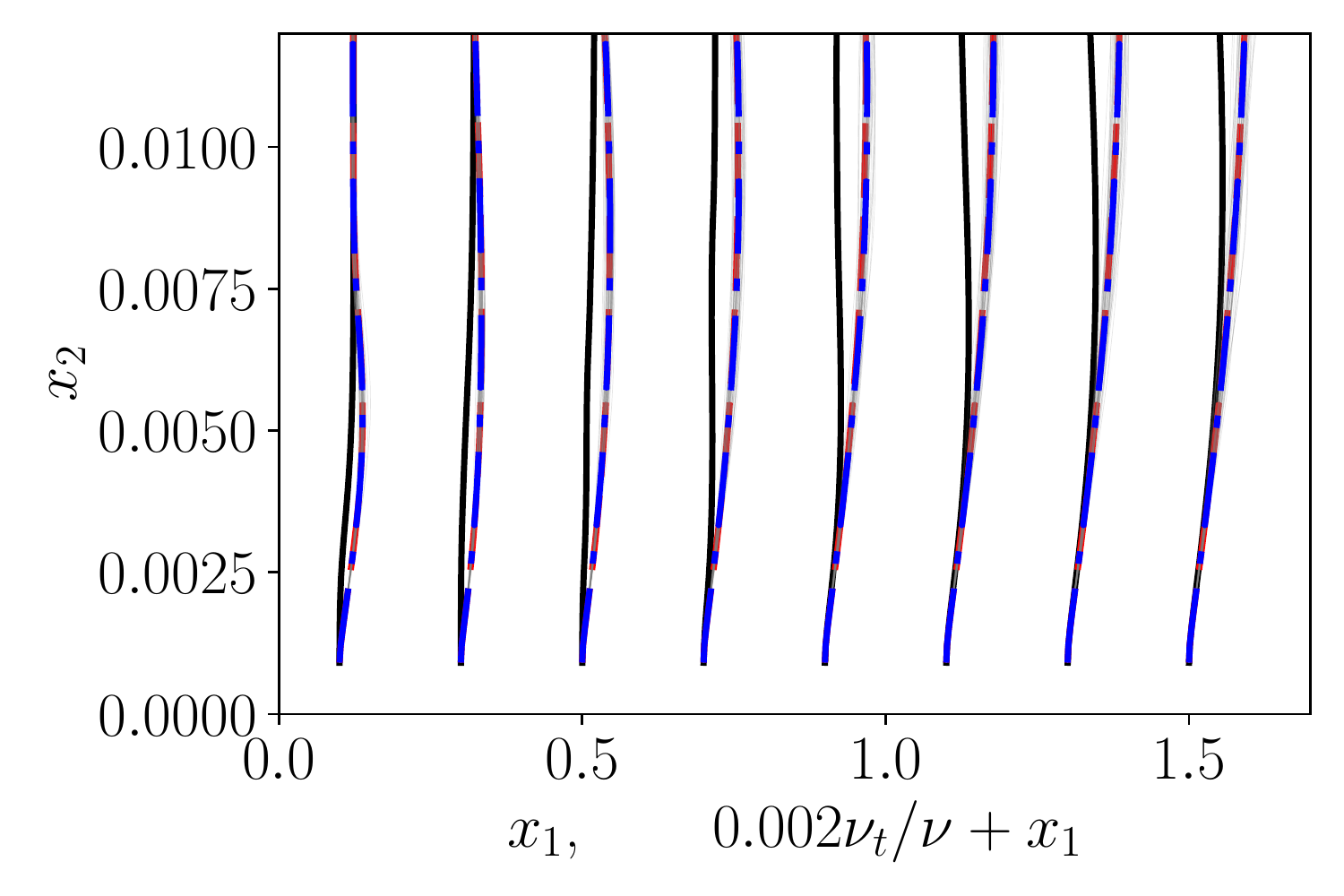}}
    \subfloat[velocity]{\includegraphics[width=0.33\textwidth]{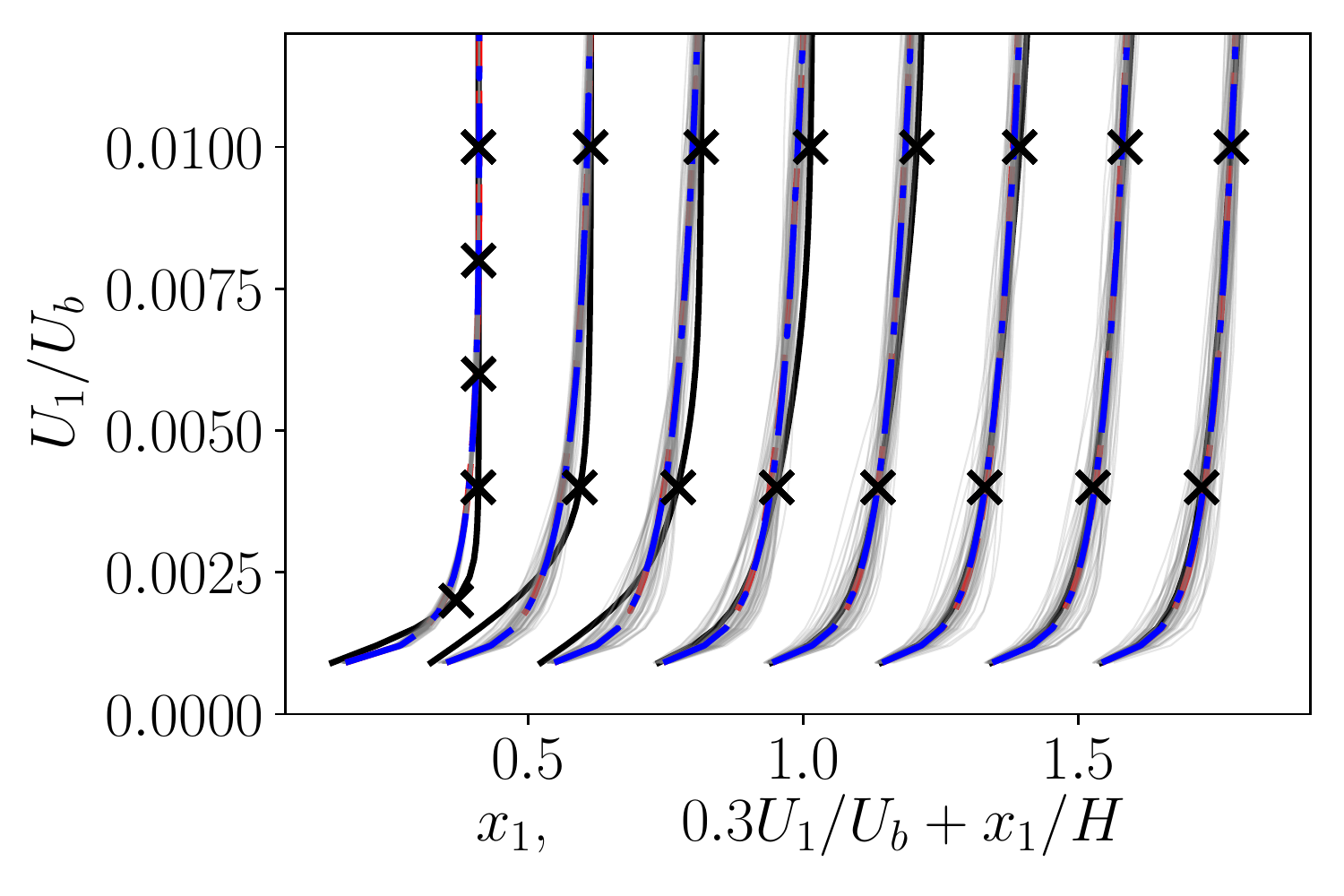}}
    \subfloat[friction coefficient]{\includegraphics[width=0.33\textwidth]{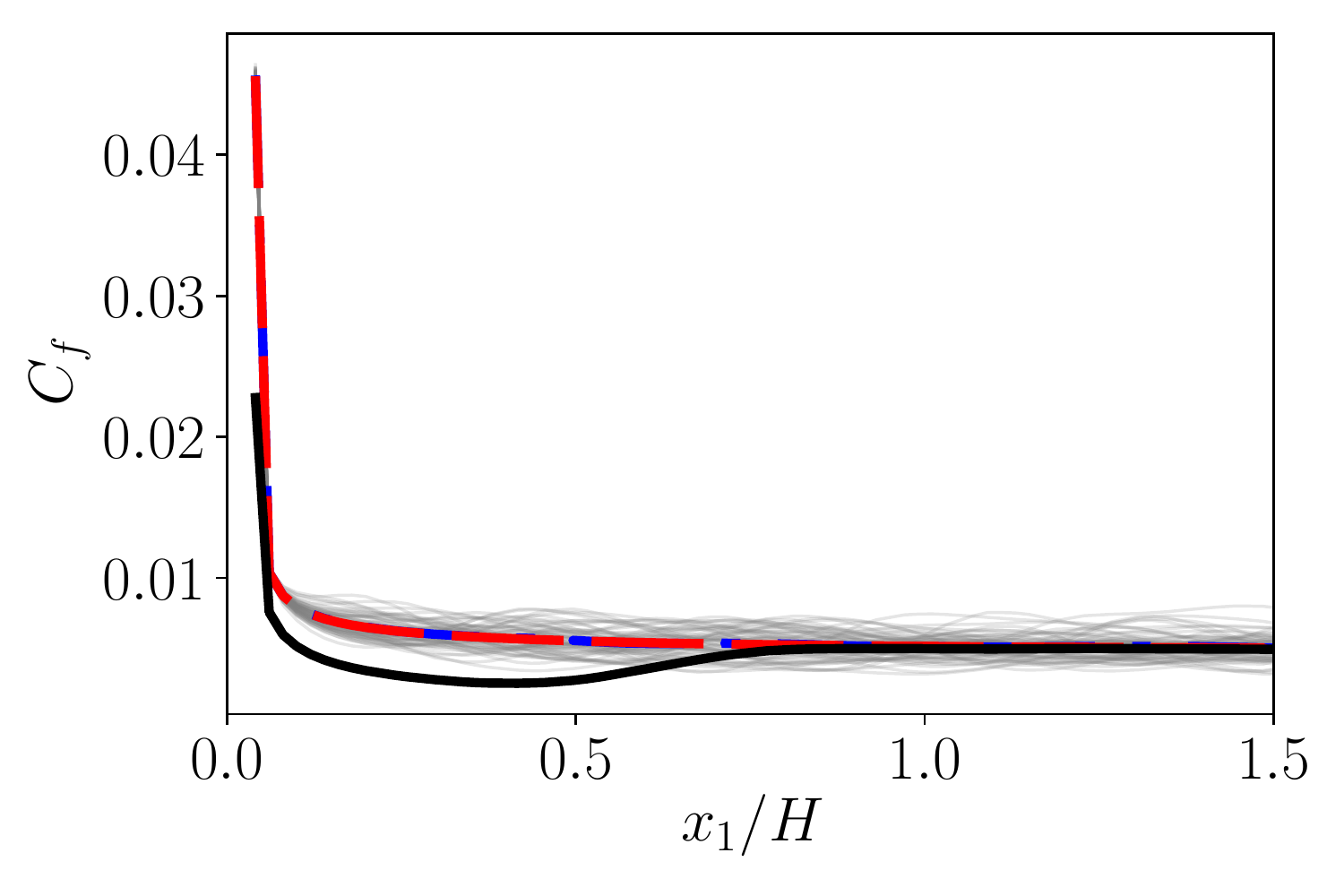}}
    \caption{Plots of the prior samples of \added[id=R1]{the eddy viscosity,} the propagated velocity, and wall friction coefficient for the flat plate case. The observed position of sparse velocity $\oy_1$ is indicated  with crosses ($\times$).}
    \label{fig:T3A_prior}
\end{figure}

In the second case, we demonstrate how to enforce wall friction information to improve the reconstruction of turbulent mean flows in a 2D flow: the turbulent flow over flat plate, which is a canonical case for investigations of the by-pass transition problem~\cite{duraisamy2015new}.
The inflow turbulence intensity is $0.033$.
The inlet bulk velocity~$U_b$ is $5.4$~m/s.
The kinetic viscosity $\nu$ is $1.5 \times 10^{-5}$~m$^2$/s.
The viscosity ratio $\nu_t/\nu$ is set as $12$.
The mesh is constructed with $10010$ cells.
The computational setup and the close-up view of the mesh around the plate are shown in Fig.~\ref{fig:mesh_T3A}.
The inlet is imposed with the uniform velocity, and the outlet is applied with zero-gradient condition for velocity.
The top boundary is set as the free stream, and the plate is solid wall with the no-slip condition.
In this case, we regard the $k$--$\omega$ model as the baseline.
The $k$--$\omega$ SST model with $Re_\theta$--$\gamma$ transition model~\cite{menter2006transition} is used as a synthetic truth, since this model has been validated to simulate the by-pass transition flow accurately.
The length scale $l$ for the streamwise and wall-normal direction is chosen as $0.1$ and $0.003$, respectively, and the standard deviation~$\sigma$ is $0.1$ in this case.
The number of modes used for generating the samples is $500$ to cover more than $99 \% $ of the variance.
The number of samples is $100$.
The plot of the initial samples of \added[id=R1]{the eddy viscosity} and the propagated velocity and friction coefficient is provided in Fig.~\ref{fig:T3A_prior}.
The $19$ observation positions of sparse velocity are placed along three straight lines, as shown in Fig.~\ref{fig:T3A_prior}\replaced[id=R1]{b}{a}.
The relative error of observation~$\oy_1$ is~$0.001$.
The friction coefficient~$C_f$ along the wall is used as the disparate data~$\oy_2$ with the relative error of~$0.01$.
The ten observed positions of the wall friction coefficient are evenly distributed along the plate. 
The friction coefficient is defined as
\begin{equation}
    C_f = \frac{\tau_w}{\frac{1}{2}\rho U_b^2} \text{,}
\end{equation}
where $U_b$ is bulk velocity, and the $\tau_w$ is the wall shear stress which is defined as
\begin{equation}
    \tau_w = \mu \left( \frac{\partial U_1}{\partial y} \right)_{y=0} \text{.}
\end{equation}

\begin{figure}[!htb]
    \centering
    \includegraphics[width=0.8\textwidth]{RANS-legend_Ux_profile.pdf}
     \subfloat[$U_1$: assimilate $C_f$]{\includegraphics[width=0.33\textwidth]{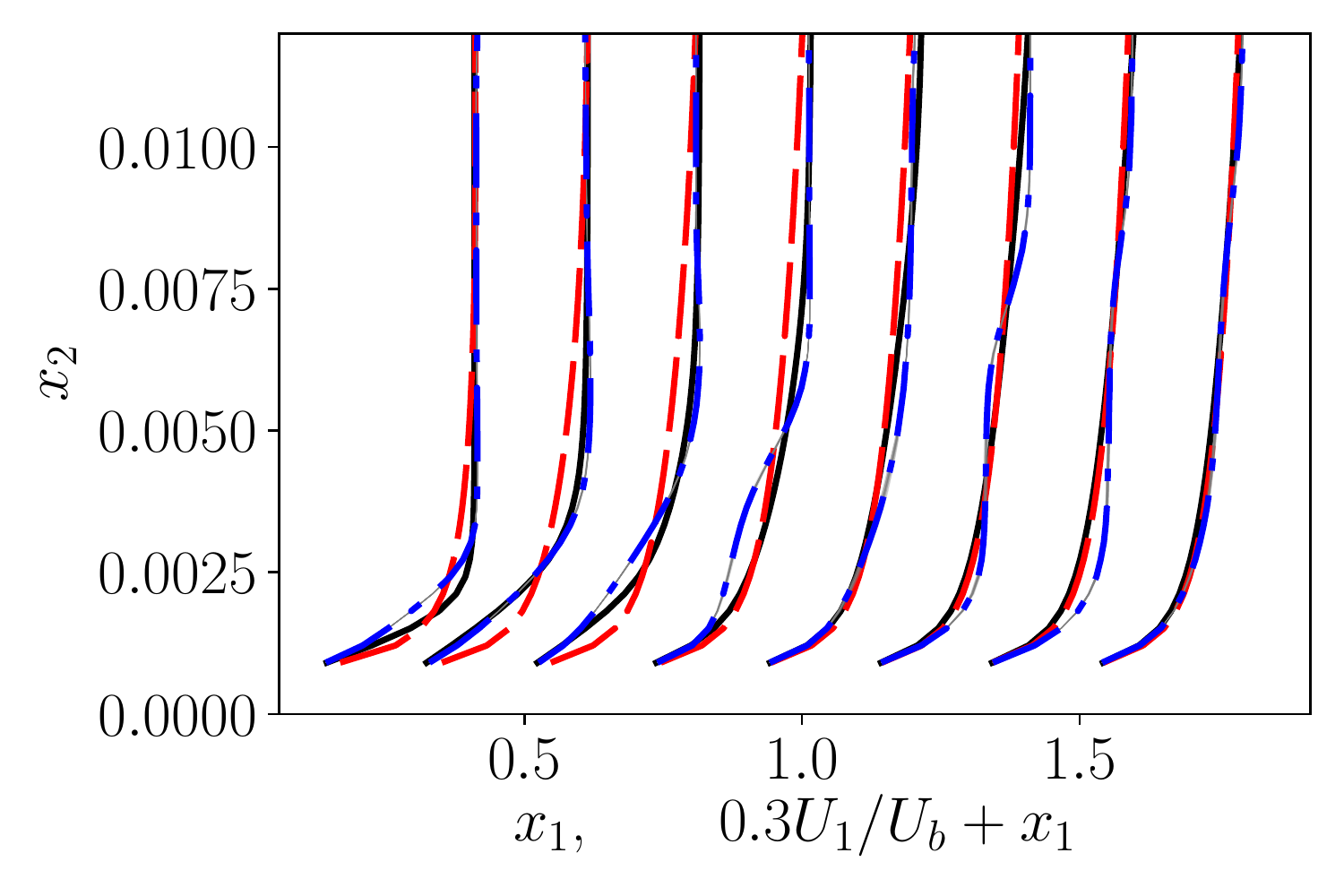}}
    \subfloat[$U_1$: assimilate $U_1$]{\includegraphics[width=0.33\textwidth]{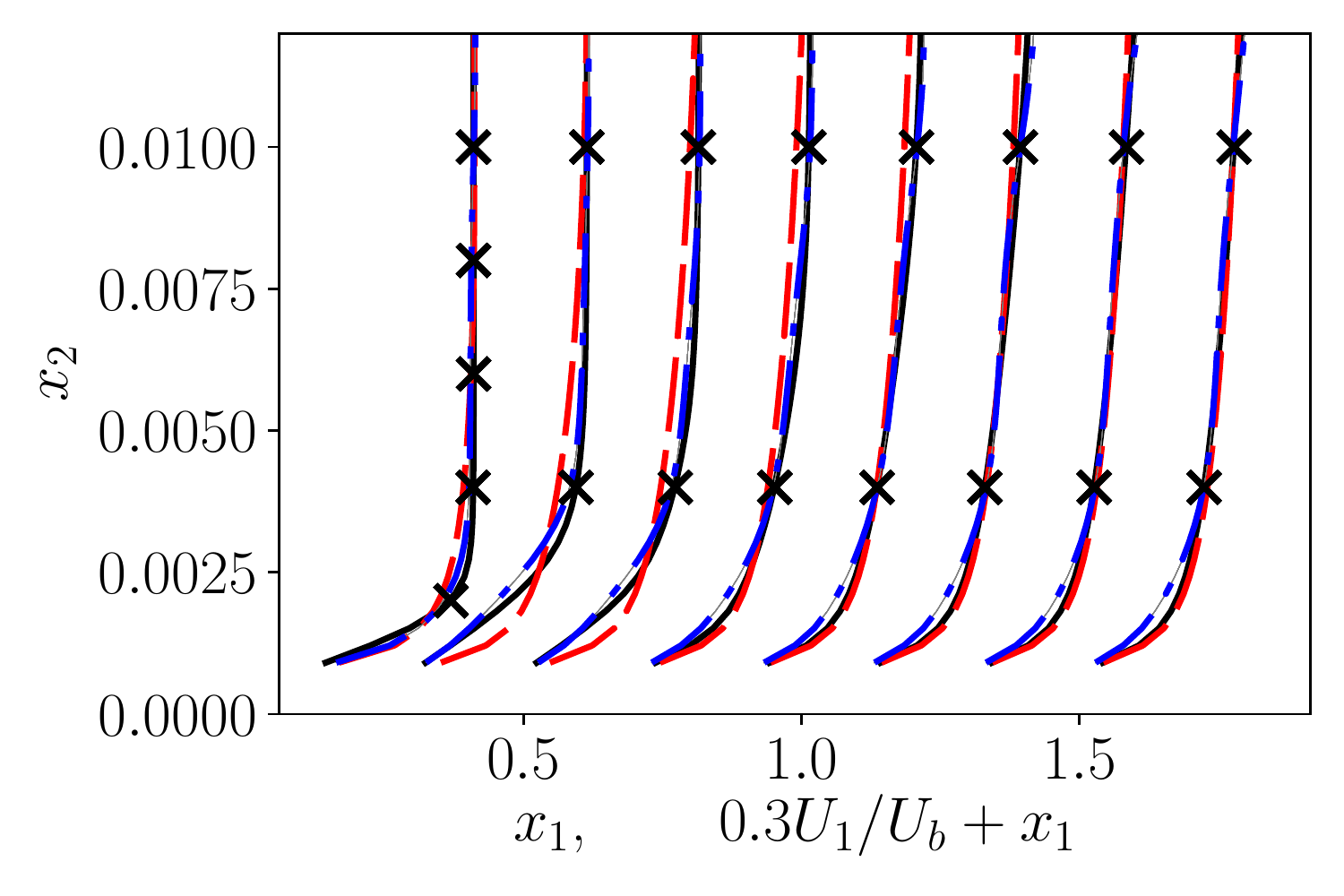}}
    \subfloat[$U_1$: assimilate $C_f$ and $U_1$]{\includegraphics[width=0.33\textwidth]{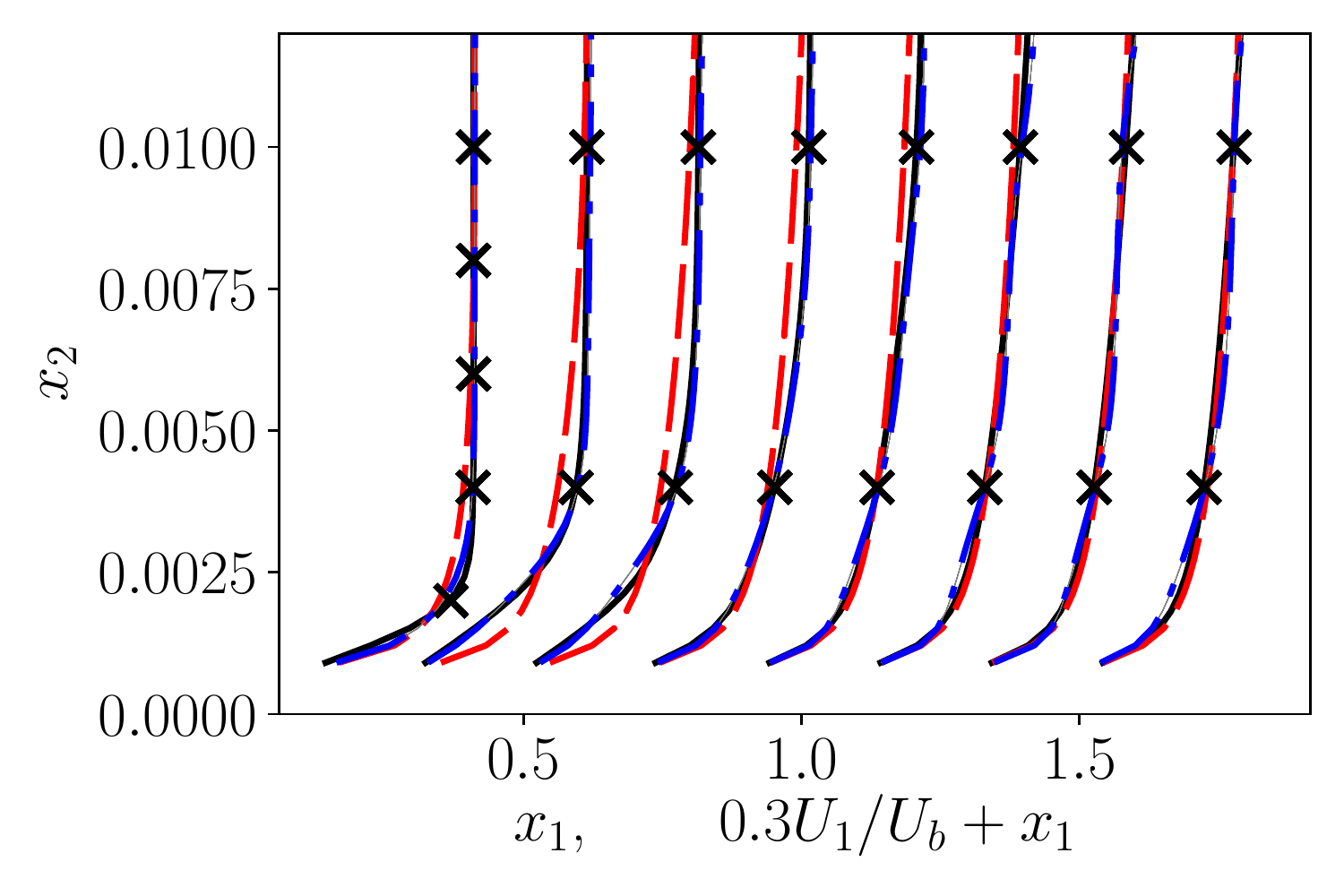}}\\
    \subfloat[$C_f$: assimilate $C_f$]{\includegraphics[width=0.33\textwidth]{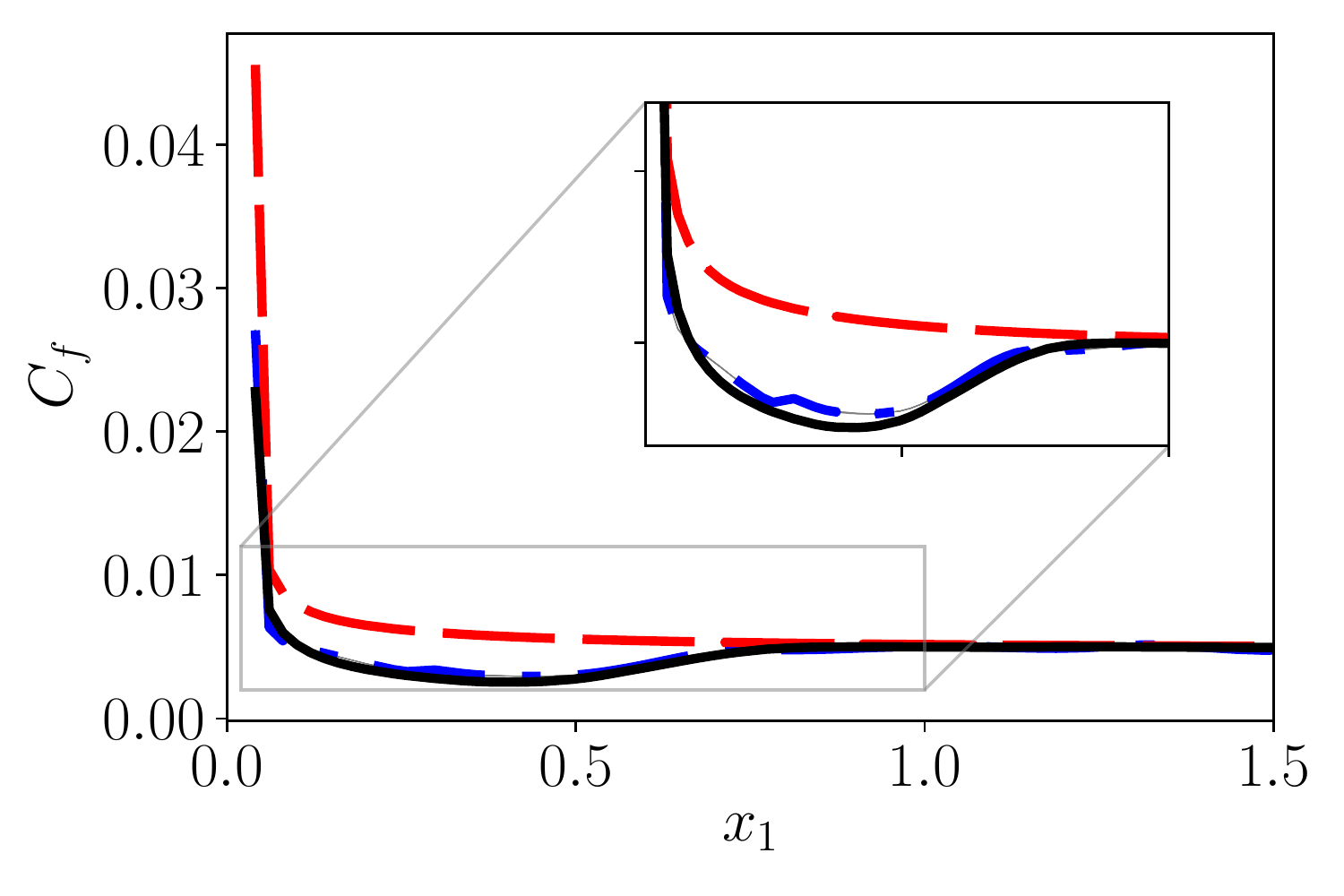}}
    \subfloat[$C_f$: assimilate $U_1$]{\includegraphics[width=0.33\textwidth]{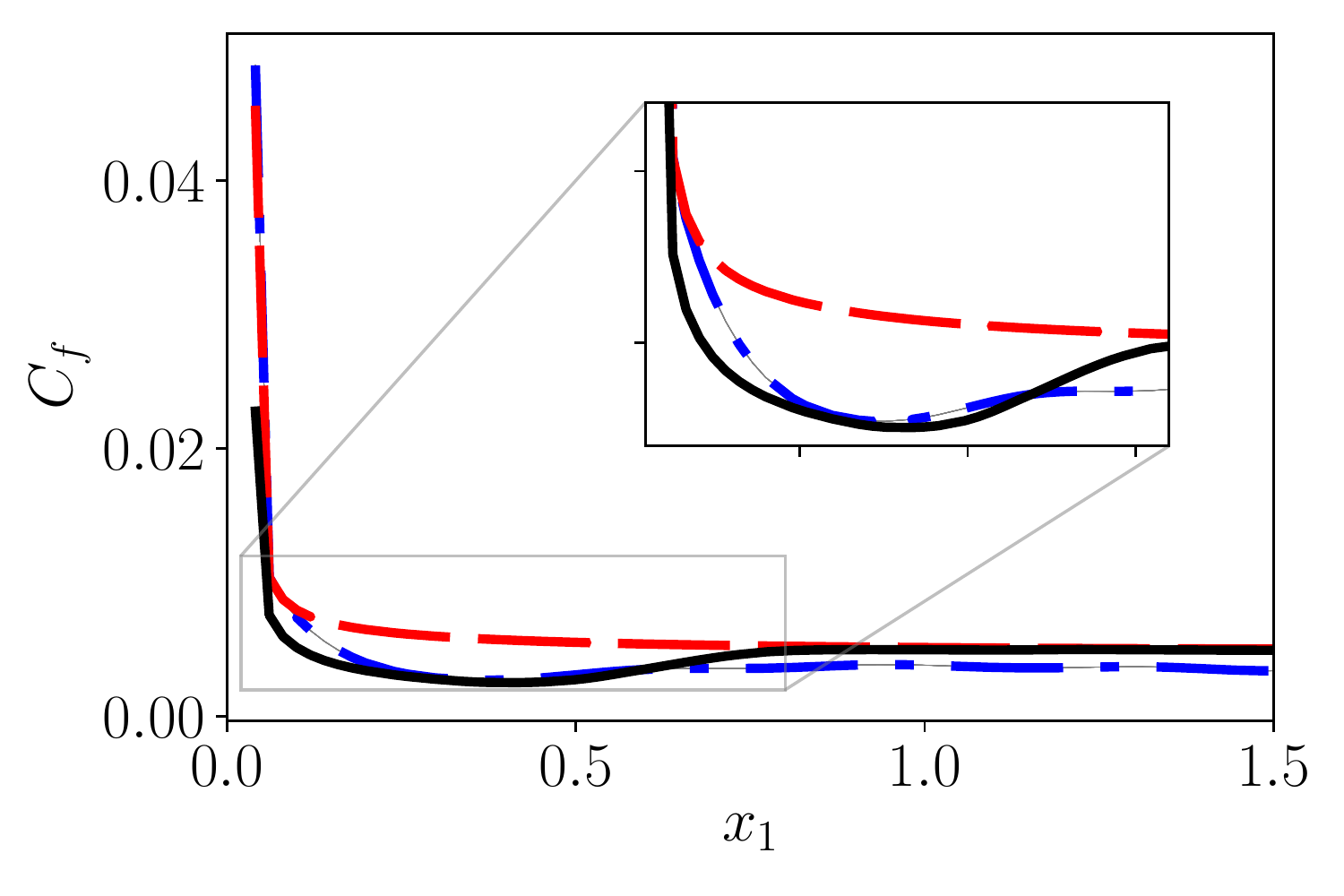}}
    \subfloat[$C_f$: assimilate $C_f$ and $U_1$ ]{\includegraphics[width=0.33\textwidth]{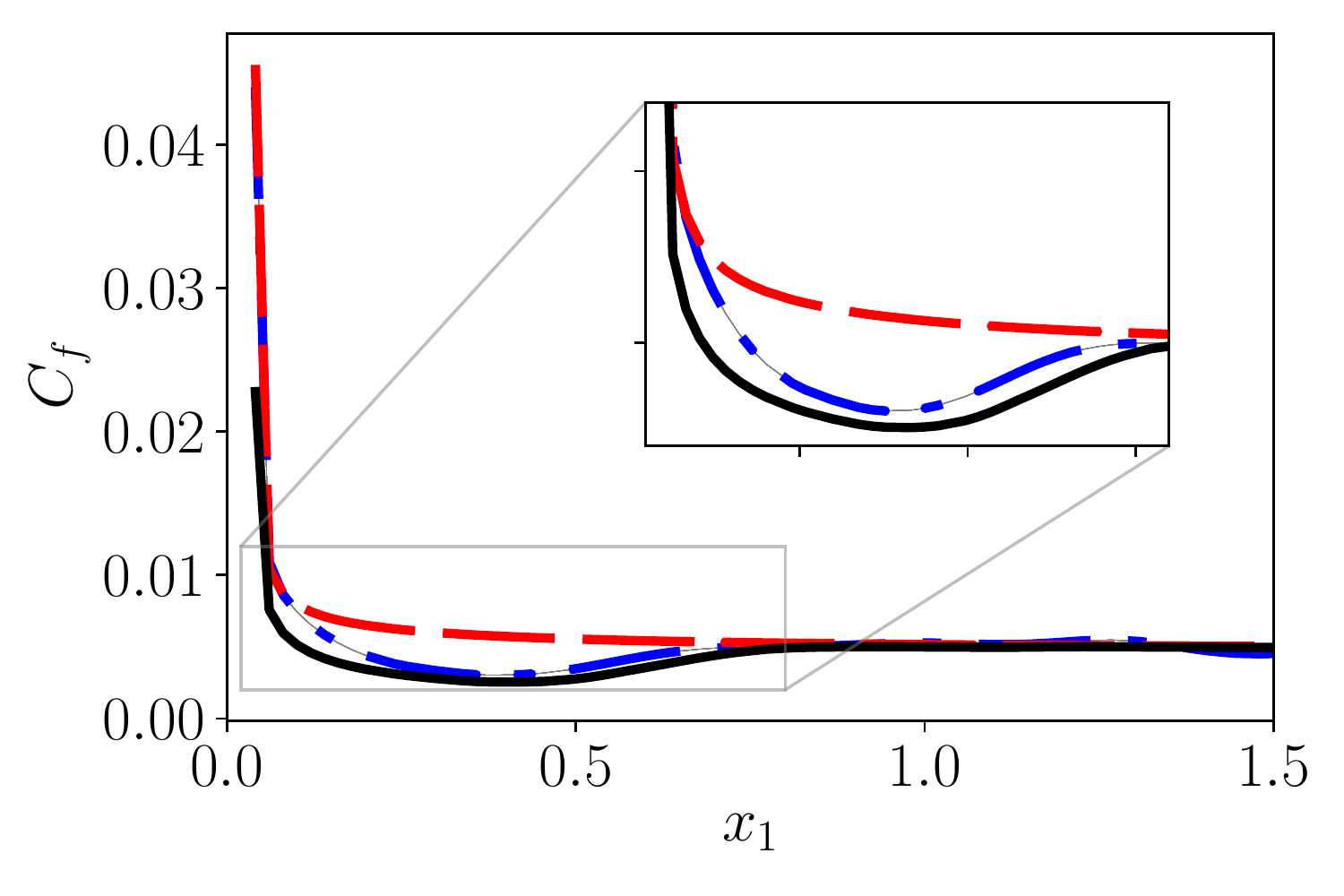}}\\
    \subfloat[$\nu_t$: assimilate $C_f$]{\includegraphics[width=0.33\textwidth]{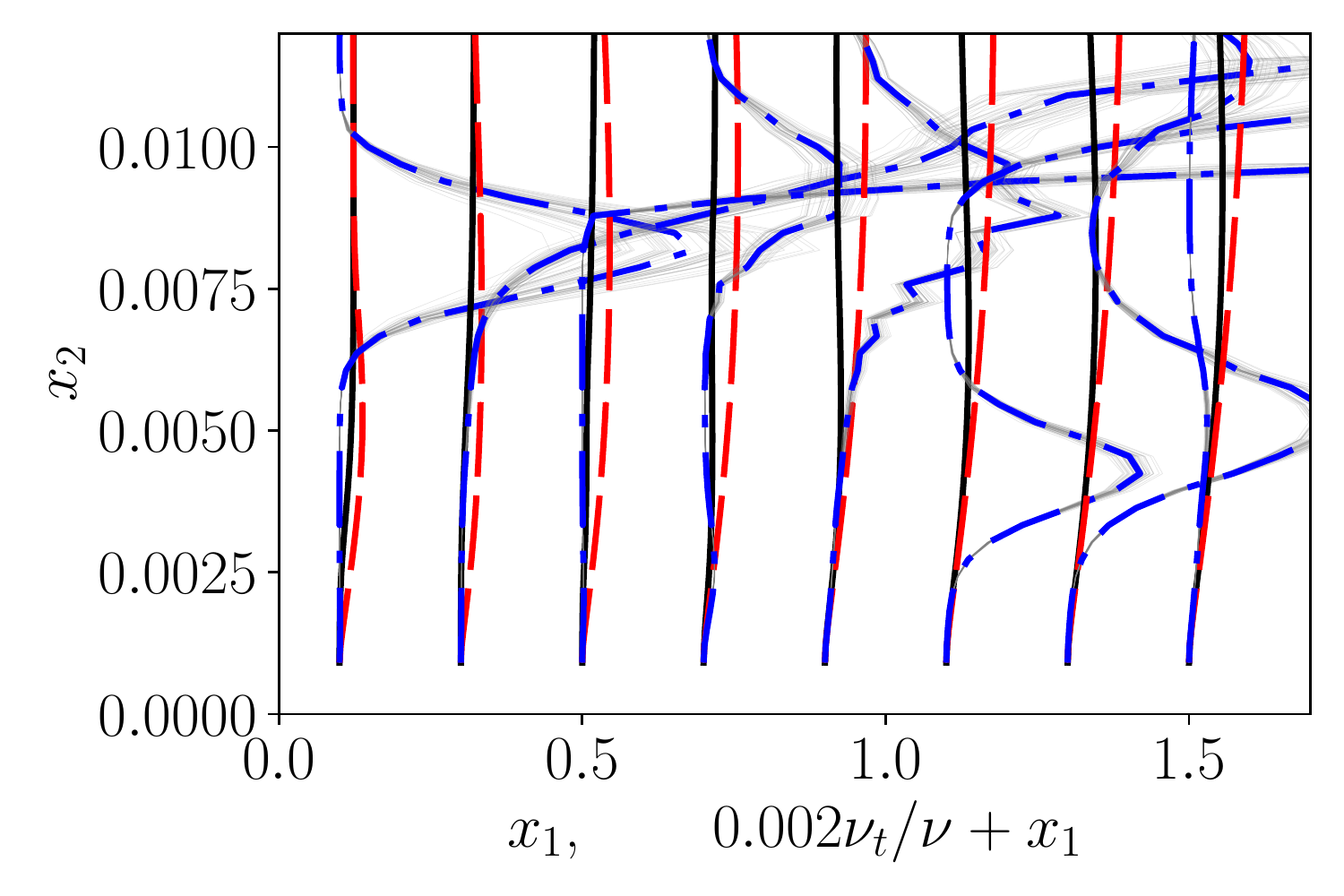}}
    \subfloat[$\nu_t$: assimilate $U_1$]{\includegraphics[width=0.33\textwidth]{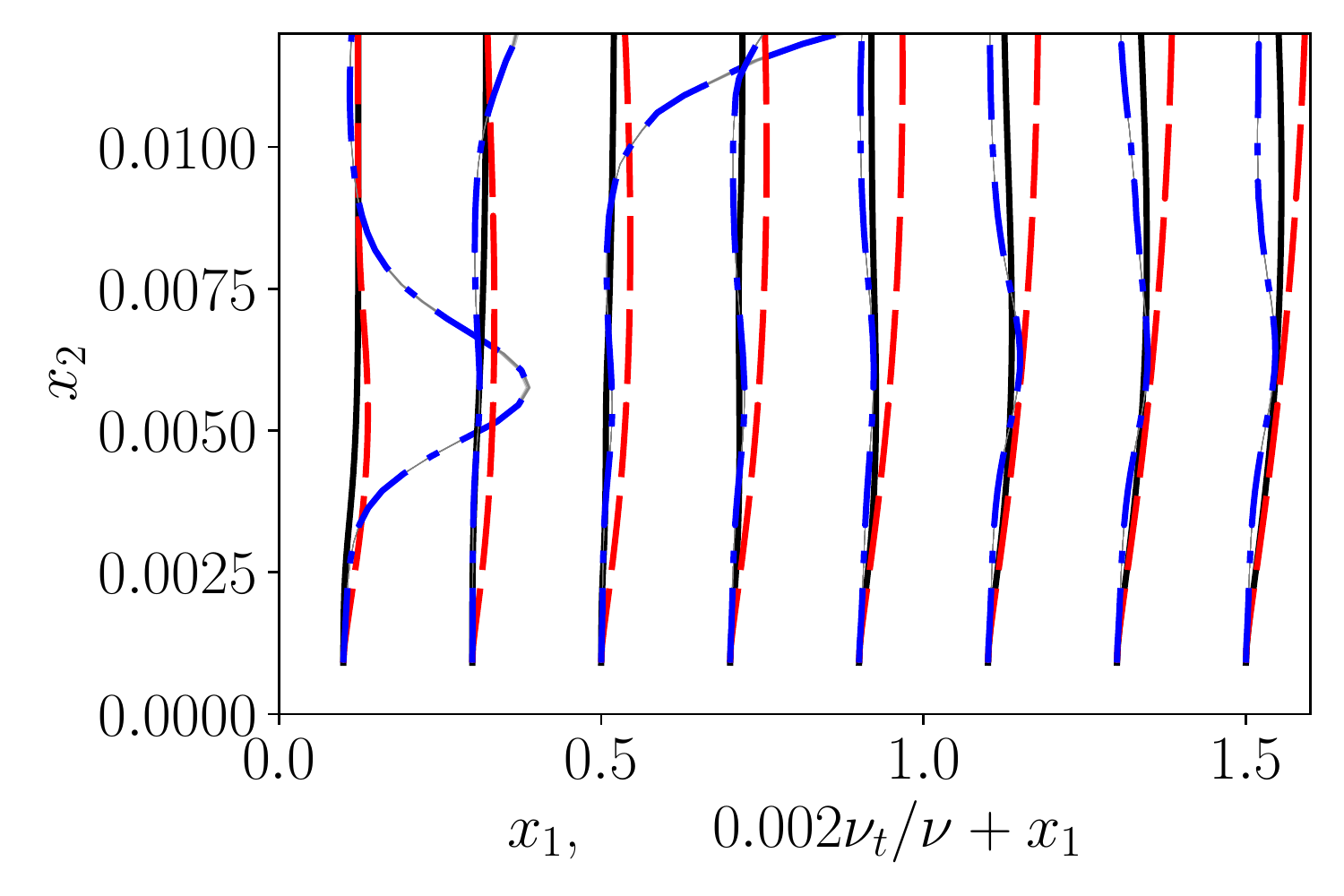}}
    \subfloat[$\nu_t$: assimilate $C_f$ and $U_1$]{\includegraphics[width=0.33\textwidth]{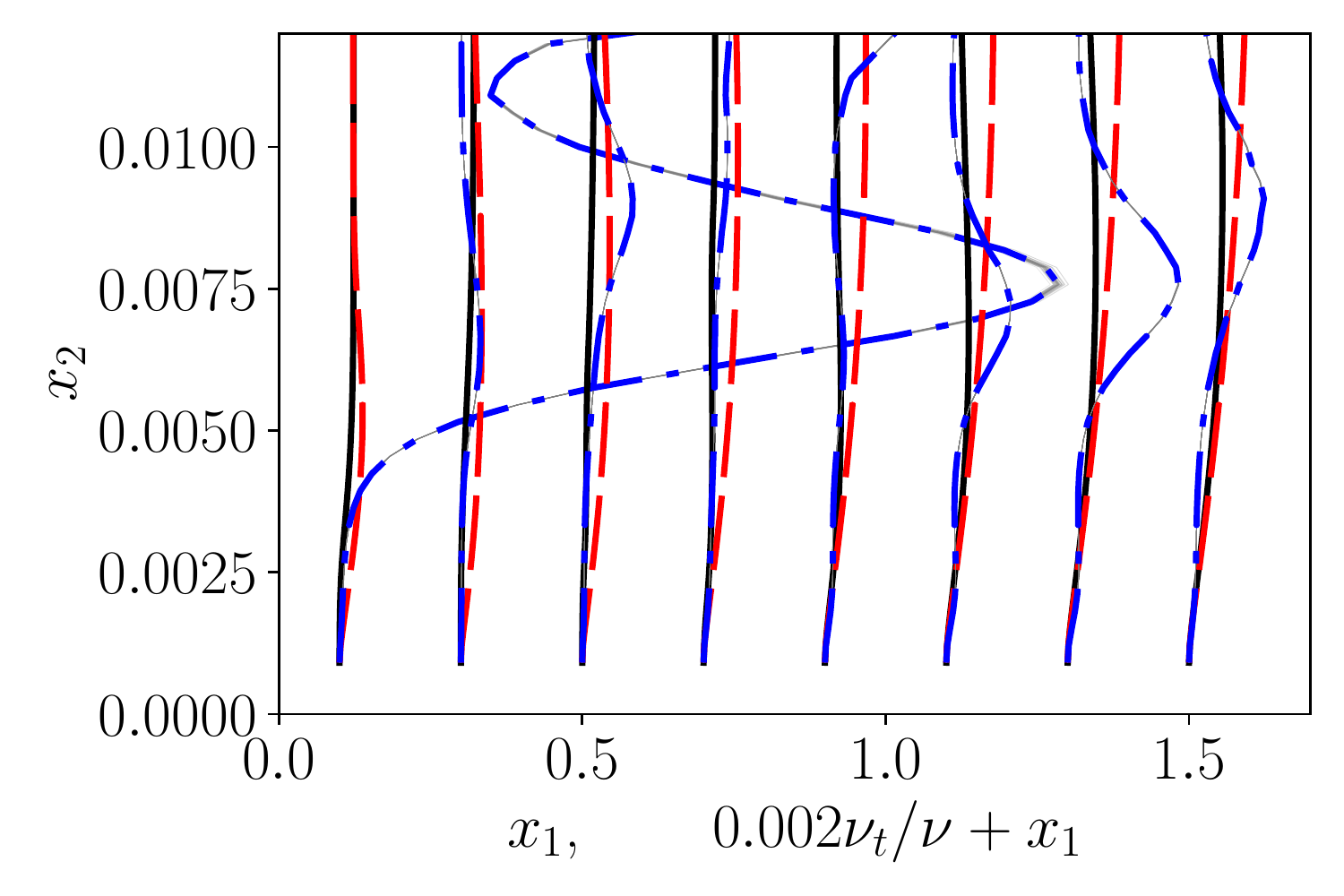}}\\
    \caption{The data assimilation results of  streamwise velocity, friction coefficient, \added[id=R1]{and eddy viscosity} by incorporating different observation data for T3A plate case}
    \label{fig:T3A_results}
\end{figure}

We first perform the EnKF with only the observation of wall friction coefficient.
The results are shown in Figs.~\ref{fig:T3A_results}a and d.
It is noticeable that EnKF assimilating only the friction coefficient can recover the turbulent flow velocity near the wall but results in the inferior mean velocity profile, especially away from the wall.
Conversely, assimilating only the sparse velocity can replicate the flow away from the wall but lead to a large discrepancy in the region adjacent to the wall as presented in Figs.~\ref{fig:T3A_results}b and e.
Further, we employ the REnKF method that incorporates both the friction coefficient and sparse velocity, and the results are shown in Figs.~\ref{fig:T3A_results}c and f.
It is clear that the assimilation of both the friction coefficient and the sparse velocity can enhance the turbulent flow reconstruction in the velocity field, exhibiting a good agreement with reference data in both the friction coefficient and the velocity.
The friction coefficient is essentially related to the velocity gradient adjacent to the wall.
As in the channel case, with only the wall friction measurement, it can offer good velocity profiles in the viscous layer adjacent to the wall but not in the outer region, while the sparse velocity observation can only improve the local flow estimate.
By contrast, the reconstruction can be enhanced by combining the wall friction velocity and the sparse velocity with the REnKF method.
\added[id=R1]{The inferred eddy viscosity in this case are shown in Figs.~\ref{fig:T3A_results}g, h and i.}
\added[id=All]{It can be seen that the inferred eddy viscosity by assimilating only the friction coefficient lead to the largest discrepancy, showing that the data assimilation with only wall measurements are ill-conditioned.
}
\added[id=R1]{The other two cases, i.e., of assimilating only $U_1$ and assimilating both $C_f$ and $U_1$, provide better results compared to that of assimilating $C_f$.}

\begin{figure}[!htb]
    \centering
    \includegraphics[width=0.6\textwidth]{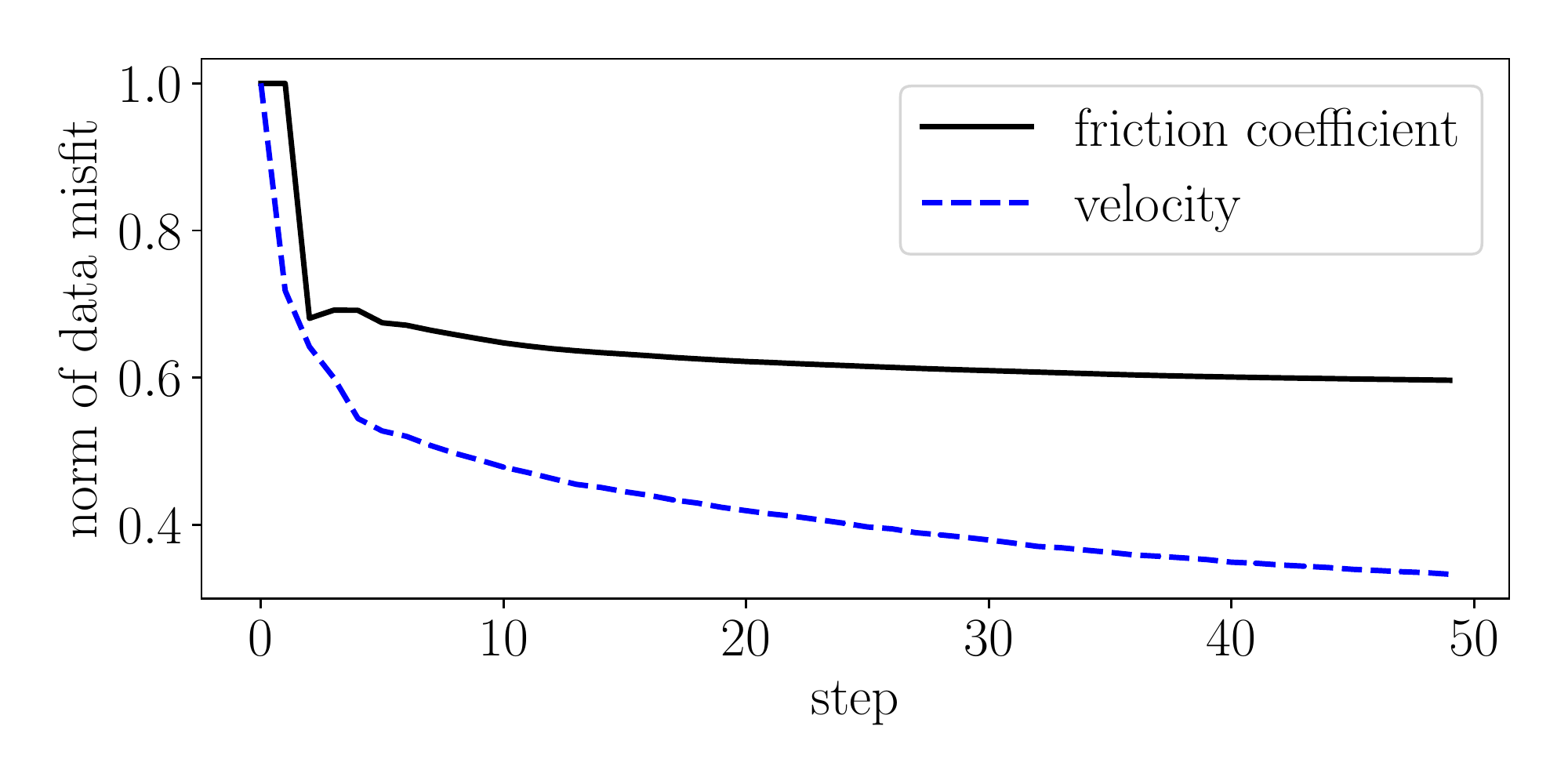}
    \caption{Convergence feature of REnKF in the normed data misfit for flat plate case. The data misfit is normed by the initial misfit to keep the convergence curve in the same magnitude.}
    \label{fig:convergence_T3A}
\end{figure}

The discrepancy between the reconstructed quantities (i.e., velocity and wall friction coefficient) and the reference is summarized in Table.~\ref{tab:summary_results} based on Eq.~\eqref{eq:error_def}.
It shows clearly that the disparate data assimilation can provide better agreement in $U_1$ compared to assimilating only friction coefficient and in both $U_1$ and $C_f$ compared to assimilating only sparse velocity.
The convergence history in this case is shown in Fig.~\ref{fig:convergence_T3A}, which demonstrates that the method can reduce the misfit of the disparate data simultaneously and robustly.

\subsection{Flow over periodic hills}

\begin{figure}[!htbp]
    \centering
    \subfloat[computational domain]{\includegraphics[width=0.4\textwidth]{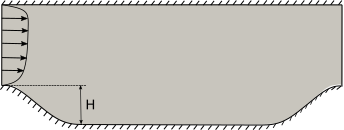}}
    \subfloat[mesh]{\includegraphics[width=0.4\textwidth]{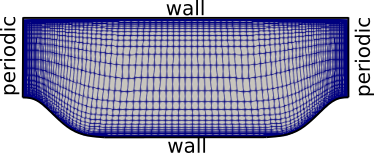}}
    \caption{Computational setup of periodic hill case. Figure (a) presents the computational domain where the crest height is $H$. Figure (b) shows the mesh and boundary setup.}
    \label{fig:mesh_pehills}
\end{figure}

\begin{figure}[!htb]
    \centering
    \includegraphics[width=0.8\textwidth]{RANS-legend_Ux_profile.pdf}\\
    \subfloat[eddy viscosity]{\includegraphics[width=0.33\textwidth]{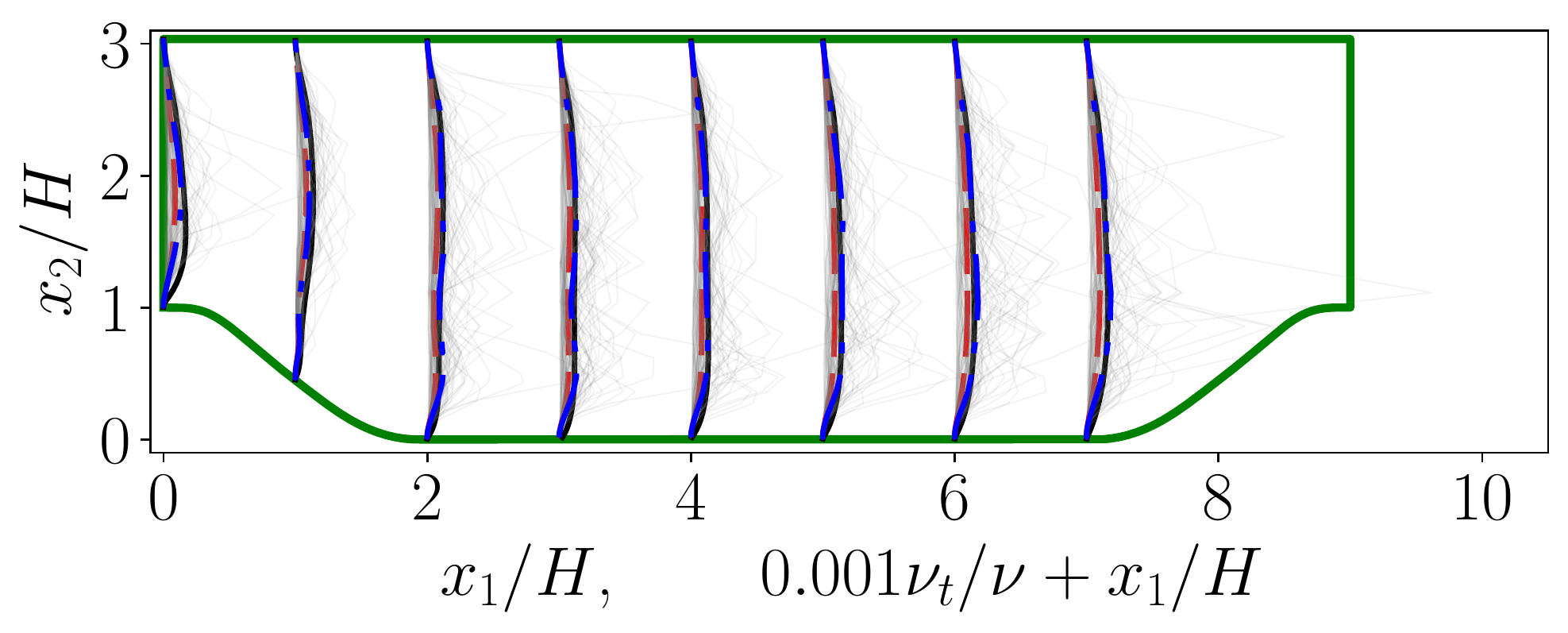}}
    \subfloat[streamwise velocity]{\includegraphics[width=0.33\textwidth]{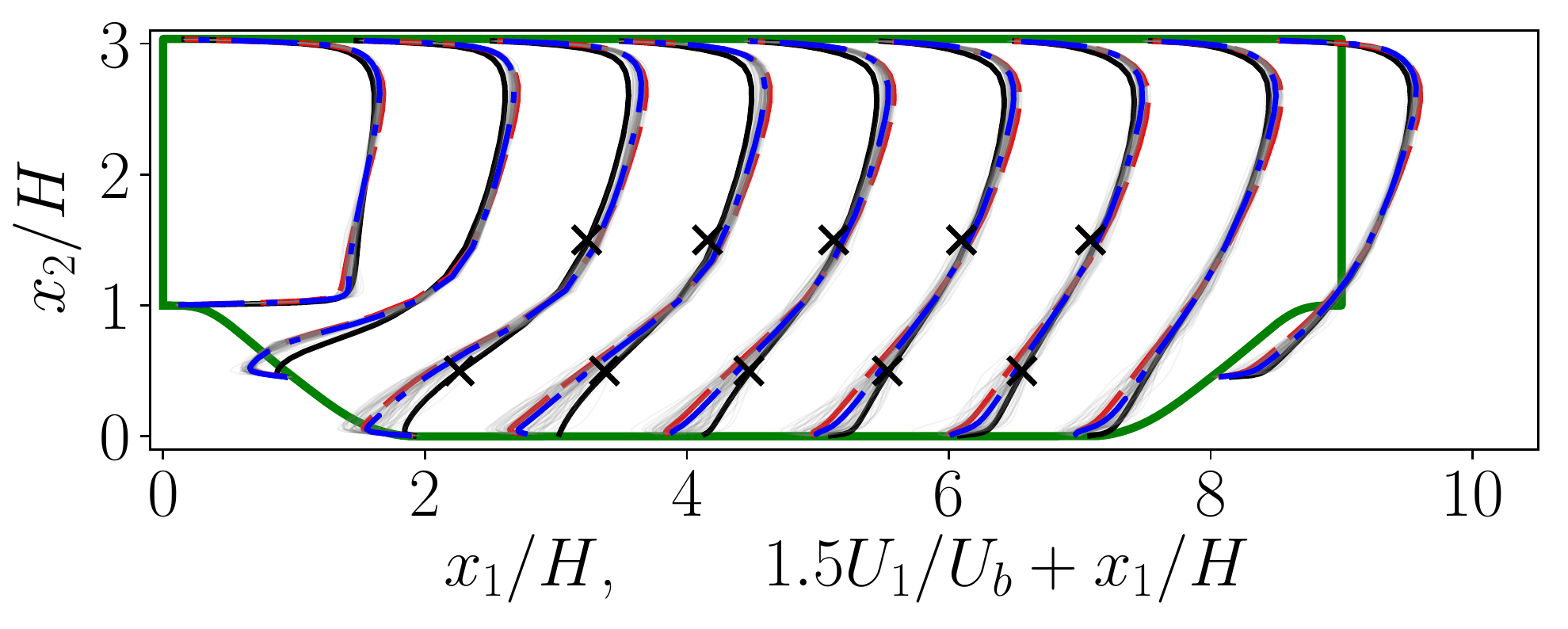}}
    \subfloat[wall pressure distribution]{\includegraphics[width=0.3\textwidth]{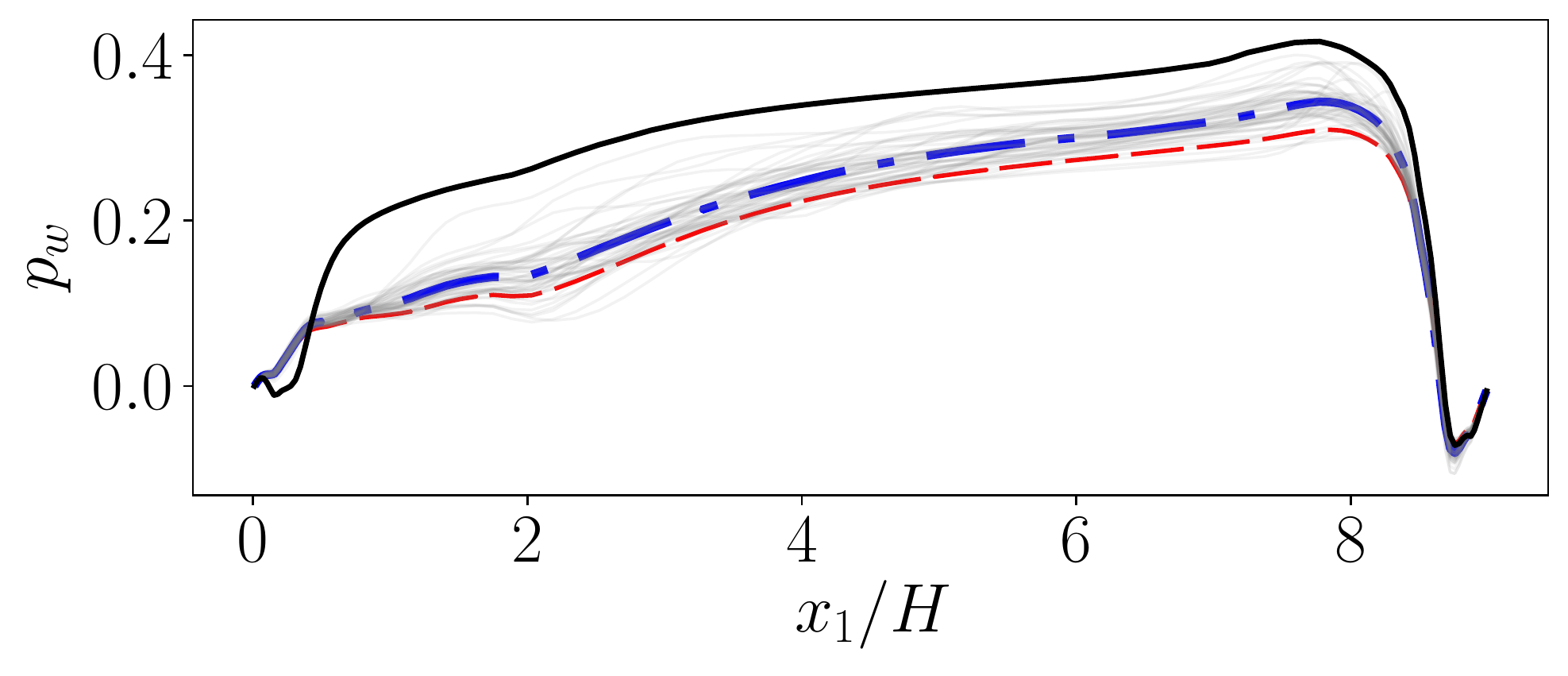}}
    \caption{Prior samples of \added[id=R1]{the eddy viscosity and} the propagated velocity and wall pressure distribution for periodic hill case. The observed position of sparse velocity~$\oy_1$ is indicated with crosses ($\times$).}
    \label{fig:pehills_prior}
\end{figure}

The above two cases both have zero pressure gradient, and we used the wall shear stress related information as disparate data to enforce the flow reconstruction.
Both the observed quantities are associated with the velocity gradient adjacent to the wall.
For the third case, we incorporate a different data source, i.e., pressure along the wall, to reconstruct the pressure and velocity fields simultaneously.
We choose the turbulent flow over periodic hills~\cite{frohlich2005highly}, which is an internal flow widely used for evaluating turbulence models~\cite{xiao2012consistent} and has been extended for a wide range of Reynolds numbers~\cite{breuer2009flow} and different hills geometries~\cite{xiao2012consistent}.
In this case, the Reynolds number based on the bulk velocity~$U_b$ and crest height~$H$ is $5600$.
The inlet and outlet are imposed as the periodic boundary condition.
The bottom and top are solid walls with no-slip condition.
The mesh is constructed with $100$ in the streamwise direction, $30$ in the normal to wall direction.
The computational setup is shown in Fig.~\ref{fig:mesh_pehills}.
In this case we regard the RANS results with Spalart-Allmaras model as the baseline.
The RANS results with $k$--$\varepsilon$ model are used as a synthetic truth to test the proposed framework.
The measurement data are taken from the synthetic truth, while the results from the baseline are referred to as a prior.
The length scale and variance in Eq.~\eqref{eq:kernel} is set as $0.25H$ and $1.0$, respectively.
The number of samples is $100$, and the first~$300$ modes are used to cover more than~$99\%$ variance.
The plots of the prior realizations are shown in Fig.~\ref{fig:pehills_prior}.
The $10$ sparse velocity observations are considered as shown in Fig.~\ref{fig:pehills_prior} \replaced[id=R1]{b}{a}.
The relative error is~$0.1\%$.
In this case, the disparate data from other physical fields, i.e., the pressure coefficient along the bottom wall, is regarded as another data with a relative error of $1.0\%$ to enhance the turbulent flow reconstruction.
The ten observed positions of wall pressure is evenly distributed along the wall. 

\begin{figure}[!htb]
    \centering
    \includegraphics[width=0.8\textwidth]{RANS-legend_Ux_profile.pdf}\\
    \subfloat[$U_1$: assimilate $p_w$ ]{\includegraphics[width=0.33\textwidth]{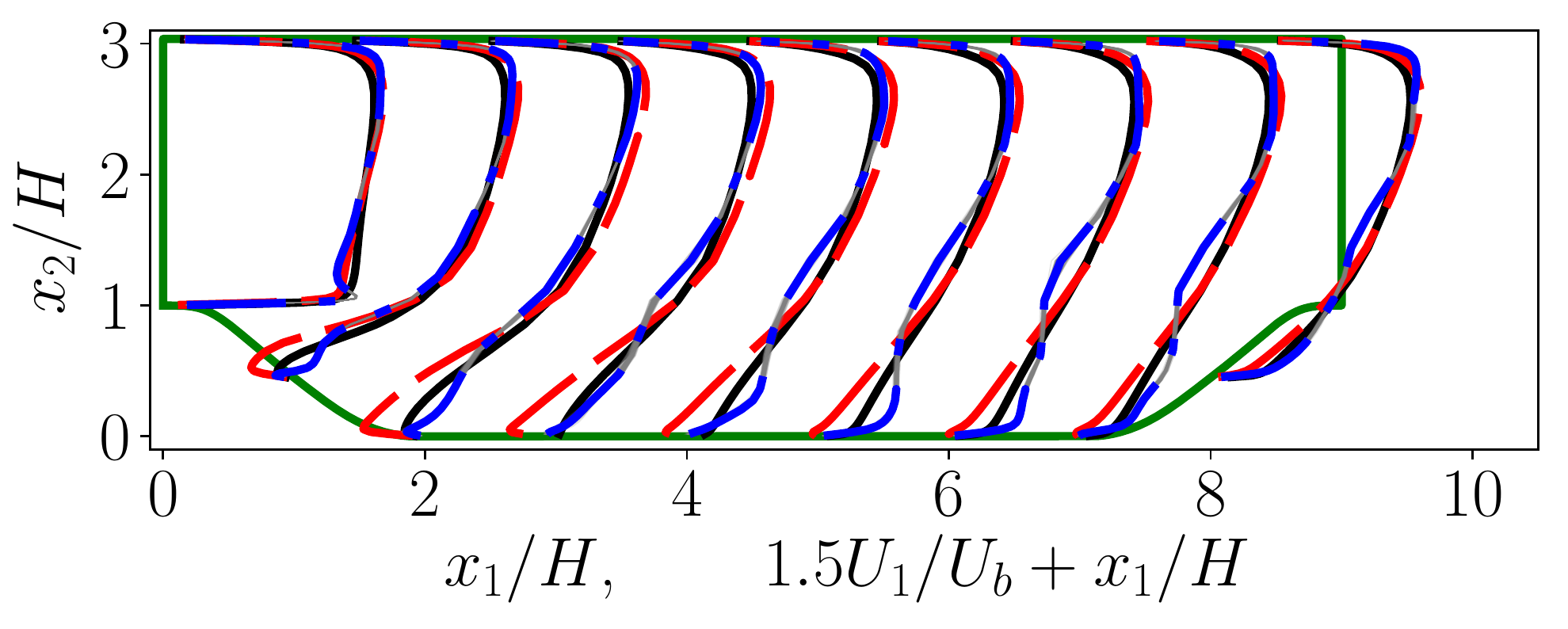}}
    \subfloat[$U_1$: assimilate $U_1$]{\includegraphics[width=0.33\textwidth]{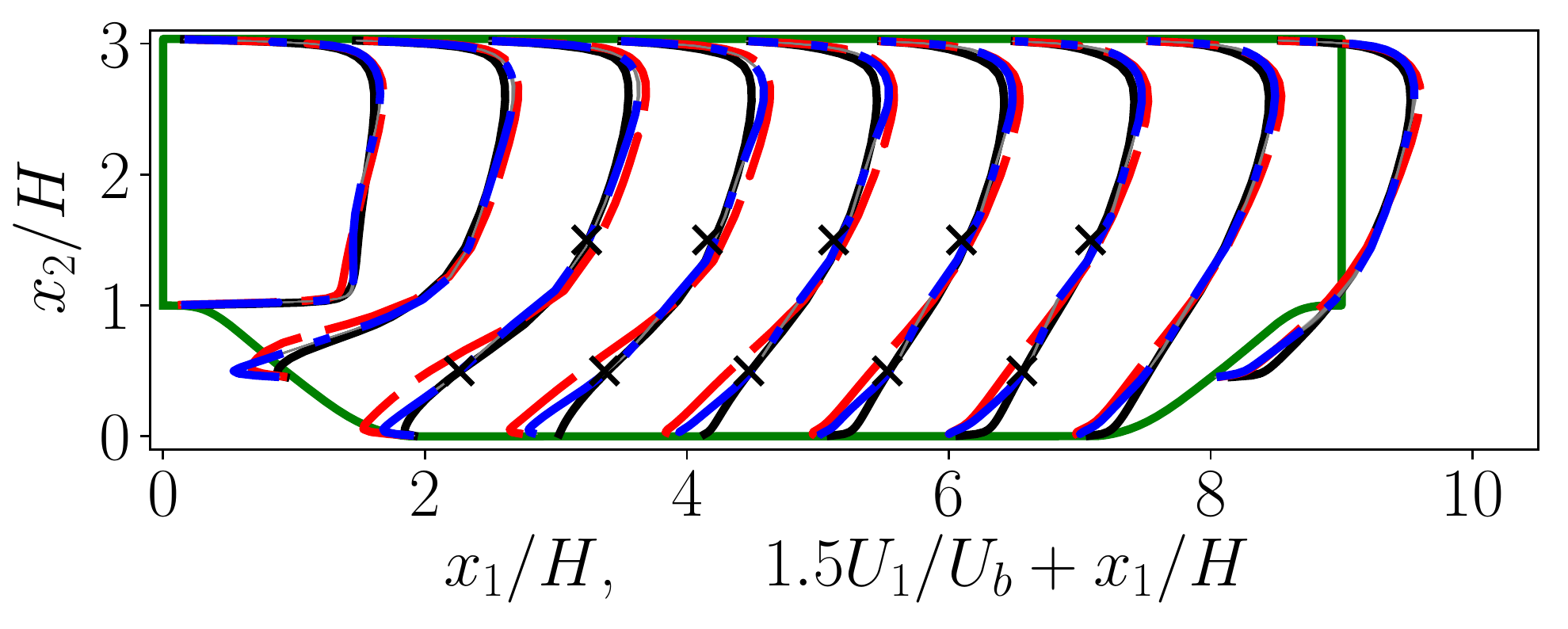}}
    \subfloat[$U_1$: assimilate $p_w$ and $U_1$]{\includegraphics[width=0.33\textwidth]{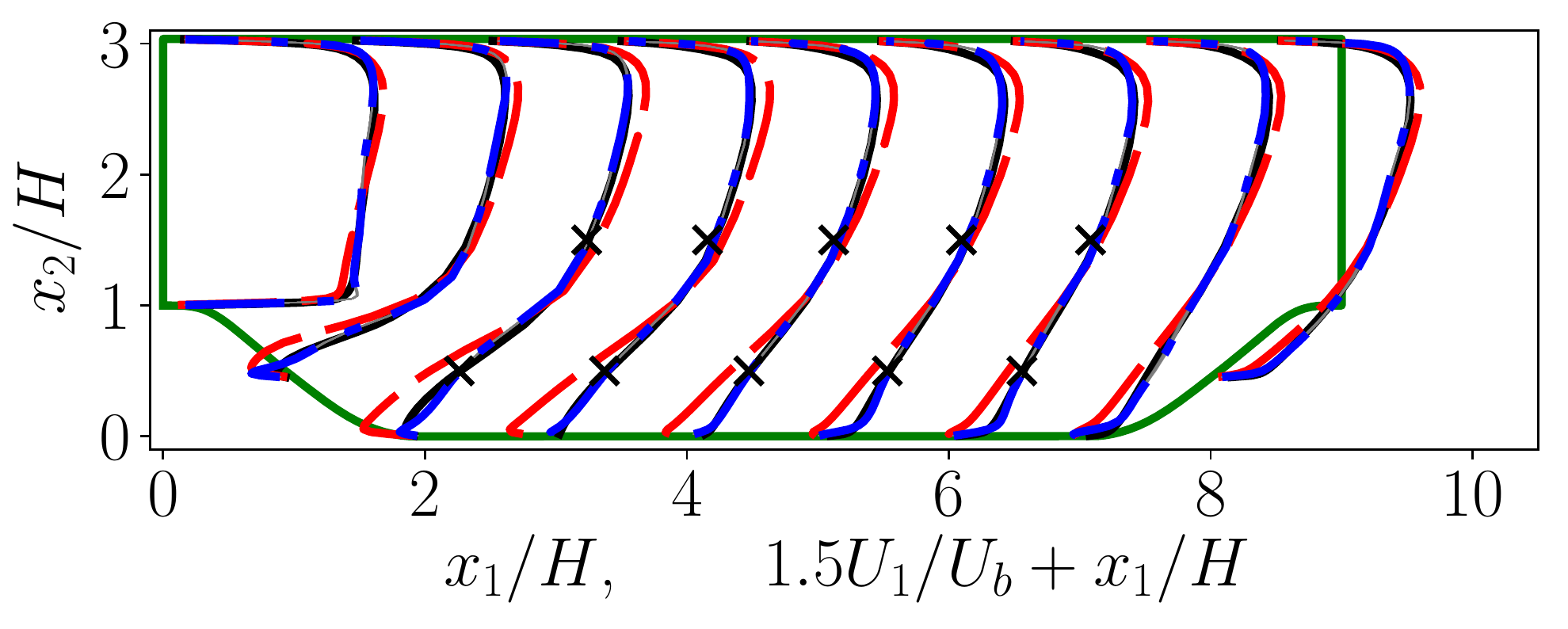}}\\
    \subfloat[$p^*$: assimilate $p_w$]{\includegraphics[width=0.33\textwidth]{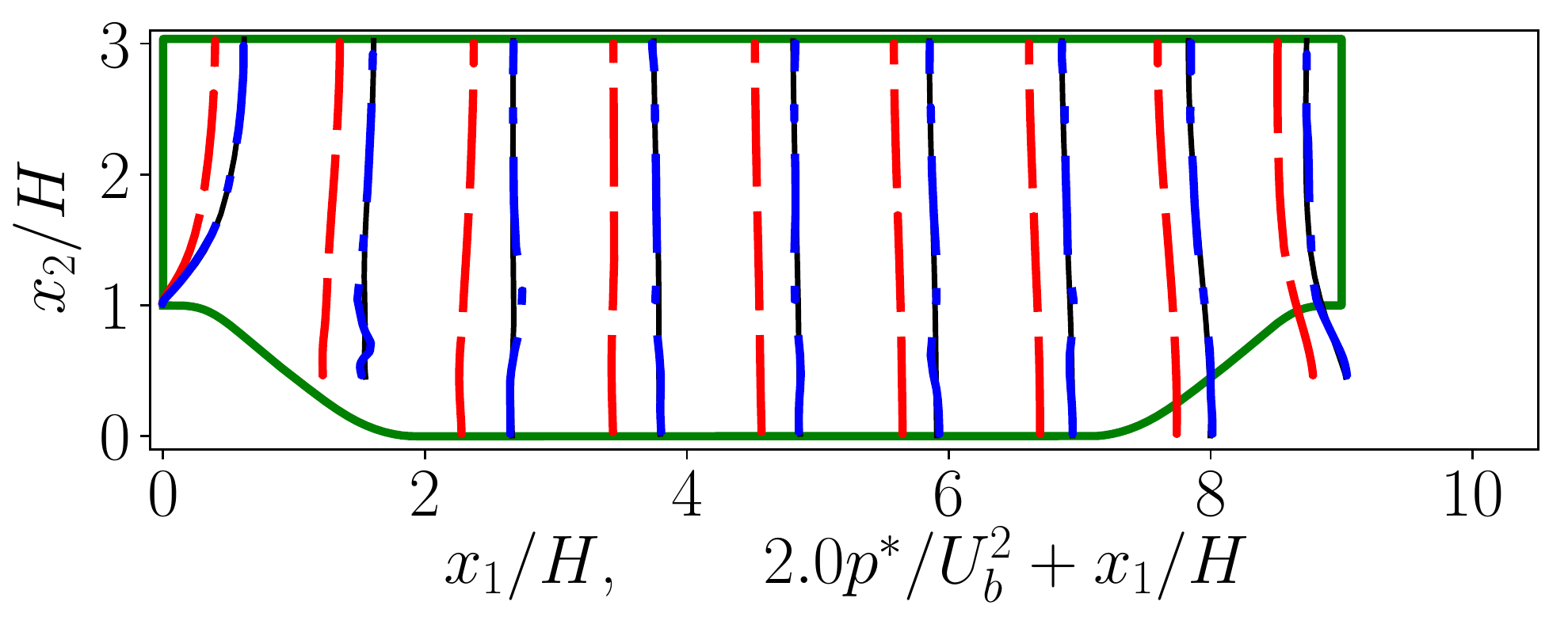}}
    \subfloat[$p^*$: assimilate $U_1$]{\includegraphics[width=0.33\textwidth]{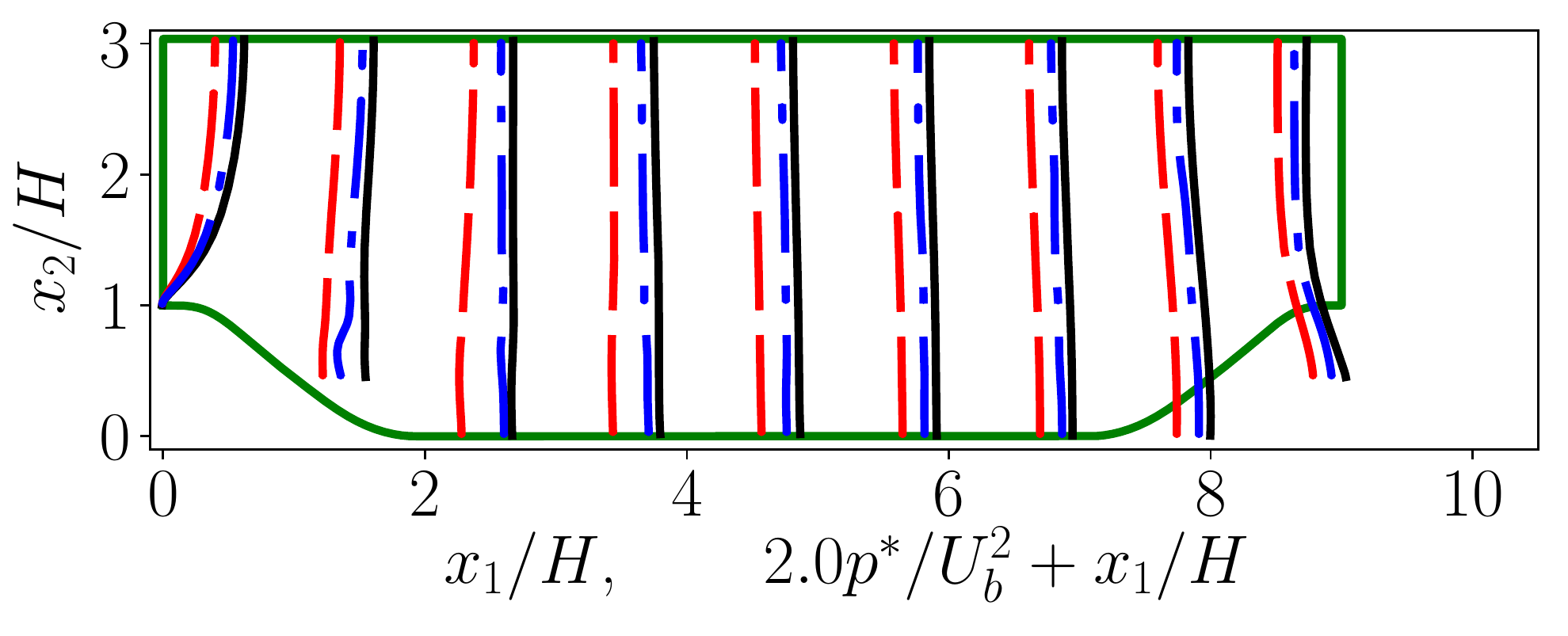}}
    \subfloat[$p^*$: assimilate $p_w$ and $U_1$]{\includegraphics[width=0.33\textwidth]{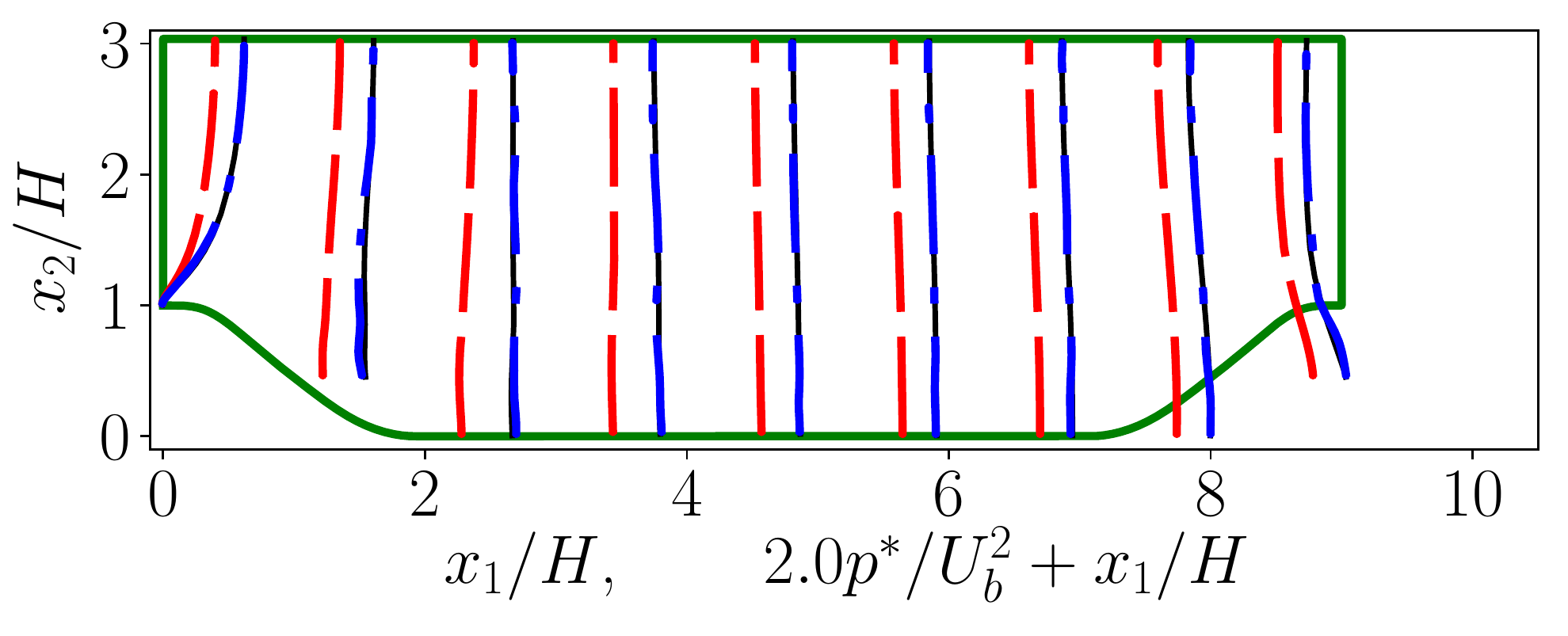}}\\
    \subfloat[$p_w$: assimilate $p_w$]{\includegraphics[width=0.33\textwidth]{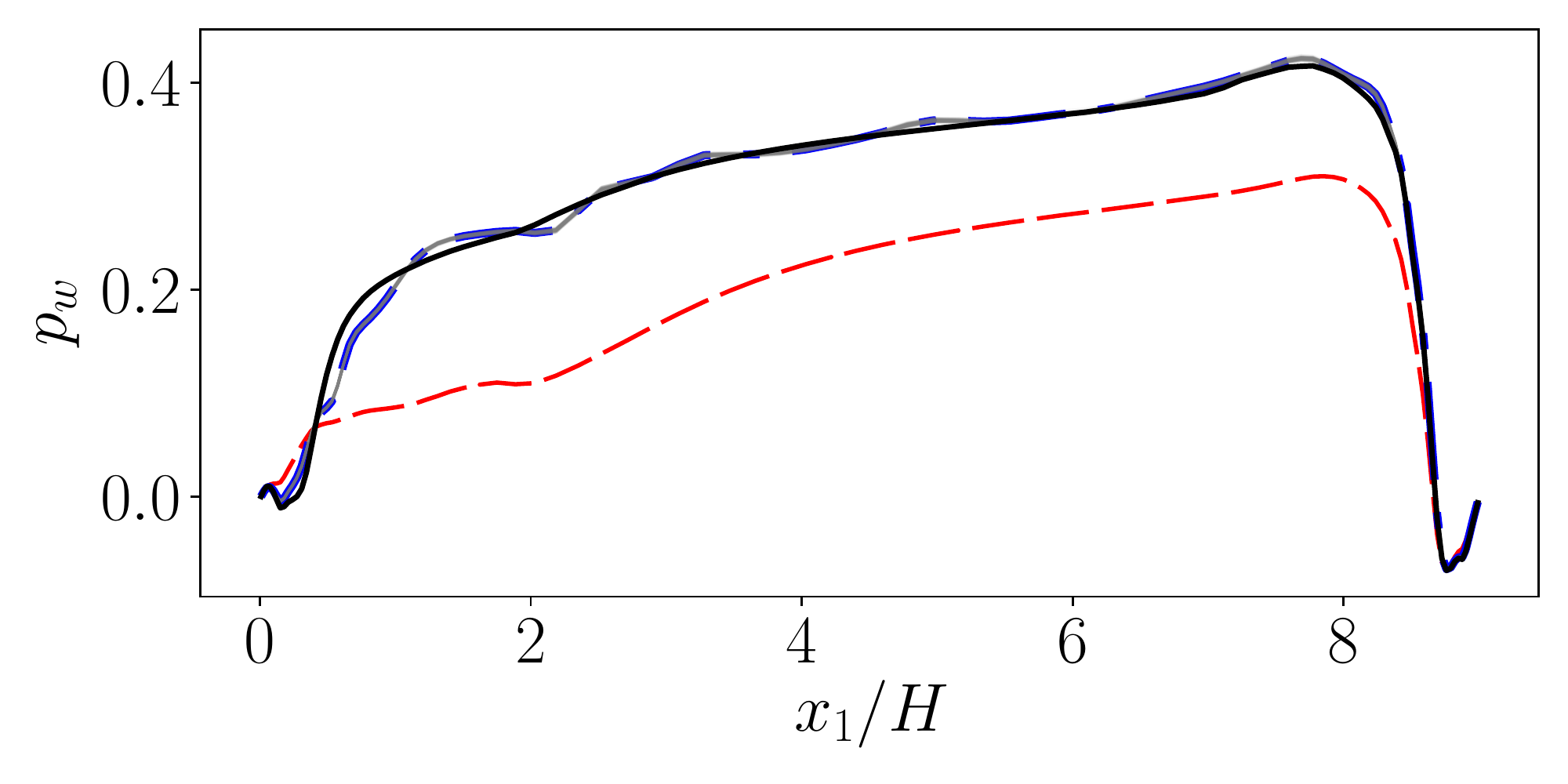}}
    \subfloat[$p_w$: assimilate $U_1$]{\includegraphics[width=0.33\textwidth]{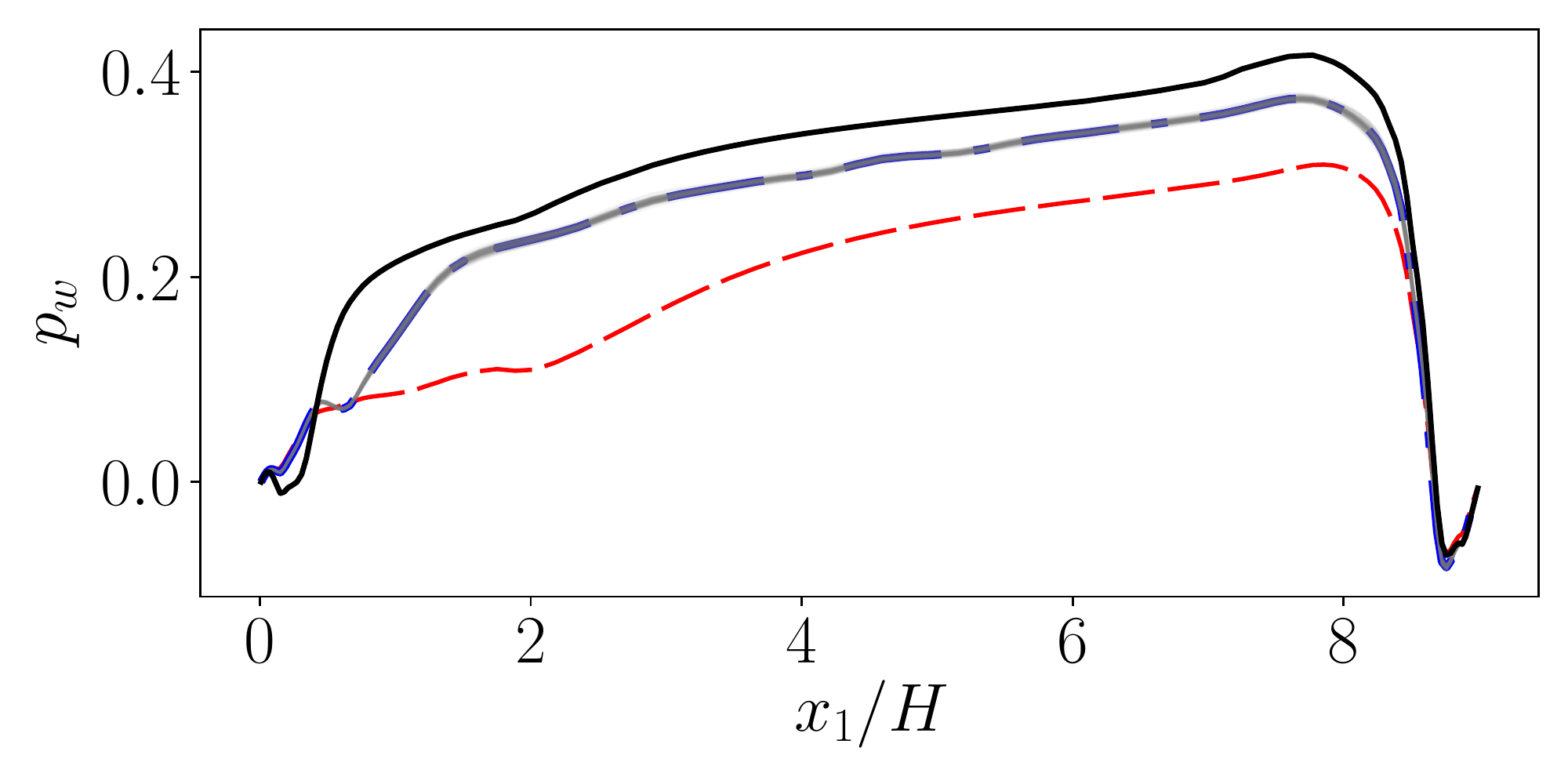}}
    \subfloat[$p_w$: assimilate $p_w$ and $U_1$]{\includegraphics[width=0.33\textwidth]{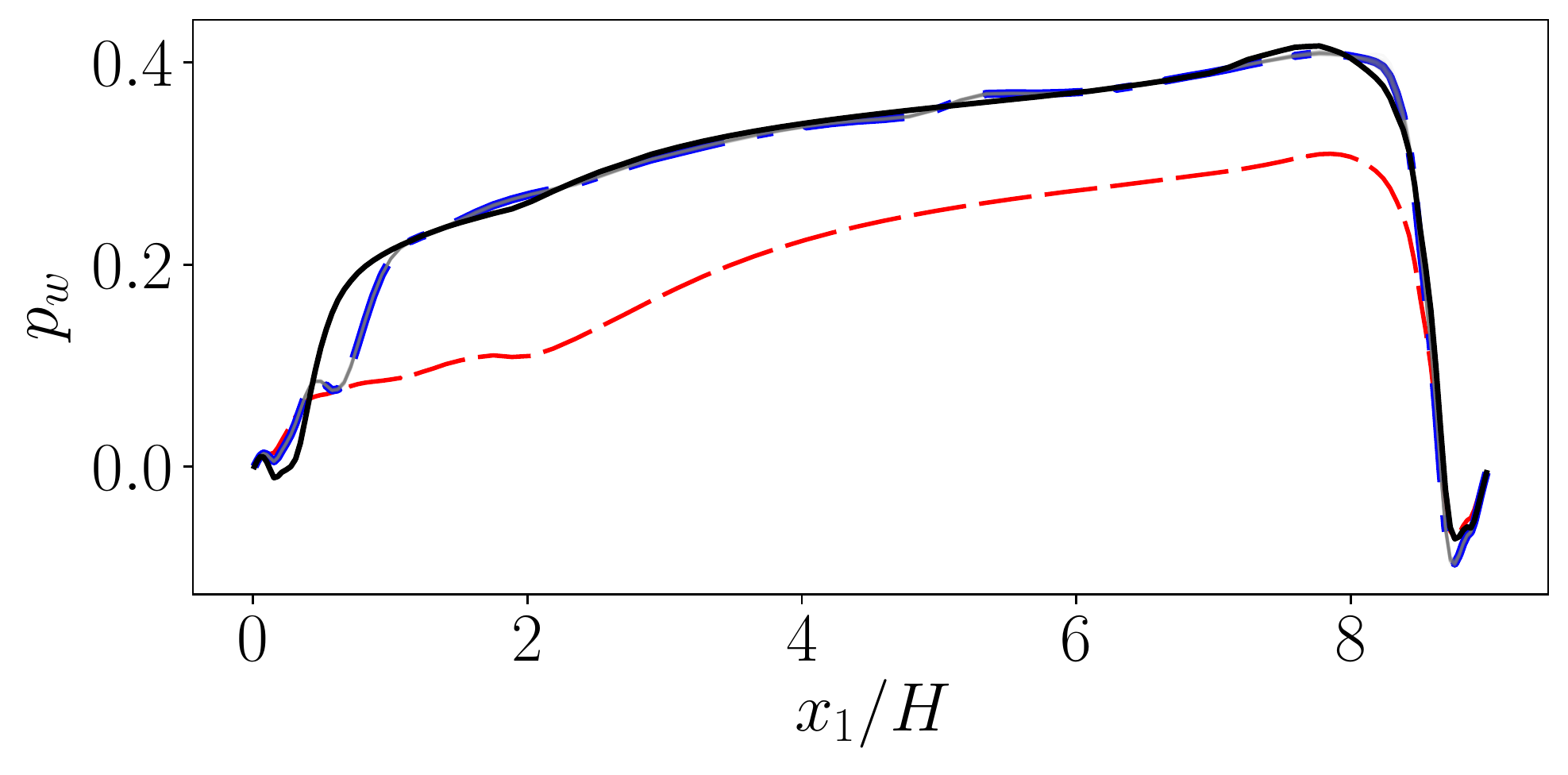}} \\
    \subfloat[$\nu_t$: assimilate $p_w$]{\includegraphics[width=0.33\textwidth]{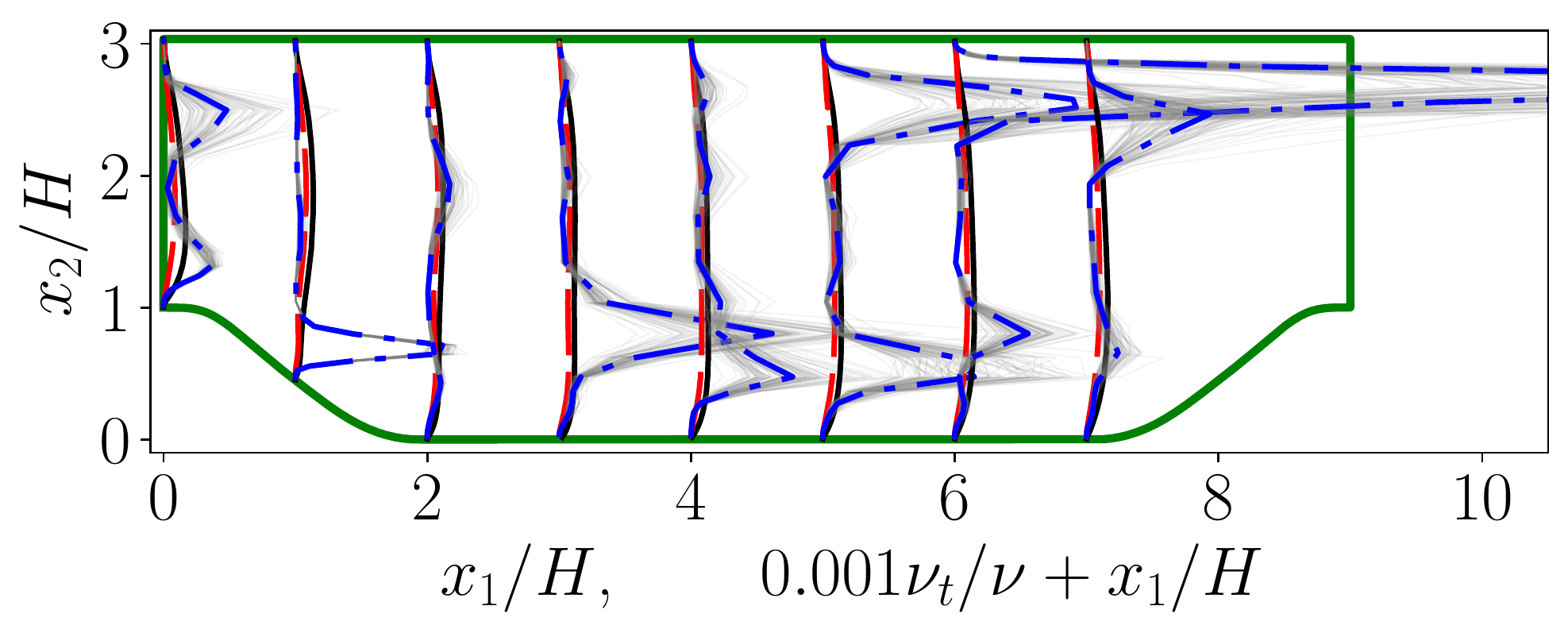}}
    \subfloat[$\nu_t$: assimilate $U_1$]{\includegraphics[width=0.33\textwidth]{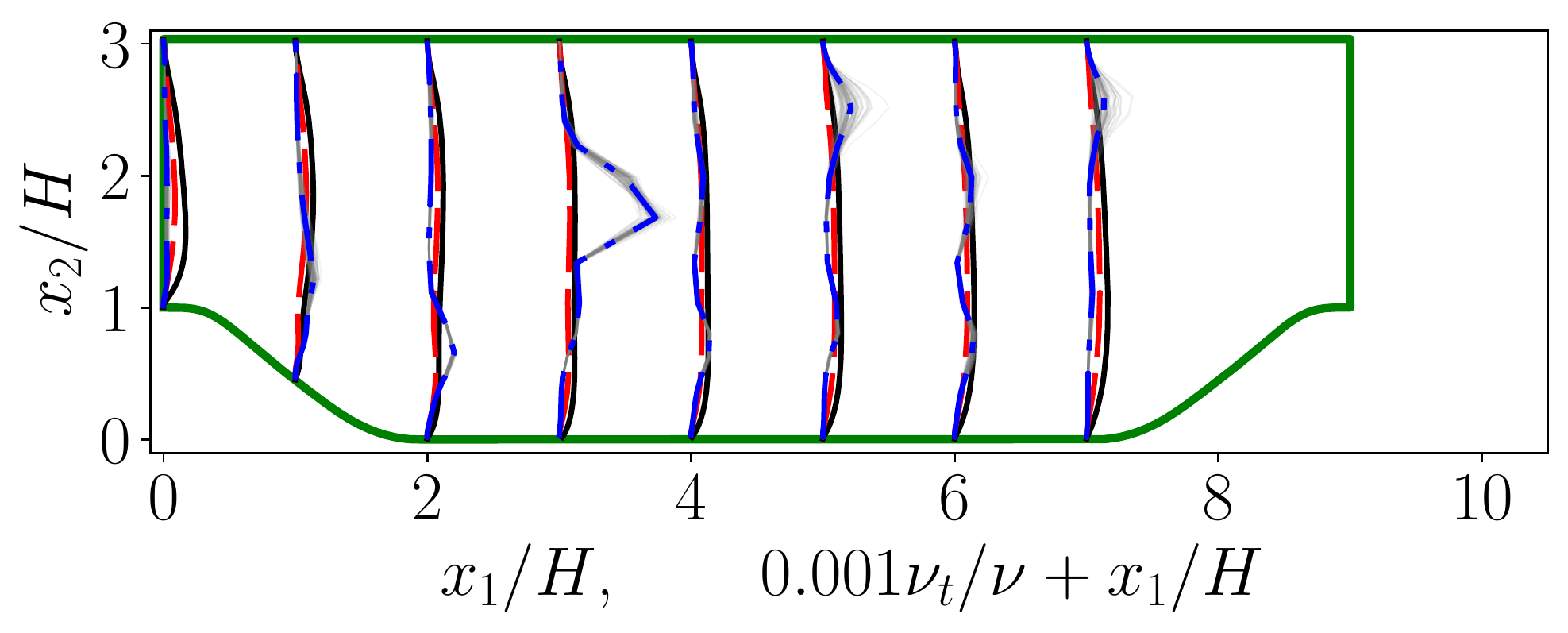}}
    \subfloat[$\nu_t$: assimilate $p_w$ and $U_1$]{\includegraphics[width=0.33\textwidth]{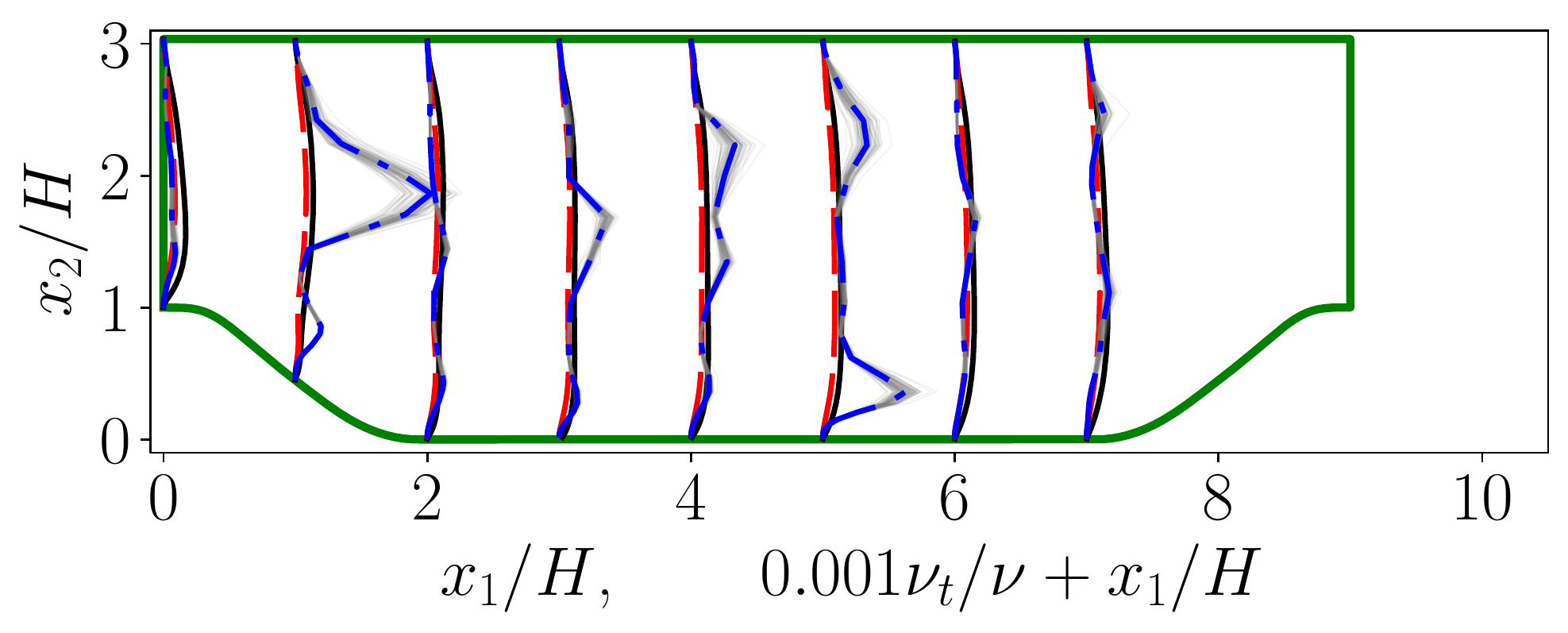}}
    \caption{Data assimilation results of velocity, pressure, \replaced[id=R1]{wall pressure, and eddy viscosity}{and wall pressure} by incorporating different observation data for periodic hill case}
    \label{fig:pehills}
\end{figure}

\begin{figure}[!htb]
    \centering
    \includegraphics[width=0.25\textwidth]{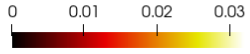}\\
    \subfloat[truth $U_1$]{\includegraphics[width=0.3\textwidth]{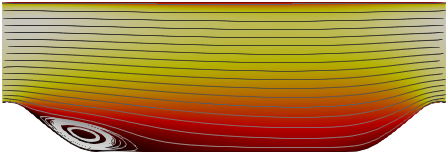}}
    \subfloat[prior $U_1$]{\includegraphics[width=0.3\textwidth]{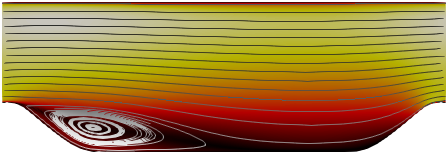}}\\
    \subfloat[posterior $U_1$: assimilate $p_w$]{\includegraphics[width=0.3\textwidth]{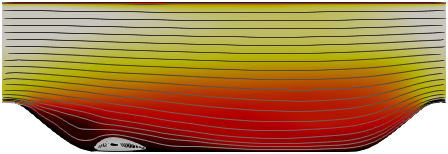}}
    \subfloat[post. $U_1$: assimilate $U_1$]{\includegraphics[width=0.3\textwidth]{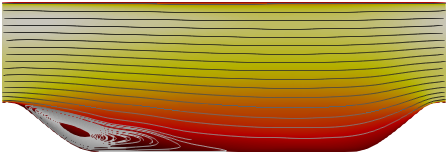}}
    \subfloat[post. $U_1$: assimilate~ $U_1$, $p_w$]{\includegraphics[width=0.3\textwidth]{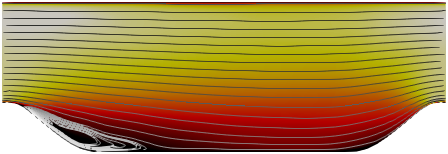}} \\
    \caption{Contour plots of streamwise velocity with comparison among the truth, prior mean, and posterior mean by assimilating different observation data for periodic hill case.}
    \label{fig:pehills_contour_vel}
\end{figure}

\begin{figure}[!htb]
    \centering
    \includegraphics[width=0.3\textwidth]{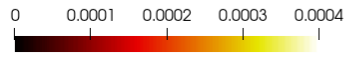}\\
    \subfloat[truth $p^*$]{\includegraphics[width=0.3\textwidth]{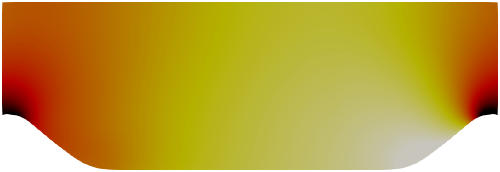}}
    \subfloat[prior $p^*$]{\includegraphics[width=0.3\textwidth]{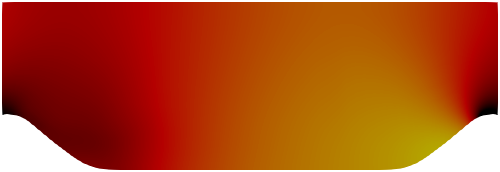}} \\
    \subfloat[posterior $p^*$: assimilate $p_w$]{\includegraphics[width=0.3\textwidth]{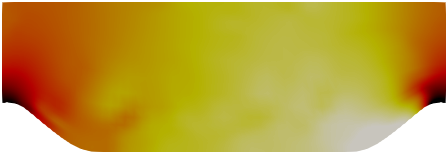}}
    \subfloat[posterior $p^*$: assimilate $U_1$]{\includegraphics[width=0.3\textwidth]{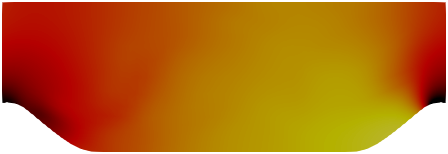}}
    \subfloat[posterior $p^*$: assimilate $U_1$, $p_w$]{\includegraphics[width=0.3\textwidth]{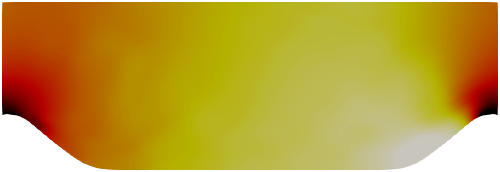}}
    \caption{Contour plots of pressure with comparison among the truth, prior mean, and posterior mean by assimilating different observation data for periodic hill case.}
    \label{fig:pehills_contour_pressure}
\end{figure}

The results for the periodic hill case with different observations are shown in Fig.~\ref{fig:pehills}.
As a baseline case, we only consider the wall pressure distribution with standard EnKF, and the results are plotted in Figs.~\ref{fig:pehills}a, d, g.
It can be seen that by only assimilating the wall pressure distribution, the result of wall pressure is in good agreement with data, and the entire pressure field can be reconstructed very well simultaneously.
However, the reconstructed velocity would have a large difference from the reference, as shown in Fig.~\ref{fig:pehills}a.
On the other hand, when only assimilating the sparse velocity, the results show that the reconstructed velocity away from the wall can be well recovered, but the near-wall velocity and the pressure field have relatively large discrepancy from the reference data, as presented in Figs.~\ref{fig:pehills}b, e, h.
Further, we assimilate both the sparse velocity and the wall pressure coefficient simultaneously.
The results show that both the pressure and velocity fields are in good agreement with the reference data.
That is likely because the pressure information is able to improve the estimation of the adverse pressure gradient, which is beneficial to the velocity reconstruction near the wall in this case.
\added[id=R1]{
The plots of the inferred eddy viscosity are shown in Figs.~\ref{fig:pehills}j, k and l.
The inferred eddy viscosity by assimilating only wall pressure results in very large discrepancy from truth in roughness and magnitude. This is the same as the T3A plate case, indicating that data assimilation with only wall measurement are not well-conditioned.
In the cases of assimilating $U_1$ and assimilating both $p_w$ and $U_1$, the inferred eddy viscosity gets smoother and has the smaller magnitudes, compared to the case of assimilating only~$p_w$.
}
The contour plots of the velocity and pressure field for the case of assimilating both velocity and wall pressure are provided in Figs.~\ref{fig:pehills_contour_vel} and~\ref{fig:pehills_contour_pressure}, respectively.
It is observed that the separation bubble size of the posterior mean by assimilating both $U_1$ and $p_w$ are closer to the synthetic truth compared to the prior mean. 
\added[id=All]{
However, the flow structures of the assimilated results are topologically different from the truth as shown in Fig.~\ref{fig:pehills_contour_vel}.
That is likely because of the ill-posed nature of inferring the full field with limited observation.
To address this issue, optimal sensor placements are of significant interest for further investigation by placing the sensor in the position where the observation data can identify the structure of recirculation zone.
}
Moreover, the pressure field is significantly improved by considering the wall pressure as shown in Fig.~\ref{fig:pehills_contour_pressure}.

\begin{figure}[!htb]
    \centering
    \includegraphics[width=0.6\textwidth]{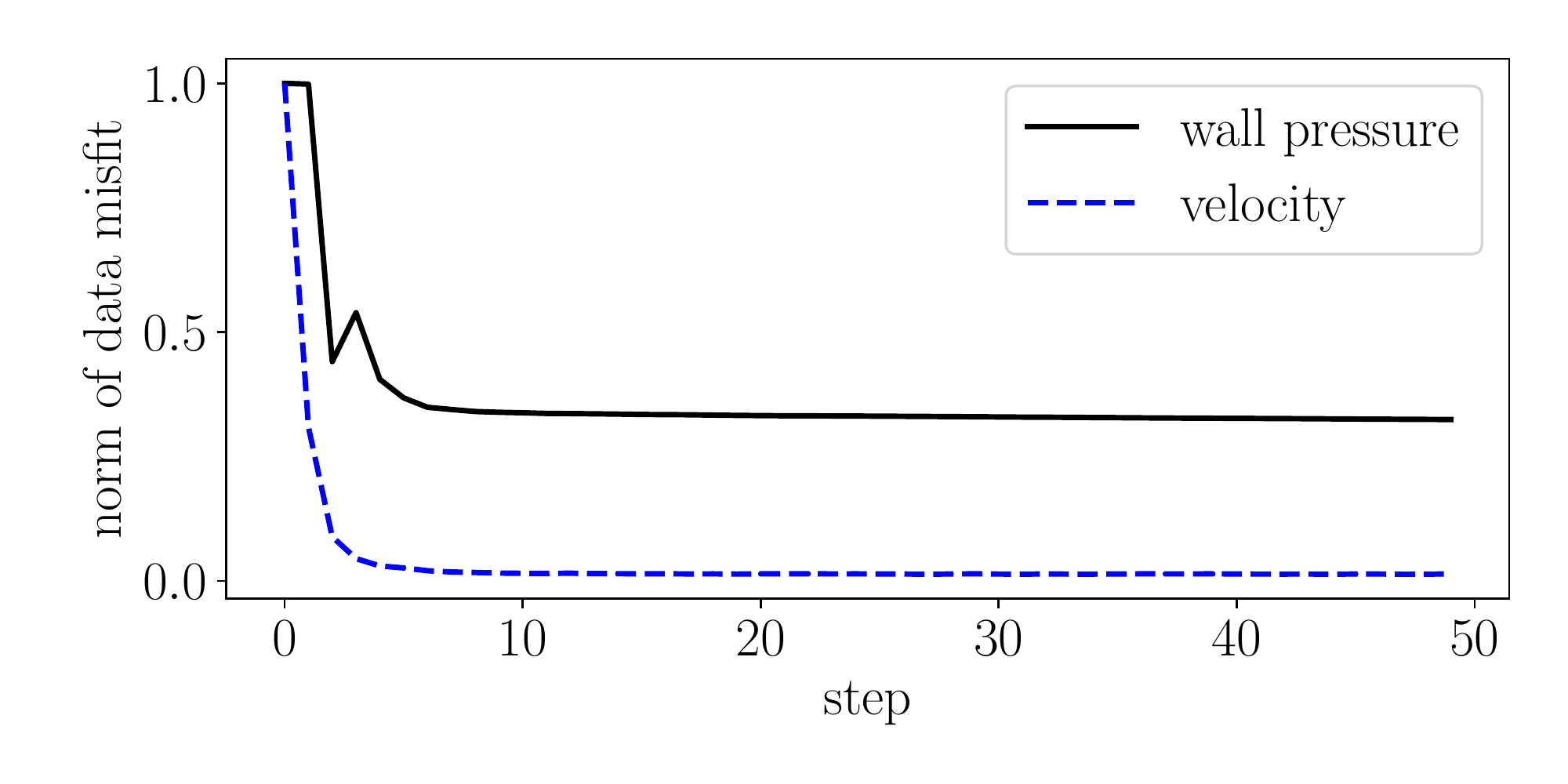}
    \caption{Convergence feature of REnKF in the normed data misfit for periodic hills case. The data misfit is normed by the initial misfit to keep the convergence curve in the same magnitude.}
    \label{fig:convergence_pehill}
\end{figure}

The convergence plot of the case is provided in Fig.~\ref{fig:convergence_pehill}.
It can be seen that the REnKF method can reduce the data misfit of both velocity and pressure robustly.
The discrepancy of the sparse velocity and wall pressure reduce significantly in the first 10 steps.
At around the fourth iteration step, the data misfit in wall pressure increase slightly, likely due to that the regularization term is negligible and the misfit of velocity is dominant.
Afterwards, the regularization term is strengthened to be dominant, and the misfit of the wall pressure keep decreasing.

\begin{figure}[!htb]
    \centering
    \includegraphics[width=0.8\textwidth]{RANS-legend_Ux_profile.pdf}\\
    \subfloat[$U_1$]{\includegraphics[width=0.45\textwidth]{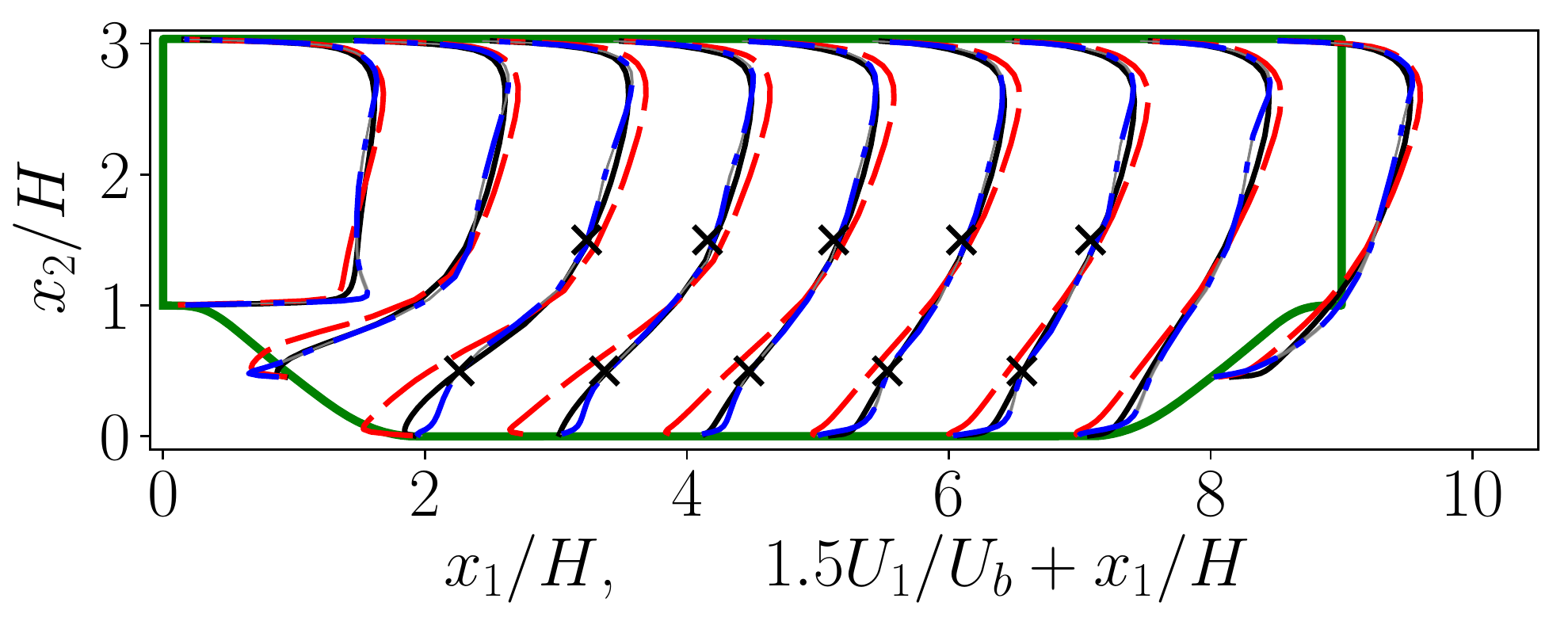}}
    \subfloat[$p^*$]{\includegraphics[width=0.45\textwidth]{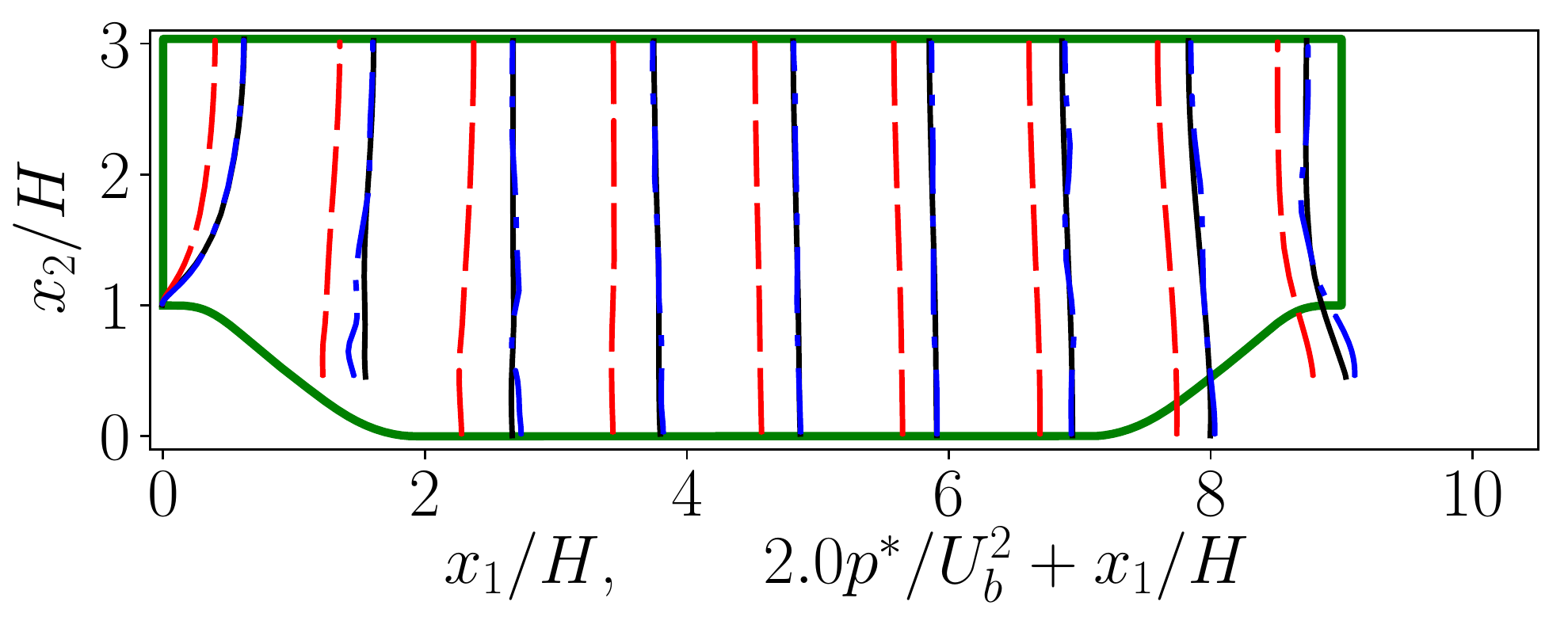}} \\
    \subfloat[$p_w$ ]{\includegraphics[width=0.45\textwidth]{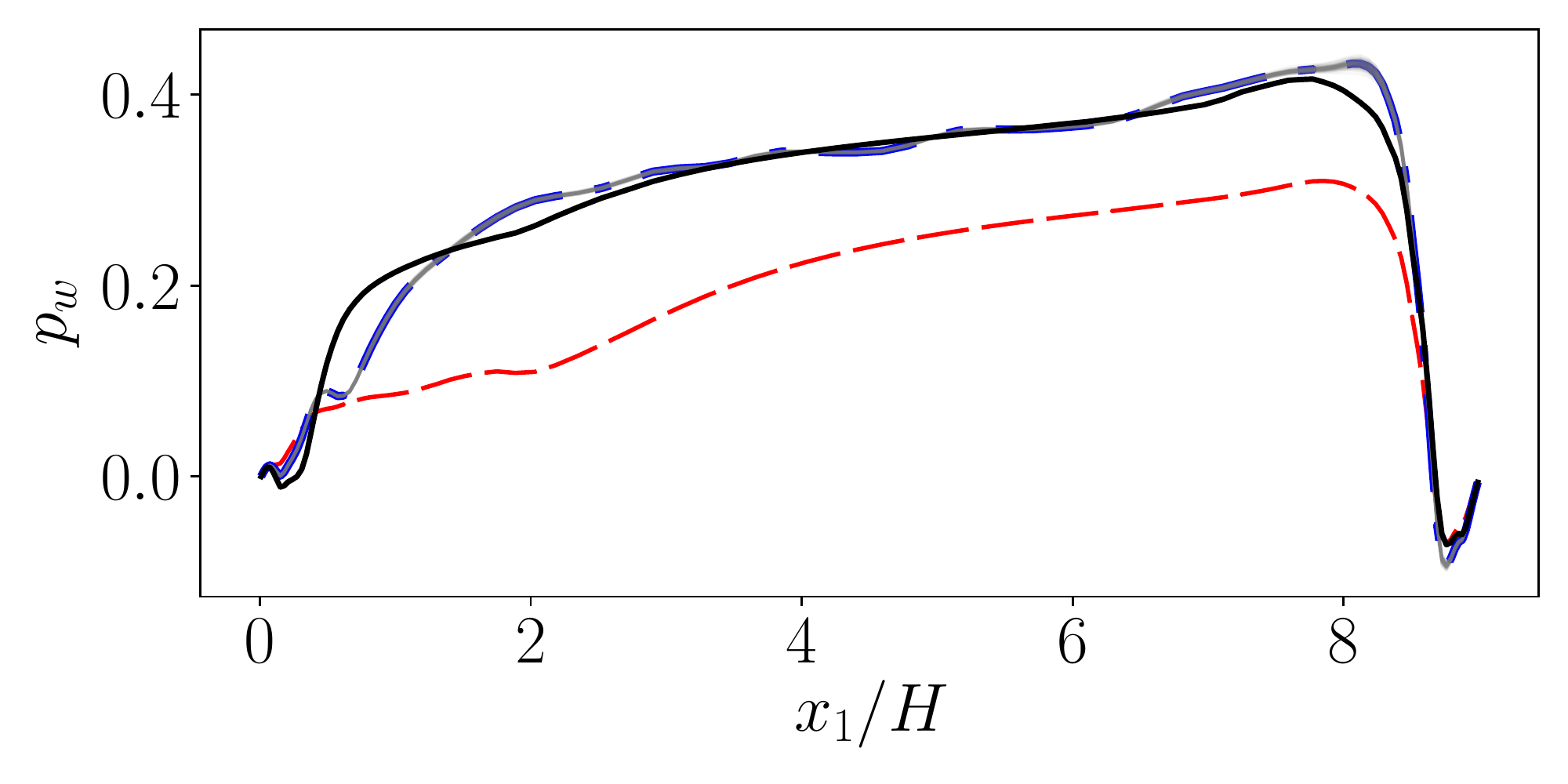}}
    \caption{Data assimilation results of velocity, pressure and wall pressure by incorporating sparse velocity and wall pressure integration for periodic hill case.}
    \label{fig:pehills_integral}
\end{figure}

Further, we investigate to use only the integral data, i.e., the integration of pressure along the wall or lift and drag force, to improve the reconstruction.
The pressure integration is defined as
\begin{equation}
    \mathbf{F} = \int p_w \vec{n} d s \text{,}
\end{equation}
where $\vec{n}$ is the unit vector in the normal direction and $s$ is the wall surface area.
The results of REnKF assimilating both sparse velocity and pressure integration are shown in Fig.~\ref{fig:pehills_integral}.
It can be seen that the integral information can also significantly improve the entire pressure distribution along the wall.
Both the velocity and pressure fields are improved, while only some local regions near the hill have noticeable discrepancies from the reference data.

The error in the reconstructed velocity field and wall pressure for this case is summarized in Table.~\ref{tab:summary_results}, which shows the superiority of disparate data assimilation clearly.
\added[id=R1]{
Moreover, it can be seen that the discrepancies are similar between the last two cases where the wall pressure distribution and the integrated pressure are regarded as disparate data.
It shows that the integral data, i.e., the lift and drag force, and the surface data source, i.e., the wall pressure distribution, can achieve similar performance of flow reconstruction. 
In this specific case, what affects the flow reconstruction seems not the disparity of data but the position of the data.
From this viewpoint, the sensor placements are substantially important for the data assimilation in the applications of flow reconstruction.
However, from another viewpoint, at different regions different types of sensors are usually used to provide the disparate observation data having different physical quantities.
For instance,  pressure sensors are placed at the wall to measure the pressure fluctuations; LDV or hot wires are usually used to measure the velocity within the flow field; microphones are often deployed at the far-field to measure the noise.
Hence, here we demonstrate that using these disparate data which are often obtained from different sensors can enhance the flow reconstruction. 
The optimal placement of these sensors is of significant interest for future investigations.
}

\begin{table}[!htbp]
    \centering
    \begin{tabular}{c|c|c|c|c}
    \hline
        Geometry & Filter & Observation & Error($\mathsf{Hx}$) & Error($\mathsf{Dx}$)  \\
        \hline
        Channel &  EnKF & \replaced[id=R1]{$u_\tau$}{$U^*$} & $3.27 \%$  & $6.68 \%$ \\
        & EnKF & $U_1$ & $3.01 \%$ & $46.9 \%$ \\
        & REnKF & \replaced[id=R1]{$u_\tau$}{$U^*$}, $U_1$ & $1.41 \%$ & $0.93 \%$ \\
        \hline
         T3A plate &  EnKF & $C_f$ &  $8.38\%$ & $11.2\%$ \\
        & EnKF & $U_1$ & $7.08\%$ & $62.4 \%$ \\
        & REnKF & $C_f$, $U_1$ & $6.33\%$ & $50.7\%$ \\
        \hline
       Periodic hills  &  EnKF & $p_w$ & $9.84 \%$ & $4.55 \%$ \\
        & EnKF & $U_1$ & $8.37 \%$ & $15.7 \%$ \\
        & REnKF & $p_w$, $U_1$ & $7.24\%$ & $9.26\%$ \\
        & REnKF & $\mathbf{F}$, $U_1$ & $7.34 \%$ & $10.2 \%$ \\
    \hline
    \end{tabular}
    \caption{Summary of data assimilation results. The error for $\mathsf{Hx}$ and $\mathsf{Dx}$ is computed based on~\eqref{eq:error_def}. $\mathsf{Hx}$ represents~$U_1$. $\mathsf{Dx}$ indicates~\replaced[id=R1]{$u_\tau$}{$U^*$} in channel case, $C_f$ in T3A plate case, and~$p_w$ and~$\mathbf{F}$ in periodic hill case. 
    }
    \label{tab:summary_results}
\end{table}

\section{Conclusion}
Reconstruction of turbulent flow based on data assimilation methods is of significant interest to improve flow-field estimation from limited observations.
Various disparate data sources such as velocity, wall shear stress, and wall pressure are measurable with different measurement techniques.
Incorporating these disparate data sources is a promising method to enhance the reconstruction of turbulent flows.
This work investigates the disparate data assimilation with ensemble methods to enhance the reconstruction of turbulent mean flows.
A regularized ensemble Kalman method is employed to incorporate both the sparse velocity and the wall measurements.
Three numerical examples are used to demonstrate the capability of the proposed framework to assimilate different data sources, including wall friction velocity, friction coefficient, wall pressure distribution, and lift and drag force.
One-dimensional channel flow and two-dimensional flow over flat plate are first investigated to incorporate the sparse velocity and wall friction.
Further, the reconstruction of separated flows over periodic hills is explored with the proposed ensemble method.
The streamwise velocity and the wall pressure are regarded as the disparate observation data to reconstruct both the velocity and pressure fields.
The results show that incorporating the disparate data sources is capable of improving the accuracy of the flow reconstruction.
The ensemble method is non-intrusive and robust for reconstructing turbulent flows.
The proposed method is a promising tool for the reconstruction of turbulent flow fields to assimilate disparate data sources from experiments.

\section*{Acknowledgment}
XLZ, GWH, and SZW are supported by the NSFC Basic Science Center Program for ``Multiscale Problems in Nonlinear Mechanics'' (No. 11988102), the Strategic Priority Research Program of the Chinese Academy of Sciences (XDB22040104) and the Key Research Program of Frontier Sciences of the Chinese Academy of Sciences (QYZDJ-SSWSYS002).
\added[id=All]{
The authors would like to thank the reviewers for their constructive and valuable comments, which helped improve the quality and clarity of this manuscript.
}

\appendix

\clearpage
\section{Equivalence of EnKF and REnKF for disparate data assimilation}
\label{sec:Append_A}
To illustrate the connection between the EnKF and REnKF for disparate data assimilation,
we reformulate EnKF with two different data sources as two Kalman update steps by use of the saddle point matrix inverse~\cite{gupta2007kalman}.
The details of the derivation are presented as follows.

The conventional EnKF for disparate data assimilation is to augment the observation $\oy_1$ with the additional observation~$\oy_2$.
The augmented observation can be written as
\begin{equation}
    \oy_\text{aug} = 
    \begin{bmatrix} 
    \oy_1 \\
    \oy_2
    \end{bmatrix} \text{.}
\end{equation}
The Kalman update scheme need to be modified accordingly as
\begin{equation}
    \sx^\text{a} = \sx^\text{f} + \mathsf{K}_\text{aug}(\oy_\text{aug} - \mathsf{H}_\text{aug}[\sx^\text{f}]) \text{,}
    \label{eq:analysis_org}
\end{equation}
where
\begin{equation}
    \mathsf{K}_\text{aug} = \mathsf{PH}_\text{aug}^\top (\mathsf{H}_\text{aug}^\top \mathsf{P} \mathsf{H}_\text{aug} + \mathsf{R}_\text{aug})^{-1} \text{,}
\end{equation}
\begin{equation}
    \mathsf{R}_\text{aug} = 
    \begin{bmatrix}
    \mathsf{R} & 0\\
    0 & \mathsf{Q}
    \end{bmatrix} \text{, and}
    \quad
    \mathsf{H}_\text{aug} = 
    \begin{bmatrix}
    \mathsf{H} \\
    \mathsf{D}
    \end{bmatrix} \text{.}
\end{equation}
The augmented Kalman gain matrix can be reformulated as
\begin{equation}
\begin{aligned}
    \mathsf{K}_\text{aug} &= \mathsf{PH}_\text{aug}^\top (\mathsf{H}_\text{aug} \mathsf{P}_\text{aug} \mathsf{H}_\text{aug}^\top + \mathsf{R}_\text{aug})^{-1} \\
    &= \mathsf{P} \mathsf{H}_\text{aug}^\top
    \begin{bmatrix}
        \mathsf{HPH}^\top + \mathsf{R} & \mathsf{HPD}^\top \\
        \mathsf{DPH}^\top & \mathsf{DPD}^\top + \mathsf{Q}
    \end{bmatrix}^{-1} \\
    &= \mathsf{P} \mathsf{H}_\text{aug}^\top
    \begin{bmatrix}
        S_a^{-1} & S_b^{-1} \\
        S_c^{-1} & S_d^{-1}
    \end{bmatrix} \\
    &=
    \begin{bmatrix} 
    \mathsf{K}_a \; \mathsf{K}_b
    \end{bmatrix}
    \label{eq:K_block}
\end{aligned}
\end{equation}
where
\begin{equation}
\begin{aligned}
    \mathsf{K}_a &= \mathsf{PH}^\top S_a^{-1} + \mathsf{PD}^\top S_c^{-1} \\
    \mathsf{K}_b &= \mathsf{PH}^\top S_b^{-1} + \mathsf{PD}^\top S_d^{-1} \\
    S_a &= \mathsf{HPH}^\top + \mathsf{R} \\
    S_b &= \mathsf{HPD}^\top \\
    S_c &=\mathsf{DPH}^\top \\
    S_d &= \mathsf{DPD}^\top + \mathsf{Q} \text{.}
\end{aligned}
\label{eq:K_component}
\end{equation}
The analysis step~\eqref{eq:analysis_org} can be rewritten as
\begin{equation}
\begin{aligned}
    \mathsf{x}^a &= \mathsf{x}^f + 
    \begin{bmatrix} 
    \mathsf{K}_a \; \mathsf{K}_b
    \end{bmatrix}
    \begin{bmatrix}
    \oy_1 - \mathsf{Hx}^f\\
    \oy_2 - \mathsf{Dx}^f
    \end{bmatrix}
    \\
    &= \mathsf{x}^f + \mathsf{K}_a (\oy_1 - \mathsf{Hx}^f) + \mathsf{K}_b (\oy_2 - \mathsf{Dx}^f)
\end{aligned}
\label{eq:analysis_step}
\end{equation}
Now consider a matrix $S$ is a symmetric saddle point matrix of 
\begin{equation}
    S = 
    \begin{bmatrix}
    A & B^\top \\
    B & -C
    \end{bmatrix} \text{.}
\end{equation}
Its inverse can be written as 
\begin{equation}
    S^{-1} = 
    \begin{bmatrix}
    A^{-1} + A^{-1}B^\top F^{-1} B A^{-1} & -A^{-1} B^\top F^{-1} \\
    - F^{-1}BA^{-1} & F^{-1}
    \end{bmatrix}
\end{equation}
where $F=-(C + BA^{-1}B^\top)$~\cite{benzi2005numerical}.
Based on that, we assume
\begin{equation}
\begin{aligned}
    A &= (\mathsf{HPH}^\top + \mathsf{R}) \\
    B &= \mathsf{DPH}^\top \\
    C &= -\mathsf{DPD}^\top - \mathsf{Q} \\
    F &= (\mathsf{DPD}^\top + \mathsf{Q} - \mathsf{DPH}^\top(\mathsf{HPH}^\top + \mathsf{R})^{-1}(\mathsf{DPH}^\top)^\top) \\
      &=  (\mathsf{D} (\mathsf{P} - \mathsf{PH}^\top(\mathsf{HPH}^\top + \mathsf{R})^{-1}\mathsf{HP})\mathsf{D}^\top + \mathsf{Q} \\
      &= \mathsf{D} (I - \mathsf{KH}) \mathsf{P} \mathsf{D}^\top + \mathsf{Q}\\
      &= \mathsf{D} \mathsf{P}^* \mathsf{D}^\top + \mathsf{Q} \\
    \mathsf{K} &= \mathsf{PH}^\top(\mathsf{HPH}^\top + \mathsf{R})^{-1} \text{,}
\end{aligned}
\end{equation}
where $\mathsf{P}^* = (\mathsf{I} - \mathsf{KH})\mathsf{P}$ is the updated error covariance after assimilating the data $\oy_1$.
Under the assumption that both $A$ and $F$ are nonsingular, we can formulate the blocks in Eq.~\eqref{eq:K_block} as
\begin{equation}
\begin{aligned}
    S_a^{-1} &=  (\mathsf{HPH}^\top + \mathsf{R})^{-1} + (\mathsf{HPH}^\top + \mathsf{R})^{-1} \mathsf{HP}^\top D^\top (\mathsf{D}\mathsf{P}^*\mathsf{D}^\top + \mathsf{Q})^{-1} \mathsf{DK}\\
             &= (\mathsf{HPH}^\top + \mathsf{R})^{-1} + \mathsf{K}^\top \mathsf{D}^\top (\mathsf{DP}^* \mathsf{D}^\top + \mathsf{Q})^{-1}\mathsf{DK} \\
    S_b^{-1} &= - (\mathsf{HPH}^\top + \mathsf{R})^{-1} (\mathsf{DPH}^\top)^\top (\mathsf{DP}^*\mathsf{D}^\top + \mathsf{Q})^{-1} \\
        &= - \mathsf{K}^\top \mathsf{D}^\top (\mathsf{D P}^* \mathsf{D}^\top + \mathsf{Q})^{-1} \\
    S_c^{-1} &= - (\mathsf{DP}^* \mathsf{D}^\top + \mathsf{Q})^{-1}\mathsf{DK} \\
    S_d^{-1} &= (\mathsf{DP}^*\mathsf{D}^\top + \mathsf{Q})^{-1}\text{.}
\end{aligned}
\end{equation}
Further,
$\mathsf{K}_a$ and $\mathsf{K}_b$ in Eq.~\eqref{eq:K_block} can be reformulated as
\begin{equation}
\begin{aligned}
    \mathsf{K}_a &= \mathsf{PH}^\top(\mathsf{HPH}^\top + \mathsf{R})^{-1} + \mathsf{PH}^\top \mathsf{K}^\top \mathsf{D}^\top (\mathsf{DP}^*\mathsf{D}^\top + \mathsf{Q})^{-1}\mathsf{DK} - \mathsf{PD}^\top(\mathsf{DP}^*\mathsf{D}^\top + \mathsf{Q})^{-1}\mathsf{DK}\\
    &= \mathsf{K} - \mathsf{P}^* \mathsf{D}^\top (\mathsf{DP}^*\mathsf{D}^\top + \mathsf{Q})^{-1} \mathsf{DK} \\
    \mathsf{K}_b &= - \mathsf{PH}^\top \mathsf{K}^\top \mathsf{D}^\top (\mathsf{DP}^* \mathsf{D}^\top + \mathsf{Q})^{-1} + \mathsf{PD}^\top (\mathsf{DP}^*\mathsf{D}^\top + \mathsf{Q})^{-1} \\
    &= \mathsf{P}^* \mathsf{D}^\top (\mathsf{DP}^*\mathsf{D}^\top + \mathsf{Q})^{-1}
    \label{eq:KaKb}
\end{aligned}
\end{equation}
Finally, by substituting $\mathsf{K}_a$ and $\mathsf{K}_b$ in~\eqref{eq:analysis_step} with~\eqref{eq:KaKb}, we have
\begin{equation}
    \mathsf{x}^a = \mathsf{x}^f + \mathsf{K}(\oy_1 - \mathsf{Hx}^f) - \mathsf{P}^* \mathsf{D} (\mathsf{DP}^*\mathsf{D}^\top + \mathsf{Q})^{-1}\mathsf{DK}(\oy_1 - \mathsf{Hx}^f) + \mathsf{P}^* \mathsf{D}^\top (\mathsf{DP}^*\mathsf{D}^\top + \mathsf{Q})^{-1}(\oy_2 - \mathsf{Dx}^f) 
\end{equation}
By setting
\begin{equation}
    \tilde{\mathsf{x}}^f = \mathsf{x}^f +  \mathsf{K}(\oy_1 - \mathsf{Hx}^f) \text{,}
\end{equation}
we have 
\begin{equation}
\begin{aligned}
    \mathsf{x}^a &= \tilde{\mathsf{x}}^f + \mathsf{P}^* \mathsf{D}^\top (\mathsf{DP}^*\mathsf{D}^\top + \mathsf{Q})^{-1}(\oy_2 - \mathsf{Dx}^f - \mathsf{DK}(\oy_1 - \mathsf{Hx}^f)) \\
    &= \tilde{\mathsf{x}}^f + \mathsf{P}^* \mathsf{D}^\top (\mathsf{DP}^*\mathsf{D}^\top + \mathsf{Q})^{-1}(\oy_2 - \mathsf{Dx}^f - \mathsf{D}(\tilde{\sx}^f - \mathsf{x}^f)) \\
    &= \tilde{\mathsf{x}}^f + \mathsf{P}^* \mathsf{D}^\top (\mathsf{DP}^*\mathsf{D}^\top + \mathsf{Q})^{-1}(\oy_2 - \mathsf{D}\tilde{\sx}^f) \text{.}
\end{aligned}
\end{equation}
Conclusively, the analysis scheme~\eqref{eq:analysis_org} of conventional EnKF with two disparate data sources can be rewritten as
\begin{equation}
\begin{aligned}
    \tilde{\mathsf{x}}^\text{f} &= \mathsf{x}^\text{f} +  \mathsf{K}(\oy_1 - \mathsf{Hx}^\text{f})  \\
     \mathsf{x}^\text{a} &=  \tilde{\mathsf{x}}^\text{f} + \tilde{\mathsf{K}}(\oy_2 - \mathsf{D}\tilde{\sx}^\text{f}) \text{,}
\end{aligned}
\label{eq:EnKF_twostep}
\end{equation}
where the Kalman gain matrix~$\tilde{\mathsf{K}}$ in the second step is expressed as
\begin{equation}
    \tilde{\mathsf{K}} = \mathsf{P}^* \mathsf{D}^\top (\mathsf{DP}^* \mathsf{D}^\top + \mathsf{Q})^{-1} \text{.}
    \label{eq:tilde_kalman_gain_enkf}
\end{equation}
It can be seen that EnKF for assimilating two different data sources is equivalent to perform two standard EnKF steps sequentially.

Regarding the REnKF method, the analysis step is formulated as
\begin{subequations}
    \begin{align}
     \tilde{\sx}_j^\text{f} & = \sx_j^\text{f}  - \mathsf{P} \mathcal{G'}[\sx_j^\text{f}]^\top \mathsf{Q}^{-1} \mathcal{G}[\sx_j^\text{f}] \text{,}
    \label{eq:pre-correction-A} \\
    \sx_j^\text{a} & = \tilde{\sx}_j^\text{f} + \mathsf{K} (\oy_j - \mathsf{H} \tilde{\sx}_j^\text{f}) \text{.}
    \end{align}
    \label{eq:renkf_A}
\end{subequations}
To compare the update scheme of REnKF with EnKF, we reformulate the analysis step~\eqref{eq:renkf_A} of REnKF as a post-processing scheme as
\begin{equation}
\begin{aligned}
    \hat{\mathsf{x}}^\text{f} &= \mathsf{x}^\text{f} +  \mathsf{K}(\oy_1 - \mathsf{Hx}^\text{f}) \\
     \mathsf{x}^\text{a} &=  \hat{\mathsf{x}}^\text{f} + \mathsf{P}^* \mathsf{D}^\top  \mathsf{Q}^{-1}(\oy_2 - \mathsf{D}\hat{\sx}^\text{f}) \text{.}
\end{aligned}
\label{eq:renkf_post}
\end{equation}
The first step is equivalent to the conventional EnKF where only the observation data~$\oy_1$ is considered.
The second step is to analyze the updated state~$\hat{\sx}$ with observation data~$\oy_2$.
By comparing Eq.~\eqref{eq:EnKF_twostep} and Eq.~\eqref{eq:renkf_post},
the main difference between EnKF and REnKF is that the REnKF omits the $\mathsf{DP^*D}^\top$ in the inverse of Eq.~\eqref{eq:tilde_kalman_gain_enkf}.
In practice,
the iterative ensemble Kalman method usually leads to sample collapses when used for steady cases~\cite{zhang2020evaluation}, which means that the method gives a small $\mathsf{P}^*$.
For this reason, the omitting of the term related to $\mathsf{P}$ in the Kalman gain matrix of EnKF would not have significant effects on the results.
Hence, EnKF and REnKF are equivalent for disparate data assimilation with the practical implementation.

\section{Comparison of EnKF and REnKF for disparate data assimilation}
\added[id=R1]{
To order to assess the performance of the proposed method, we conduct the disparate data assimilation with the EnKF and REnKF, respectively. 
The comparison between EnKF and REnKF in terms of the $\text{Error}(\mathsf{Hx})$ and $\text{Error}(\mathsf{Dx})$ is shown in Table.~\ref{tab:summary_results_EnKF_REnKF}.
For the channel case, the disparate data are the friction velocity~$u_\tau$ and the sparse velocity~$U_1$; for the T3A plate case, the disparate data are the friction coefficient~$C_f$ and the sparse velocity~$U_1$; for the periodic hills case, the disparate data are the wall pressure~$p_w$ and the sparse velocity~$U_1$. 
It can be seen clearly that the REnKF method and the EnKF method achieve the similar results in the reconstructed flow field in the three test cases.
}
\begin{table}[!htbp]
    \centering
    \begin{tabular}{c|c|c|c}
    \hline
        Geometry & Filter & Error($\mathsf{Hx}$) & Error($\mathsf{Dx}$)  \\
        \hline
        Channel &  EnKF  & $1.38 \%$  & $1.60 \%$ \\
        & REnKF &  $1.41 \%$ & $0.93 \%$ \\
        \hline
         T3A plate &  EnKF &   $5.16\%$ & $48.9\%$ \\
        & REnKF & $6.33\%$ & $50.70\%$ \\
        \hline
       Periodic hills  &  EnKF & $7.31 \%$ & $8.28 \%$ \\
        & REnKF &  $7.24 \%$ & $9.26 \%$ \\
    \hline
    \end{tabular}
    \caption{Summary of data assimilation results with EnKF and REnKF. The error for $\mathsf{Hx}$ and $\mathsf{Dx}$ is computed based on~\eqref{eq:error_def}. $\mathsf{Hx}$ represents~$U_1$. $\mathsf{Dx}$ indicates~$u_\tau$ in channel case, $C_f$ in T3A plate case, and~$p_w$ in periodic hill case. 
    }
    \label{tab:summary_results_EnKF_REnKF}
\end{table}

\end{document}